\documentclass[12pt]{article}
\usepackage{graphicx}
\usepackage{lscape}
\usepackage{amssymb}
\usepackage{pazh}
\tightenlines

\def\*{$^{*}$}

\def\a{$^{\mbox{\scriptsize a}}$}
\def\b{$^{\mbox{\scriptsize b}}$}
\def\c{$^{\mbox{\scriptsize c}}$}
\def\d{$^{\mbox{\scriptsize d}}$}

\def\deg{$^\circ$}
\def\etal{{\it et~al.\,}}

\def\arcmin{$^{\prime\,}$}

\begin{document}

{\footnotesize Astronomy Letters, Vol. 30, No. 12, 2004, pp. 824-833. Translated from Pis'ma v Astronomicheskii
Zhurnal, Vol. 31, No. 11, 2005, pp. 729-747. Original Russian Text Copyright \copyright\, 2005 by Filippova,
Tsygankov, Lutovinov, Sunyaev.}

\title{\bf Hard Spectra of X-ray Pulsars from INTEGRAL Data}

\author{\bf \hspace{-1.3cm} \ \
E.V. Filippova\affilmark{1}$^{\,*}$, S.S. Tsygankov\affilmark{1}
A.A. Lutoviniv\affilmark{1}, R.A. Sunyaev\affilmark{1,2} }
                                                                                
\affil{
$^1$ {\it Space Research Institute, Russian Academy of Sciences, Profsoyuznaya ul. 84/32, Moscow 117810, Russia } \\
$^2$ {\it Max Planck Institut fur Astrophysik, Karl-Schwarzschild-Str. 1, Postfach 1317, D-85741 Garching, Germany}
}
 
\vspace{2mm}

{\bf Abstract\/}.We present spectra for 34 accretion-powered X-ray and one millisecond pulsars that were within
the field of view of the INTEGRAL observatory over two years (December 2002--January 2005) of its
in-orbit operation and that were detected by its instruments at a statistically significant level ($>$ 8$\sigma$ in
the energy range 18--60 keV). There are seven recently discovered objects of this class among the pulsars
studied: 2RXP~J130159.6-635806, IGR/AX~J16320-4751, IGR~J16358-4726, AX~J163904-4642,
IGR~J16465-4507, SAX/IGR~J18027-2017 and AX~J1841.0-0535. We have also obtained hard X-ray ($>$ 20 keV) spectra
for the accretion-powered pulsars A~0114+650, RX~J0146.9+6121,  AX~J1820.5-1434, AX~J1841.0-0535 
and the millisecond pulsar XTE~J1807-294 for the first time. We analyze the evolution of spectral
parameters as a function of the intensity of the sources and compare these with the results of previous
studies.
\copyright\,{\it 2004 MAIK "Nauka/Interperiodica"}.


\vfill
 
{$^{*}$ E-mail: kate@hea.iki.rssi.ru}

\newpage
\thispagestyle{empty}
\setcounter{page}{1}

\section*{INTRODUCTION}

    Accretion-powered X-ray pulsars were discovered
more than 30 years ago (Giacconi \etal 1971),
and some 100 such objects are known to date. A
qualitative understanding of the nature of the 
observed pulsed emission came almost at once (see,
e.g., Pringle and Rees 1972; Lamb \etal 1973). 
X-ray pulsars are rapidly rotating neutron stars with a
strong magnetic field ($B > 10^{11}$ G) that are members
of binary systems and that accrete matter from their
stellar companion. As the plasma approaches a 
neutron star, it is stopped by the pressure of the magnetic
field (which, to a first approximation, is in the form of
a dipole), is frozen into the latter, and moves along
the field lines toward the magnetic poles of the star
to produce two hot spots (at these locations, the
captured matter releases its gravitational energy in
the form of X-ray and gamma-ray radiation). If the
rotation axis of the neutron star does not coincide
with its magnetic axis (an oblique rotator), then these
spots will periodically cross the line of sight at a
certain orientation of the binary relative to a remote
observer and, thus, give rise to pulsed emission.

   X-ray pulsars are a unique laboratory for studying
the behavior of matter under extreme conditions --
at high temperatures and in strong magnetic and
gravitational fields. Analysis of their energy spectra
gives an idea of the physical processes in the emitting
region, the structure of the accreting matter, and
the parameters of the compact object. For example,
the detection of cyclotron lines in the spectrum of a
pulsar allows the magnetic field of a neutron star to be
measured (Gnedin and Sunyaev 1974). The magnetic
field was first measured by this method for the pulsar
Her X-1 (Truemper \etal 1978).

    Many papers devoted to various sources of this
class have been published over the period of research
on X-ray pulsars; here, we mention only a few 
review articles in which particular properties of these
objects were discussed. Such an attempt was first
made by White \etal (1983), who summarized the
spectra and pulse profiles for the then known 
X-ray pulsars and suggested an empirical model to 
describe their spectra. Subsequently, Nagase (1989)
gave an overview of accretion-powered pulsars 
using new Hakucho, TENMA, EXOSAT, and GINGA
observations. Based on data from the KVANT 
module, Gilfanov \etal (1989) analyzed the evolution
of the pulsation periods for pulsars. GRANAT 
(Lutovinov \etal 1994) and ComptonGRO (Bildsten et
al. 1997) data were used to investigate in detail the
pulse profiles and the evolution of the 
pulsation periods. Coburn \etal (2002) 
and Orlandini and Dal Fiume (2001) used, respectively, RXTE and BeppoSAX
data to analyze the pulsars whose spectra exhibited
the cyclotron lines attributable to the resonant 
scattering of photons by electrons in a magnetic field; in
certain cases, several harmonics were detected from
objects.

    Despite the long period of research, as yet there is
no convincing theoretical model that would describe
the spectra of accretion-powered X-ray pulsars. The
most popular model used to fit the spectra yields a
power-law spectral shape with an exponential cutoff
(models (1) and (2) in the Section "Data Analysis").
For most sources, the photon index, the cutoff energy,
and the e-folding energy lie within the ranges 0.3-2,
7-30 keV, and 9-40 keV, respectively. The sensitivity
of the IBIS instrument at these energies is well suited
for determining the e-folding energy in the spectrum.
The spectra also often exhibit the following: low-energy 
absorption that can be attributed both to the
interstellar medium and to internal absorption in the
binary system with the column density $N_{H}$ varying
within the range $10^{21}-10^{24} cm^{-2}$ ; 
iron emission lines at 6-7 keV; and gyrolines at the energies
corresponding to the electron cyclotron frequency and
its harmonics.
    The INTEGRAL observatory, whose main 
instruments have a high sensitivity and a large fields of view,
allows one not only to study known sources, but also
to discover new objects, including X-ray pulsars, and
to analyze their behavior over a wide energy range.
In this paper, we provide an overview of the X-ray
pulsars observed by the INTEGRAL observatory and
construct their broadband spectra. Some 70 known
and recently discovered pulsars were within the field
of view of the INTEGRAL instruments. For 35 of
these, we were able to reconstruct their energy 
spectra; the remaining sources were either not detected
or detected, but the data on these are not publicly
accessible at present (e.g., the pulsar 4U~0115+63
from which the INTEGRAL observatory detected an
outburst in August 2004; Lutovinov \etal 2004a).

\section*{OBSERVATIONS}

    The international gamma-ray observatory 
INTEGRAL (Winkler \etal 2003) was placed in orbit
by a Russian Proton launcher on October 17, 2002
(Eismont \etal 2003). It carries four instruments: the
SPI gamma-ray spectrometer, the IBIS gamma-ray
telescope, the JEM-X X-ray monitor, and the OMC
optical monitor, which allow the emission from 
astrophysical objects to be analyzed over a wide 
wavelength range. In this paper, we use data from the
ISGRI detector of the IBIS telescope (Ubertini et
al. 2003) and from the JEM-X monitor. Both 
instruments operate on the principle of a coded aperture.
The ISGRI effective energy range is 20--200 keV (the
energy resolution is $\sim7$\% at 100 keV), the field of view
is 29$^{\circ}\times$29$^{\circ}$  (the fully coded zone 
is 9$^{\circ}\times$9$^{\circ}$ ), and the     
nominal space resolution is $\sim12$\arcmin the angular size of
the mask element). A more detailed description of the
detector can be found in Lebrun \etal (2003). The
JEM-X monitor consists of two identical modules,
JEM-X1 and JEM-X2 (Lund \etal 2003). Each of
the modules of the telescope has the following 
technical characteristics: the energy range is \mbox{$3-35$} keV,
the field of view (the fully coded zone) is 13.2\deg\, (4.8\deg)
in diameter, the geometrical area of the detector is
500 cm$^2$ , and the nominal space resolution is $\sim3$ \arcmin.
   Here, we used the INTEGRAL observations from
orbit 23 (MJD 52629, December 21, 2002) to 
orbit 239 (MJD 53276, September 28, 2004); these
are the currently publicly accessible data and the
data of the Russian quota obtained as part of the
Galactic plane scanning (GPS), the Galactic center
deep exploration (GCDE), and the observations in
the General Program. Only the publicly accessible
observations of the X-ray pulsar V0332+53 that were
performed from orbit 272 (MJD 53376, January 6,
2005) to orbit 278 (MJD 53394, January 24, 2005)
constitute an exception.

\newpage

\section*{DATA ANALYSIS}

     For all of the detected X-ray pulsars, we 
constructed light curves in the energy range $18-60$ keV
and analyzed their variability. We constructed 
average spectra for persistent sources and analyzed the
dependence of the spectrum on the source's state
for pulsars with variable fluxes: if the spectrum did
not change, we also provided an average spectrum;
otherwise, we gave the spectra of different states.
The fluxes from the pulsars determined from these
light curves are given in mCrabs (1 mCrab=$1.36\times10^{-11}$ 
erg cm$^{-2}$ s$^{-1}$ in the energy range $18-60$ keV
under the assumption of a power-law spectrum for the
Crab Nebula with an index of 2.1 and a normalization
of 10 at 1 keV).
     The image reconstruction method and the spectral
analysis of the ISGRI/IBIS data used here were 
described by Revnivtsev \etal (2004) and Lutovinov et
al. (2003a). Analysis of a large number of calibration
observations for the Crab Nebula revealed that the
method yields a systematic error in measuring the
absolute flux from the source of  10\% over a wide
energy range and that the spectral shape is 
reconstructed with an accuracy up to $2-5$\%. To take into
account this peculiarity, we added a systematic error
of 5\% when analyzing the spectra in the XSPEC
package. As an example, Fig. 1 shows the energy
spectrum for the Crab Nebula reconstructed by this
method from the data of orbit 170. The response
matrix was constructed from the data of orbit 102. In
fitting this spectrum by a power law, we added a 
systematic error of 2\% and obtained the following 
parameters: foton index  $\Gamma=2.13\pm0.02$ and $Norm=11.27\pm0.35$. All
of the errors given here are statistical and correspond
to one standard deviation.

    The data of the Russian quota for the pulsar were
arbitrarily divided into two groups, before and after
orbit 200, because our ISGRI response matrix was
constructed from calibration measurements of the
Crab Nebula. After orbit 200, the flux (in counts)
from the Crab Nebula increased due to a change in
the ISGRI operating parameters; our response matrix
constructed from the data of orbit 239 changed 
accordingly. We reconstructed the spectra separately for
each data group and analyzed the variability of their
shape.
    For our spectral analysis of the JEM-X data,
we used the standard OSA 4.2 software package
provided by the INTEGRAL Science Data Center
(http://isdc.unige.ch). It should be borne in mind that
the JEM-X field of view is considerably smaller than
the IBIS one. Therefore, the effective exposure for
the observations of sources by this instrument is also
shorter and, in certain cases, no sources fell within
its field of view or the sensitivity of the instrument
was not enough for their detection at a statistically
significant level.
    Since the absolute fluxes from the sources are
reconstructed from the JEM-X data not quite 
accurately, the normalization of the JEM-X data was
left free when simultaneously fitting the JEM-X and
ISGRI spectra of the sources in the XSPEC package.
It is also worth mentioning that there are a number of
features near energies $5-7$ keV in the spectra 
reconstructed from the JEM-X data that are attributable to
the flaws in the current response matrix of the 
instrument. These features make it difficult to study in detail
a source's spectrum, in particular, to identify the iron
emission line and to determine its parameters.\footnote{Private 
correspondence with Carol Anne Oxborrow and Peter Kretschmar.}

   To fit the spectra, we used a standard (for pulsars)
empirical model that includes a power law with a
high-energy cutoff (White \etal 1983):

\begin{equation}
PLCUT(E)=AE^{-\Gamma}\times \left\{ \begin{array}{ll}
                       1 & \mbox{($E \leq E_{cut}$)}\\
                       exp^{-(E-E_{cut})/E_{fold}} & \mbox{($E > E_{cut}$),} \end{array} \right.
\end{equation}

where $\Gamma$  is the photon index,$E_{cut}$ and $E_{fold}$ are the
cutoff energy and the e-folding energy, respectively.
For several pulsars for which we failed to set a 
reasonable limit on the parameter $E_{cut}$ when fitting their
spectra by model (1), we used the following model:

\begin{equation}
CUTOFF(E)=AE^{-\Gamma}\times exp^{-E/E_{fold}}.
\end{equation}

   In certain cases, the standard model did not
describe the pulsar's spectral shape quite accurately.
Therefore, we introduced additional components
when fitting the spectrum:

 -- low-energy photoelectron absorption described
by the formula
\begin{equation}
WABS(E)=exp(-N_H\times\sigma(E)),
\end{equation}
 $\sigma(E)$ - is the cross sectionfor the process (Morrison and McCammon 1983);

 -- an iron emission line described by a Gaussian
profile
 
\begin{equation}
GAUS(E)=\frac{A}{\sqrt{(2\pi)\sigma_{Fe}}}\times exp(-((E-E_{Fe})/2\sigma_{Fe})^2),
\end{equation}
where $E_{Fe}$ is the line center, $\sigma_{Fe}$ is the line width;

 -- a resonance cyclotron absorption line whose
model is
  
\begin{equation}
CYCL(E)=\frac{exp(-\tau_{cycl}(\sigma_{cycl}/E_{cycl})^2)}{(E-E_{cycl})^2+\sigma_{cycl}^2},
\end{equation}
where $E_{cycl}$ is the line center, $\tau_{cycl}$ is the line depth, and 
$\sigma_{cycl}$ is the line width.

   To find features related to the resonance cyclotron
absorption line in the radiation spectrum for those
sources in which this feature was not observed 
previously, we added the corresponding component to the
model fit (see above). The energy of the line center
$E_{cycl}$ was varied over the range $20-90$ keV at 5 keV
steps, while the line width was fixed at 5 keV. The
choice of an energy range for the search of these lines
was dictated by the presence of systematic features
in the JEM-X response matrix and by the fact that
the standard X-ray range has been well studied using
data from other missions. We found the most probable
position of the possible cyclotron line and its 
significance in units of the standard deviation using the $\Delta\chi^2$
test.

\newpage

\section*{RESULTS}

    The X-ray pulsars that fell within the field of view
of the INTEGRAL observatory and that were 
detected at a statistically significant level by its 
instruments are listed in \mbox{Table 1}. This table also gives their
parameters, the parameters of the corresponding 
binaries, and references to the papers from which these
were taken ($P$ is the spin period of the neutron star,
$P_{orb}$ is the orbital period of the binary, HMXB stands
for a high-mass X-ray binary, and LMXB stands for
a low-mass X-ray binary). The spectra of the sources
are shown in Fig. 2, and the best-fit parameters for the
spectra are presented in Tables 2 and 3. Table 2 gives
the parameters for the standard model and lists the
exposures of the observations from which the average
spectrum was constructed and the fluxes from the
pulsars in the energy ranges $6-20$ and $18-60$ keV
determined by analyzing the spectra. Table 3 gives the
parameters of the models that describe the features in
the spectra of the pulsars, more specifically, the iron
emission line and the cyclotron absorption line and
its harmonics. Below, we briefly describe the results
obtained for each of the pulsars.

{\it A~0114+650}.
    The X-ray pulsar A~0114+650 is
one of the longest-period accretion-powered pulsars. 
The data from January through July 2003
(MJD 52653-52835) were used to analyze the
source. The source was not detected in the observations 
performed from MJD 52653 to MJD 52655; the
upper (1$\sigma$) limit on its $18-60$ keV flux was 1 mCrab.
In the remaining time, the mean flux in the same
energy range was  8 mCrab. The pulsar's average
spectrum was reconstructed from these observations.
Since JEM-X did not detect the source, we were able
to reconstruct its spectrum only in the hard X-ray
energy range. The pulsar's spectrum is typical of this
class of objects and is described in the soft X-ray
energy range by a power law with a high-energy cutoff
with the following parameters: $\Gamma\sim1.3$, $E_{cut}\sim8$ keV,
and $E_{fold}\sim20$ keV (Hall \etal 2000). Since the
IBIS exposure for the source was short, we failed to fit
the spectrum obtained by models (1) or (2); therefore,
it was fitted by a simple power law with an index of
2.3$\pm$0.4.

{\it SMC~X-1 (4U~0115-73)}. 
   The source SMC~X-1
was within the JEM-X/IBIS field of view during the
observations of the Small Magellanic Cloud region
from July 24 through July 27, 2003 (MJD 52843--
52846).

   We used the standard model (1) in our spectral
analysis. Fitting the spectrum by this model yielded
the parameters given in Table 2. Moon \etal (2003)
showed the constancy of the spectral parameters with
the source's flaring activity, which is also confirmed
by their agreement with our parameters. However,
note a slightly higher e-folding energy in the source's
spectrum obtained from the INTEGRAL data. Based
on our spectral analysis, we also obtained an upper
limit on the presence of a resonance cyclotron 
absorption line in the source's spectrum by the method
described in the section "Data Analysis". No such
feature was found at a confidence level higher than
$\sim1\sigma$.

{\it RX~J0146.9+6121}. 
   To analyze the source
RX~J0146.9+6121, we used the publicly accessible
INTEGRAL observations covering the period from
MJD 52636 to MJD 53018. Because of the low flux
from the source (in the energy range $18-60$ keV, it
was about 3 mCrab), the JEM-X monitor did not detect
the pulsar in this period. Since the source is weak, we
used a simple power law with the estimated photon
index $\Gamma=2.9^{+1.1}_{-0.8}$ to fit its spectrum. It should be
noted that the hard X-ray spectrum of the source has
not been analyzed up until now.

 {\it V~0332+53}. 
  As part of the Galactic plane scanning 
by the INTEGRAL observatory, its instruments 
detected an intense X-ray outburst from
the source V~0332+53 that began at the very end
of 2004 (Swank \etal 2004). In this paper, we 
analyze the observations performed by the INTEGRAL
observatory from January 6 through January 24,
2005, (MJD 53376--53394) with a total exposure
of 180 ks. In this period, the source was in a very
bright state, and its $18-60$ keV flux did not fall below
$\sim350$ mCrab.

    In our analysis, the model fit was chosen using 
the results of previous studies and consisted of
a power law with low-energy absorption (the hydrogen 
column density was taken from Makishima
\etal 1990), a high-energy cutoff, and three resonance 
cyclotron absorption lines ($E_{cycl1}=24.25$ keV, 
$E_{cycl2}=46.8$ keV, $E_{cycl3}=67.9$ keV). The derived
positions of the cyclotron lines are confirmed by the
analysis of RXTE data (Coburn \etal 2005) and
the first 100 ks of INTEGRAL observations of the
pulsar under study (Kreykenbohm \etal 2005).

    As the X-ray luminosity of the pulsar decreased,
we found a change in the shape of its radiation 
spectrum. Thus, for example, when the mean $3-100$ keV
luminosity of the source fell from $14.9\times10^{37}$ 
to $5.2\times10^{37}$ erg s$^{-1}$ 
,the photon index in the model used
slightly decreased (from $0.76\pm0.03$ to
$0.59\pm0.03$), but the cutoff parameters remained the same, within
the error limits. More detailed temporal and spectral
analyses will be performed in a special paper.

{\it 4U~0352+309 (X~Per)}.
    The pulsar was within
the field of view of the IBIS X-ray telescope during
the calibration observations of the Crab Nebula 
performed on August 14, 2003 (MJD 52865). The mean
$20-100$ keV flux from the source was $\sim40$ mCrab.
We used model (1) with the inclusion of a resonance
cyclotron absorption line to fit the pulsar's radiation
spectrum. The source's spectrum (Fig. 2) and the
best-fit parameters (Tables 2 and 3) were taken from
Lutovinov \etal (2004b).

{\it LMC~X-4 (4U~0532-664)}. 
  The X-ray pulsar
LMC~X-4 was observed by the INTEGRAL observatory 
in January 2003 as part of the General 
program. The total exposure was more than a million
seconds for both instruments of the observatory (IBIS
and JEM-X). These observations covered almost the
entire superorbital period ($\sim30.5$ days) related 
to accretion disk precession.

   Irrespective of the state in which the object under
study was, its radiation spectrum was constant, within 
the error limits, although all of the main parameters 
are slightly lower than those obtained by other
authors (see, e.g., La Barbera \etal 2001). Based on
INTEGRAL data, Tsygankov and Lutovinov (2005a)
studied the spectral properties of the pulsar LMC X-4
in more detail.

{\it A~0535+260}.
   The pulsar A~0535+260 is a transient 
source. On October 26, 2003 (MJD 52938), the
INTEGRAL observatory detected an increase in its
$18-60$ keV flux to  $\sim10$ mCrab compared to the upper
limit of  $\sim2$ mCrab during previous observations. On
October 28, 2003, (MJD 52940), the flux reached
 $\sim40$ mCrab in the same energy range. The 
subsequent observations of the source were interrupted due
to solar flaring activity. Because of the short IBIS
exposure for the pulsar, we constructed an average
spectrum from all observations for this period during
the outburst. The JEM-X monitor did not detect the
source.

    It follows from previous studies that the pulsar's
spectrum is one of the hardest and may extend to
 $\sim200$ keV (Maisack \etal 1997). The INTEGRAL
observatory detected the source at a statistically 
significant level up to 50 keV, which can be explained by
its short exposure. The spectrum obtained was fitted
by a power law with an index of $2.81\pm0.38$, with the
reduced $\chi^2=0.6(5)$. Kendziorra \etal (1994) fitted
the source's spectrum in the energy range $3-200$ keV
by model (1) with the following parameters: $\Gamma\sim1.2$, $E_{cut}\sim24$,
 and $E_{fold}\sim20$ keV. We also fitted the
spectrum by model (1) by fixing the photon index
and the cutoff energy; the derived e-folding energy
was found to be a factor of 1.5 lower than the value
given in the above paper. For this model, the reduced
$\chi^2=0.07(5)$.

{\it Vela~X-1 (4U~0900-403)}. 
   The pulsar Vela~X-1
was regularly within the field of view of the 
instruments during the Galactic plane survey and during
the observations of the region near the source as
part of the General program. Preliminary results of
the source's study using INTEGRAL data were 
presented by Kretschmar \etal (2004). The 
observations from January through July 2003 (MJD 52644--52832) 
were used in this paper. The flux from the
pulsar is highly variable and subjected to orbital 
modulations. A flux of $\sim7$ mCrab was detected from the
pulsar at a statistically significant level during 
X-ray eclipse (much as was observed by the GRANAT
observatory; Lutovinov \etal 2000), and it reached
700 mCrab in the energy range $18-60$ keV at the
maximum. We constructed the pulsar's spectra for
the eclipse phase and for various fluxes. Since the
source was not detected by the JEM-X instrument
during eclipses, we were able to construct its 
spectrum at these times only in the hard X-ray energy
range. We fitted it by a simple power law with an index
of 3.1$\pm$0.3.

    Since our analysis of the spectra outside eclipse
revealed no marked differences in their shape, we give
the pulsar's average spectrum. A cyclotron line at
 $\sim24$ keV and its harmonic at  $\sim50$ keV were detected
in the source (Table 3), in agreement with previous
results (see, e.g., Coburn \etal 2002). A prominent
feature that can be described by a model iron emission
line is also observed near an energy of $6-7$ keV. 
However, given the peculiarities of the JEM-X response
matrix (see above), we treat this interpretation with
great caution.

{\it Cen~X-3 (3U~1118-60)} 
  We used the pointings from 
January through July 2003 (MJD 52668--52832) 
obtained both during the Galactic plane scanning 
and as part of the General program to analyze 
the pulsar's spectrum. Using the known orbital
parameters of the binary (Burderi \etal 2000), we
determined the orbital phases for our observations
and analyzed the emission from the source during
and outside X-ray eclipse. The source was not 
detected during eclipse; the upper 1$\sigma$ limit on 
its $18-60$ keV flux was 2.5 mCrab. Outside eclipse, the mean
flux from the pulsar was $\sim17$ mCrab in the same
energy range. Several outbursts during which the
flux reached  $\sim90$ mCrab were detected against this
background.

    We constructed the pulsar's radiation spectrum
averaged over all outbursts and an average persistent
spectrum for the source outside eclipse. Our study
showed that the spectrum becomes softer during 
outbursts: the photon index increases from 0.87 to 1.16
(both spectra are shown in Fig. 2). In fitting the
spectra, we attempted to introduce a component that
describes the iron emission line in the model. Since
we failed to do this properly due to the difficulties in
reconstructing the spectra from JEM-X data 
mentioned above, we described the source's spectrum in
the standard X-ray energy range by a simple power
law. Otherwise, our best-fit parameters for the 
spectrum (Table 2) are in good agreement with the values
obtained by Burderi \etal (2000) and La Barbera \etal (2004).

{\it 4U~1145-619,  1E~1145.1-614 }. 
 When analyzing
the emission from 4U~1145-619, White \etal (1978)
found pulsations from this pulsar at two close 
frequencies. This puzzle was solved using data from the
Einstein observatory, whose instruments detected the
 second source, 1E~1145.1-614, at less than 20\arcmin from
the first source (Lamb \etal 1980).

    The pulsar 4U~1145-619 is a transient from which
regular outbursts with a duration of  10 days at 
intervals of  186.5 days are observed; this is assumed
to be the orbital period in the binary. At the end of May
2003 (MJD 52788), the INTEGRAL observatory 
detected an outburst from it during which 
the mean $18-60$ keV flux was  $\sim26$ mCrab. We used the available
data in this time interval to construct the pulsar's
average spectrum. In the remaining time, the upper
(1$\sigma$ ) limit on the flux from the pulsar was 4 mCrab in
the same energy range.

    The study of 1E~1145.1-614 is severely 
complicated by the proximity of its twin, which becomes
much brighter during outbursts, and the standard
collimator X-ray instruments do not allow these
sources to be observed separately. The pulsar could be
studied in more detail after its discovery only several
years later using ART-P (Grebenev \etal 1992)
and RXTE data, when its twin was in quiescence
(Ray and Chakrabarty 2002). In our paper, we
used the observations from March 2003 through
September 2004 (MJD 52710--53276) to analyze
1E~1145.1-614. In this period, its flux was constant
and had a mean value of $\sim20$ mCrab, except the
following times: 52795 MJD, when the flux from the
pulsar increased to  $\sim100$ mCrab; 53196 MJD, when
the source flared up again (Bodaghee \etal 2004);
and the outburst time of the pulsar 4U~1145-619,
when the mean flux from 1E~1145.1-614 rose to
40 mCrab. However, we found no clear correlations
of the flux rises between the two sources; therefore,
we concluded that these events were independent.
We constructed the source's spectra during and
outside these outbursts. The source was not within
the JEM-X field of view during the outbursts and
was not detected by the instrument outside these,
except the outburst period of 4U~1145-619. Since
no clear differences were found in the spectral shape
of 1E~1145.1-614, we give here only an average
spectrum for all observations. The derived best-fit
parameters for the average spectrum of 
1E~1145.1-614 are in good agreement with the values from Ray
and Chakrabarty (2002), who analyzed the source's
spectrum using RXTE data, except the photon index,
which proved to be slightly smaller. The 
parameter $N_{H}$ was fixed at the value given in the paper
mentioned above.

    During the outbursts in 1984 and 1985, the
mean photon index for the pulsar 4U~1145-619 was
equal to one, the low-energy absorption changed
from 26$\times10^{22}$ sm$^{-2}$ to 3.1$\times10^{22}$
sm$^{-2}$, the cutoff energy
remained constant $\sim$6 keV, and the e-folding energy
rose from $\sim$12 keV during the 1984 outburst to
$\sim$17 keV during the 1985 outburst (Cook and 
Warwick 1987). Our analysis showed that the pulsar's
spectrum became softer, $\Gamma=1.5\pm0.1$, while the 
e-folding energy increased to 30$\pm4$ keV. The sensitivity
of the JEM-X detector is too low to determine the
low-energy absorption, while its spread does not
allow it to be fixed at a particular value; therefore,
we did not introduce this component in the standard
model when fitting the spectrum.

{\it GX~301-2 (3A~1223-624)}. 
   To analyze the
source, we used the publicly accessible INTEGRAL
observational data obtained from January through
July 2003. Over this period, the IBIS telescope made
about 250 pointings at the object under study, which
included two times close to the periastron passage by
the neutron star. Therefore, all of the data obtained
were arbitrarily divided into low and high (periastron
passage) states. For the high state, we had some
10 IBIS pointings at our disposal (we took into
consideration the pointings for which the orbital
phase was within the range from 0.87 to 0.92), while
for the low state, we were able to reconstruct the
broadband spectrum using also 23 JEM-X pointings.
A preliminary analysis based on the first several
IBIS pointings was performed by Kreykenbohm et
al. (2004).

    Tsygankov \etal (2004) showed that the shape of
the source's radiation spectrum and its hardness were
subjected to random variations on a time scale of the
order of several thousand seconds. In this paper, we
failed to analyze these variations due to the lack of
JEM-X data in the high state and insufficient 
statistics in the low state. In our spectral analysis of the low
state for the pulsar GX~301-2, we corrected 
significantly the model fit (1): low-energy absorption (many
authors pointed out a significant hydrogen column
density that strongly depends on the object's orbital
phase and that reaches  $\sim2\times10^{24}$ atoms cm$^{-2}$ (Endo
\etal 2002), an iron line, and a resonance cyclotron
absorption line were added to the power law with a
high-energy cutoff. The latter feature whose existence
was mentioned by various authors (see, e.g., 
Orlandini \etal 2000; Coburn \etal 2002) has a confidence
level higher than  $3\sigma$ and improves significantly the
quality of the fit. In our case, for the low state (the
mean $3-100$ keV flux from the source 
was $1.9\times10^{-9}$ erg cm$^{-2}$ s$^{-1}$ ), 
we obtained the energy $E_{cyc}=47.4\pm1.9$ keV 
for the cyclotron absorption line that
is closest to the value given by Orlandini \etal (2000)
(the line width was fixed at the value taken from this
paper). Within the error limits, our results are also in
good agreement with the results of other authors.

    As was mentioned above, we had only the IBIS
data at our disposal for the high state. Nevertheless,
the quality of the spectrum allowed us to detect a
statistically significant change in the photon index,
which slightly increased compared to the low state
(Table 2), but was slightly lower than that in Borkus
\etal (1998), while the remaining model parameters
were in good agreement. When this parameter is fixed
at 0.3 obtained in the low state, the quality of the
fit deteriorates sharply. In this case, we also added
the resonance cyclotron absorption line at energy 
$\sim49$ keV whose confidence level was about 
2$\sigma$ (Table 3) to the standard model (1).

{\it 2RXP~130159.6-635806}.
 Hard X-ray emission from the region 
of the sky containing this object
was detected by the INTEGRAL observatory 
during its outburst in late January -- early February 2004
(Chernyakova \etal 2004), when the $20-60$ keV flux
from the source reached $\sim15$ mCrab. Subsequently,
XMM-Newton data revealed X-ray pulsations from
it with a period of $\sim$700 s. The spectral and temporal
properties of the new pulsar were analyzed in detail by
Chernyakova \etal (2005) using both current XMM-Newton 
and INTEGRAL data and archival ASCA
and BeppoSAX data. The source's spectrum and its
best-fit parameters over a wide energy range 
($1-70$ keV) presented in Fig. 2 and Table 2, respectively,
were taken from the same paper.

{\it 4U~1538-522}. 
The pulsar 4U~1538-522 is a
persistent source. Over the period of our 
observations from February 2003 through September 2004
(MJD 52671--53260), its mean $18-60$ keV flux was
 $\sim15$ mCrab. We constructed the pulsar's average
spectrum from all of the available data in the energy
range $4-80$ keV. The derived best-fit parameters for
the spectrum are in good agreement with the values
from Robba \etal (2001) ($N_{H}$ was taken from the
same paper), who analyzed the pulsar's spectrum
using BeppoSAX data.

{\it 4U~1626-67}.  
  To study the pulsar 4U~1626-67,
we used the observations from March through 
October 2003 (MJD 52699--52915) performed as part
of the Galactic plane scanning and the deep Galactic
center survey. The light curve of the source 
exhibited no statistically significant variations in the flux,
whose mean value was  $\sim12$ mCrab in the energy
range $18-60$ keV; therefore, we constructed an 
average spectrum of the pulsar. Since the source was
far from the Galactic plane ($b=-13.1$), it was not
within the JEM-X field of view, which did not allow us
to reconstruct its spectrum in the soft X-ray energy
range. Preliminary results of the pulsar's analysis
based on INTEGRAL data were presented by Denis
\etal (2004), who described the source's spectrum
by a simple power law with an index of 3.4. In our
paper, the exposure of the available observations is
a factor of 10 longer; therefore, we were able to fit
the spectrum by the standard model (1) by fixing
the photon index at the value given in Orlandini et
al. (1998), who analyzed the pulsar's spectrum over a
wide energy range ($0.1-100$ keV). Our values of $E_{cut}$
and $E_{fold}$ are in good agreement with those given in
the same paper.

{\it IGR/AX~J16320-4752, IGR~J16358-4726, AX~J163904-4642, IGR~J16465-4507}.
 These recently discovered long-period pulsars (with pulsation
periods of several hundred seconds) belong to the
class of strongly absorbed sources discovered by the
INTEGRAL observatory. These are not detected by
the JEM-X monitor because of significant absorption. 
Therefore, Lutovinov \etal (2005c) analyzed
their spectra over a wide energy range using data from
the RXTE, ASCA, and XMM-Newton observatories
in the standard X-ray energy range and from the
ISGRI detector at energies above 18 keV; We used
model (2) to fit the spectra. The spectra of the sources
(Fig. 2) and their best-fit parameters (Table 2) were
taken from Lutovinov \etal (2005c).

{\it OAO~1657-415}. 
  The intensity of the pulsar
OAO~1657-415 is subjected to strong orbital 
modulations and varies between a few mCrab (during 
X-ray eclipse) and $100-150$ mCrab in the energy range
$18-60$ keV. Since our study of the source's spectrum
at various orbital phases (the orbital parameters of
the binary were taken from Baykal (2000)) revealed
no marked differences in its shape, we constructed an
average spectrum from all of the available data from
March 2003 until April 2004 (MJD 52699--53097).
The pulsar's spectrum is one of the hardest spectra
for X-ray pulsars in binaries, extending to 100 keV
(see Table 2 and Fig. 2), and strong absorption with
$N_H\sim10^{23}$ cm$^{-2}$ is observed at low energies. The
pulsar will be analyzed in detail using INTEGRAL
and RXTE data in a special paper.

{\it EXO~1722-363}. 
  The flux from the pulsar EXO~1722-363 
is subjected to orbital modulations 
(Markwardt and Swank (2003); Lutovinov et
al. 2004b) and changed from a few to 50 mCrab in
the energy range $18-60$ keV over the period of our
observations from March 2003 through April 2004
(MJD 52698--53097). Based on INTEGRAL data,
Lutovinov \etal (2004b, 2004c) improved the 
localization accuracy of the object and showed that
the shape of the hard part of the pulsar's spectrum
remains constant, although the flux is highly variable,
while the shape of its soft part analyzed using RXTE
data strongly depends on the orbital phase of the
binary, and the photoabsorption can 
reach $N_H\simeq10^{24}$ cm$^{-2}$ .

    We failed to reconstruct the source's spectrum
from JEM-X data, since it was within the field of view
of this instrument only during states with low fluxes.
Our analysis showed that the pulsar is detected at
a statistically significant level up to $\sim60$ keV, and
its hard X-ray spectrum can be described either by a
power law with an index of 3.5 or by model (2), from
which it follows that the e-folding energy is $\sim15$ keV.
However, both models describe the source's spectrum
poorly, and further studies over a wide energy range
are required to improve its parameters.

{\it GX~1+4 (4U~1728-247)}.
 To analyze the radiation spectrum 
of the X-ray pulsar GX~1+4, we
used the currently publicly accessible data that cover
the interval from late February through late 
September 2003. The total exposure for the IBIS telescope
was  2400 ks; the source was within the JEM-X
field of view much more rarely.

    Our analysis showed that the spectral 
parameters of the pulsar's radiation depend on its flux.
When constructing average spectra for the object
under study, we took into account the pulsar's
intensity in the time interval where the averaging
was performed. Therefore, we identified three 
segments (MJD 52698--52700, MJD 52710--52753,
and MJD 52874--52910) in the light curve with mean
$18-60$ keV fluxes from the source of  $\sim130$, $\sim11$ and
$\sim6$ mCrab, respectively. In the latter case (low state),
the source's spectrum was best fitted by a simple
power law. Despite significant errors, we see from
Table 2 that as the intensity of the radiation from the
object under study decreases, its spectrum becomes
slightly softer, as confirmed by the studies of other
authors (Paul \etal 1995).
 
{\it IGR/SAX~J1802.7-2017}. 
 The INTEGRAL spectrum of the new 
X-ray pulsar IGR/SAX~J1802.7-2017 was 
analyzed by Lutovinov \etal (2005a), from
which we took its best-fit parameters.

{\it XTE~J1807-294}. 
 Formally, this source is not an
accretion-powered X-ray pulsar and belongs to the
class of millisecond pulsars. Campana \etal (2003)
showed that the source's XMM-Newton radiation
spectrum at low energies is best fitted by the 
combination of an absorbed blackbody and Comptonization
models or a power law without absorption or emission
lines.

    Because of its transient nature, the pulsar 
under study was detected by the IBIS telescope at a
statistically significant level only in the period 
between February 20 and May 1, 2003, 
(MJD 52690--52760), without being detected by the JEM-X 
monitor. It should be noted that this source is the only
millisecond pulsar in our sample, which is why we
pay special attention to choosing the model fit. To
analyze the spectral properties of the pulsar's 
radiation, we used models (1) and (2) and a simple
power law; for the composite models, the power-law
index was fixed at the value taken from Campana et
al. (2003). Based on $\Delta\chi^2$ statistics, we established
that model (1) with the parameters from Table 2 is
in best agreement with the experimental data. Such
high values of $E_{cut}\sim48$ keV and $E_{fold}\sim76$ keV are
not a rarity for flaring millisecond pulsars (see, e.g.,
Heindl and Smith 1998).

{\it AX~J1820.5-1434 }. 
 The pulsar AX~J1820.5-1434
was discovered in 1997 during the Galactic plane
scanning by the ASCA observatory (Kinugasa et
al. 1998); these authors also analyzed the source's
spectrum in the soft X-ray energy range. The 
INTEGRAL observatory first detected the source in the
hard X-ray energy range (Lutovinov \etal 2003b).
The available observations can be arbitrarily divided
into two groups: from March through April 2003
(MJD 52699--52759), during which the flux from
the source did not change appreciably and was 
$\sim8$ mCrab in the energy range $18-60$ keV, and from
September through October 2003 (MJD 52909--52929)
when the source was not detected at a statistically
significant level and the upper 1$\sigma$ limit on its flux was
0.3 mCrab in the same energy range. We constructed
an average spectrum from all of the data when the
source was detected and used model (1) to fit it;
the photon index was fixed at 0.9 given in Kinugasa
\etal (1998). We see from Fig. 2 that the pulsar
was detected at a statistically significant level up to
 $\sim70$ keV.

{\it AX~J1841.0-0535}.
The pulsar AX~J1841.0-0535, discovered  by the ASCA observatory
(Bamba \etal\, 2001), was detected during observations of 
the Sagittarius arm region perfomed by the 
INTEGRAL observatory in the spring of 2003. A preliminary analysis of 
the INTEGRAL data showed that the source position is differ 
from the position obtained by the ASCA observatory,
therefore  it was named as a new source 
IGR~J18406-0539 (Molkov \etal 2004a).  A new outburst from the same 
sky region was detected in November 2004. It was again assigned to a new source 
IGR~J18410-0535 at first (Rodriguez \etal 2004), but later it was  
identified with the source IGR~J18406-0539 (Hallpen \etal 2004).

 The following analysis showed, that in the spring of 2003 the source 
was significantly detected during 2 pointings, and its flux 
reached $\sim40$ mCrab in the $18-60$ kev energy range. 
The increase of the source flux up to $\sim10$ mCrab 
in the same 
energy range was also registered in October 2003. In the 
remaining time the upper 1$\sigma$ limit on its flux was $\sim$1 mCrab ($18-60$ keV). 
Here we present for the first time a hard source's spectrum, which was 
averaged over the source high states.
Because of the low statistics of the data we fitted 
the source's spectrum by a simple 
power-law with an index of $2.2\pm0.3$.

{\it GS~1843+009}. 
 An average spectrum for the
transient pulsar GS~1843+009 in the hard X-ray
energy range ($20-100$ keV ) was constructed from
the data obtained in early May 2003 
(MJD 52759--52760), when an outburst was detected from the
source (Cherepashchuk \etal 2003) and its 
$18-60$ keV flux during the outburst was  $\sim7$ mCrab. The
JEM-X monitor did not detect the pulsar.

   Because of the lack of data in the softer part of the
spectrum, when describing it by model (1), we fixed
the following parameters obtained from BeppoSAX
data over a wide energy range during the outburst
from the source in April 1997: $\Gamma=0.34$, 
and $E_{cut}=5.95$ keV (Piraino \etal 2000). The derived e-folding
energy $E_{fold}=17.4\pm1.4$ keV is in good agreement
with the value given in Piraino \etal (2000). This
may suggest that the shape of the source's radiation
spectrum is constant irrespective of its luminosity.

{\it A~1845-024}. 
   Soffitta \etal (1998) identified the
pulsar A~1845-024 with the sources GS~1843-02
and GRO~J1849-03. To analyze its spectrum, we
used the publicly accessible observations from March
through October 2003 (MJD 52699--52930). Over
this period, one outburst was detected from the pulsar
during which the $18-60$ keV flux reached  $\sim7$ mCrab;
the outburst began approximately on MJD 52728
and lasted for $\sim40$ days. We constructed the pulsar's
average spectrum for this period from ISGRI data; the
JEM-X instrument did not detect the source. In the
remaining time, the upper 1$\sigma$ limit on the flux from the
source was 0.4 mCrab in the same energy range. The
pulsar's $18-90$ keV spectrum was fitted by a simple
power law with a photon index of $\Gamma=2.62\pm0.19$, in
agreement with that obtained by Zhang \etal 1996,
whose used ComptonGRO data.

{\it XTE~J1855-026}. 
  For the pulsar XTE~J1855-026,
we were able to construct a broadband spectrum in
the energy range $4-100$ keV using the 9-ks-averaged
JEM-X data obtained on October 18, 2003, and the
IBIS data averaged over all of the available 
observations (March 2003--April 2004).

    We used model (1) to fit the INTEGRAL 
data; because of the low flux from the source 
($2.68\times10^{-10}$ erg cm$^{-2}$ s$^{-1}$ in the energy range $3-100$ keV),
the quality of the soft ($<20$ keV ) X-ray spectrum did
not allow us to detect an iron line and low-energy 
absorption. It should be noted that there is a discrepancy
between our best-fit parameters and those given in
Corbet \etal 1999: according to the INTEGRAL
data, the spectrum is slightly softer, while the 
e-folding energy is larger by about
10 keV.

{\it XTE~J1858+034}.
The pulsar XTE~J1858+034 is a transient source, which
was discovered during outburst in 1998 by RXTE observatory 
(Remillard \etal 1998). INTEGRAL observations of the Sagittarius arm region 
(MJD 53116 -- 53128) allowed to detect a new outburst from the source and 
improve the accuracy of its localization (Molkov \etal, 2004b).

In the period MJD 53116 -- 53119 the mean $18 - 60$ keV flux from the source 
was about 6 mCrab then it increased and reached $\sim83$ mCrab 
on MJD 53128 in the same energy range. For lack of 
the INTEGRAL data, the further behaviour 
of the source was analyzed using data of the  ASM monitor of the 
RXTE observatory in the $1-12$ keV energy range,
which are available from  http://xte.mid.edu.
It was found that flux from the pulsar remained stable on the same level 
for $\sim$ 4 days, and then began to decrease.
 
 No significant dependence of the pulsar's spectrum shape 
on its flux was revealed in our analyses, 
therefore we constructed an average spectrum of the source. It was fitted by the model (1) 
with the absorption on the low energies; the following
 parameters were obtained: $N_H=(14.3\pm0.7)\times10^{22}$ sm$^{-2}$, $\Gamma=1.38\pm0.02,
E_{cut}=25.16\pm0.33$ keV, $E_{fold}=7.92\pm0.22$ keV.
Paul and Rao (1998) analyzed the pulsar's spectrum in the $2-50$ keV 
energy range using the RXTE observatory data, however they 
failed to obtain its meaningfully constrained best-fit parameters. 
Thus our results are the first  reliable measurements
of the source spectrum parametrs in the broad energy range.  
More detailed analysis of the pulsar behaviour 
during the outburst in 2004 
will be perfomed in a separate paper.

{\it X~1901+03}. 
 The X-ray pulsar X~1901+03 was
observed by the INTEGRAL observatory during its
outburst in the spring of 2003 (Galloway \etal 2003).
The spectrum and its best-fit parameters based on
model (1) were taken from Molkov \etal (2003),
who performed spectral and temporal analyses of the
pulsar's behavior.

{\it 4U~1907+097}. 
   To construct the spectrum for the
pulsar 4U 1907+097, we used the publicly accessible
observational data from March through May 2003
(MJD 52705--52762 MJD). The mean $18-60$ keV
flux from the source in this period was $\sim20$ mCrab;
however, we observed episodes with a duration of
$\sim1$ day when the flux dropped by a factor of 2 and
detected one outburst during which the flux doubled
compared to its mean value.

    Roberts \etal (2001) analyzed the source's soft
X-ray spectrum in detail using ASCA data, which
were fitted by a simple power law with low-energy
absorption, and RXTE data in the range $2.5-20$ keV,
which were fitted by model (2). This analysis showed
that the best-fit parameters for the pulsar's spectrum
change only slightly with flux, except the absorption,
which changes from  $\sim2\times10^{22}$sm$^{-2}$ 
to  $\sim8\times10^{22}$ cm$^{-2}$
throughout the orbital cycle.

    We constructed the source's average spectrum
from all of the available observations and fitted it by
model (1). Our photon index agrees with the values
given in Roberts \etal (2001) and Cusumano et
al. (1998) (in this paper, the source's spectrum was
analyzed over a wide energy range using BeppoSAX
data and was fitted by model (1)). At the same time,
the e-folding energy proved to be a factor of 1.7 lower
than the values given in these papers.

   Cusumano \etal (1998) detected a cyclotron line
at 19 keV and its harmonic in the pulsar's spectrum. 
Our analysis did not reveal these features in the
source's spectrum.

{\it KS~1947+300}. 
  The transient X-ray pulsar
KS~1947+300 was within the IBIS/INTEGRAL field
of view from December 2002 until April 2004 $\sim$700
times, with the total exposure being $\sim$1.5 million s.
Because of its flaring activity, several states differing
in intensity for which independent spectral analyses
were performed using INTEGRAL and RXTE data
(Tsygankov and Lutovinov 2005b) can be identified in
the source's light curve.

    Here, we provide the INTEGRAL spectrum of
the source obtained on April 7, 2004, when it was
detected at a statistically significant level by both the
IBIS telescope and the JEM-X monitor. The pulsar's
spectrum (Fig. 2) and its best-fit parameters (Table 2)
were taken from Tsygankov and Lutovinov (2005b).

{\it EXO~2030+375 }. 
  The INTEGRAL observations
of the transient pulsar EXO~2030+375 prior to
MJD 52650 were analyzed in several papers (see,
e.g., Kuznetsov \etal 2004; Camero Arranz et
al. 2004; and references therein). Over the period
from MJD 52650 to MJD 52838, three outbursts
(MJD 52717, MJD 52761, MJD 52805) were detected 
from the source, during which the $18-60$ keV
flux reached  $\sim80$ mCrab. In the remaining time,
the upper 1$\sigma$ limit on its flux was 2 mCrab
in the same energy range. Because of the scarcity
of data, we failed to analyze the spectra for each
outburst separately; therefore, we constructed an
average spectrum for all outbursts. Our best-fit
parameters based on model (1) agree with those
given in Kuznetsov \etal (2004) and Reynolds et
al. (1993a) for the hard and soft X-ray energy ranges,
respectively.

{\it SAX~J2103.5+4545}. 
  The source
 SAX~J2103.5+4545 was within the field of view of the
INTEGRAL instruments during the calibration 
observations of the Cyg~ X-1 region in December 2002
(MJD 52629--52632 and MJD 52636--52637); a
detailed analysis of these observations can be found in
our previous paper (Filippova \etal 2004). The data
obtained later during the Galactic plane scanning
were analyzed by Sidoli \etal (2004). The pulsar's
spectrum shown in Fig. 2 and its best-fit parameters
(Table 2) were taken from Filippova \etal (2004).

\section*{CONCLUSIONS}
    We have presented a catalog of spectra for 34
accretion-powered X-ray pulsars and one millisecond 
pulsar that were observed by the INTEGRAL
observatory and that were detected by its instruments 
at a statistically significant level in the period
from MJD 52629 to MJD 53276. For 18 of the 35
sources, we were able to reconstruct their broadband
spectra. The sources under study include one 
millisecond pulsar, XTE~J1807-294, and seven  recently
discovered X-ray pulsars: 2RXP~J130159.6-635806,
IGR/AX~J16320-4751, IGR~J16358-4726,
AX~J163904-4642, IGR~J16465-4507, 
SAX/IGR~J18027-2017 and AX~J1841.0-0535. Hard X-ray spectra have
been obtained for the pulsars A~0114+650,
RX~J0146.9+6121,  AX~J1820.5-1434, AX~J1841.0-0535 
and XTE J1807-294 for the first time.

   For variable sources, we analyzed the flux 
dependence of the spectral shape. For example, 
the spectrum of the pulsar GX~1+4 becomes harder with
increasing intensity of the source. We also compared
our best-fit parameters with the results of previous
studies and discussed their evolution.

    A hard X-ray spectrum has been obtained for the
pulsar Vela X-1 for the first time during an eclipse of
the source by its optical companion. We were able to
reconstruct it only in the hard X-ray energy range,
since the JEM-X instrument did not detect the pulsar
at this time. The spectrum was described by a simple
power law with an index of 3.1.

   Cyclotron lines and their harmonics were detected
in the spectra of several pulsars: one harmonic in
4U~0352+309, one harmonic in both low and high
states in GX~301-2, two harmonics in Vela X-1, and
three harmonics in V~0332+53.

\newpage
\section*{ACKNOWLEDGMENTS}

    We thank E.M. Churazov, who developed the 
algorithms for IBIS data analysis and provided 
the software. We also thank M.R. Gilfanov 
and P.E. Shtykovskiy for a discussion 
of the results obtained. This
work was supported by the Russian Foundation for
Basic Research (project no. 04-02-17276). We are
grateful to the European INTEGRAL Science Data
Center (Versoix, Switzerland) and the Russian 
INTEGRAL Science Data Center (Moscow, Russia)
for the data. The work was performed in part during
visits to the European INTEGRAL Science Data
Center (Versoix, Switzerland); A.A. Lutovinov and
S.S. Tsygankov thank its staff for hospitality and
computer resources. A.A. Lutovinov thanks the ESA,
and S.S. Tsygankov thanks the Russian Academy of
Sciences (the "Nonstationary Phenomena in 
Astronomy" Program) for support of these visits. We are also
grateful to the INTEGRAL maintenance service and
the software developers for the JEM-X instrument,
namely, C.A. Oxborrow and P. Kretschmar for help
in interpreting the results obtained from the data of
this instrument. The results of this work are based on
observations of the INTEGRAL observatory, an ESA
project with the participation of Denmark, France,
Germany, Italy, Switzerland, Spain, Czechia, Poland,
Russia, and USA.

\newpage

\section*{REFERENCES}

\parindent=0mm

\begin{enumerate}

\item G. Augello, R. Iaria, N. Robba, \etal, Astrophys. J.
596, 63 (2003).

\item A.Bamba, J. Yokogama, M. Ueno \etal\,
Publ. Astron. Soc. Japan\, 53, 1179 (2001)

\item A. Baykal, Mon. Not. R. Astron. Soc. 313, 637
(2000).

\item A. Baykal, M. Stark, and J. Swank, Astrophys. J.
544, L129 (2000).

\item L. Bildsten, D. Chakrabarty, J. Chiu, \etal, Astrophys. J., Suppl. Ser. 113, 367 (1997).

\item A. Bodaghee, N. Mowlavi, and J. Ballet, Astron.
Telegram 290, 1 (2004).

\item V. V. Borkus, A.`S. Kaniovsky, R. A. Sunyaev, \etal,
Pis'ma Astron. Zh. 24, 83 (1998) [Astron. Lett. 24,
60 (1998)].

\item K. Borozdin, M. R. Gilfanov, R. A. Sunyaev, \etal,
Pis'ma Astron. Zh. 16, 804 (1990) [Sov. Astron.
Lett. 16, 345 (1990)].

\item L. Burderi, T. Di Salvo, N. Robba, \etal, Astrophys.
J. 530, 429 (2000).

\item A. Camero Arranz, P. Reig, P. Connell, \etal, in
Proceedings of the 5th INTEGRAL Workshop on                                            
"The INTEGRAL Univers", Ed. by V. Schonfelder et
al. (ESA Publ. Division, Noordwijk, 2004), SP-552,
p. 279.

\item S. Campana, M. Ravasio, G.L. Israel, \etal, Astrophys. J. 594, L39 (2003).

\item D. Chakrabarty, J. Grunsfeld, A. Thomas, \etal,
Astrophys. J. 403, 33 (1993).

\item D. Chakrabarty, L. Homer, P. Charles, \etal, Astrophys. J. 562, 985 (2001).

\item D. Chakrabarty, Z. Wang, A. Juett, \etal, Astrophys. J. 573, 789 (2002).

\item A. Cherepashchuk, S. Molkov, L. Foschini, \etal,
Astron. Telegram 159, 1 (2003).

\item M. Chernyakova, P. Shtykovskiy, A. Lutovinov, et
al., Astron. Telegram 251, 1 (2004).

\item M. Chernyakova, A. Lutovinov, J. Rodriguez, and
M. Revnivtsev, Mon. Not. R. Astro. Soc. (accepted), astro-ph/0508515 (2005).

\item G. Clark, Astrophys. J. 542, 131 (2000).
W. Coburn, W. Heindl, R. Rothschild, \etal, Astrophys. J. 580, 394 (2002).

\item W. Coburn, P. Kretschman, I. Kreykenbohm, \etal,
Astron. Telegram 381, 1 (2005).

\item M. Coe, B. Payne, A. Longmore, \etal, Mon. Not.
R. Astron. Soc. 232, 865 (1988).

\item M. Cook and R. Warwick, Mon. Not. R. Astron. Soc.
227, 661 (1987).

\item R. Corbet, F. Marshall, A. Peele, \etal, Astrophys.
J. 517, 956 (1999).

\item R. Corbet and K. Mukai, Astrophys. J. 577, 923
(2002).

\item D. Crampton, J. Hutchings, and A. Cowley, Astrophys. J. 299, 839 (1985).

\item G. Cusumano, T. di Salvo, L. Burderi, \etal, Astron.
Astrophys. 338, L79 (1998).

\item M. Denis, J. Grygorczuk, T. Bulik, \etal, in 
Proceedings of the 5th INTEGRAL Workshop on "The                                        
INTEGRAL Univers", Ed. by V. Schonfelder \etal
(ESA Publ. Division, Noordwijk, 2004), SP-552, p.
295.

\item N. Eismont, A. Ditrikh, G. Janin, \etal, Astron.
Astrophys. 411, L37 (2003).

\item T. Endo, M. Ishida, K. Masai, \etal, Astrophys. J.
574, 879 (2002).

\item E. V. Filippova, A. A.Lutovinov, P. E. Shtykovskiy,
\etal, Pis'ma Astron. Zh. 30, 905 (2004) [Astron.
Lett. 30, 824 (2004)].

\item M. Finger, L. Bildsten, D. Chakrabarty, \etal, Astrophys. J. 517, 449 (1999).

\item J. Finley, T. Belloni, and Cassinelli, Astron. Astrophys. 262, L25 (1992).

\item W. Forman, H. Tananbaum, and C. Jones, Astrophys. J. 206, 29 (1976).

\item D. Galloway, R. Remillard, E. Morgan, \etal, IAU
Circ. 8070, 2 (2003).

\item D. Galloway, E. Morgan, and A. Levine, Astrophys.
J. 613, 1164 (2004).

\item R. Giacconi, H. Gursky, E. Kellogg, \etal, Astrophys. J. 167, L67 (1971).

\item A. Giangrande, F. Giovannelli, C. Bartolini, \etal,
Astron. Astrophys,. Suppl. Ser. 40, 289 (1980).

\item M. R. Gilfanov, R. A. Sunyaev, E. M. Churazov, et
al., Pis'ma Astron. Zh. 15, 675 (1989) [Sov. Astron.
Lett. 15, 291 (1989)].

\item Yu. Gnedin and P. Sunyaev, Astron. Astrophys. 36,
379 (1974).

\item S. A. Grebenev, M. N. Pavlinsky, and R. A. Sunyaev,
Pis'ma Astron. Zh. 18, 570 (1992) [Sov. Astron.
Lett. 18, 228 (1992)].

\item T. Hall, J. Finley, R. Corbet, \etal, Astrophys. J.
536, 450 (2000).

\item J. Halpern, E. Gotthelf, D. Helfand \etal\,
Astron. Telegram\, 289, 1 (2004)

\item W. Heindl and D. Smith, Astrophys. J. 506, L35
(1998).

\item W. Hiltner, J. Werner, and P. Osmer, Astrophys. J.
175, 19 (1972).

\item M. Iye, Publ. Astron. Soc. Jpn. 38, 463 (1986).
S. Ilovaisky, C. Chevalier, and C. Motch, Astron.
Astrophys. 114, 7 (1982).

\item G. Israel, I. Negueruela, S. Campaha, \etal, Astron.
Astrophys. 371, 1018 (2001).

\item R. Kelley, J. Jernigan, A. Levine, \etal, Astrophys.
J. 264, 568 (1983).

\item R. Kelley, S. Rappaport, G. Clark, \etal, Astrophys.
J. 268, 790 (1983).

\item E. Kendziorra, P. Kretschmar, H. Pan, \etal, Astron.
Astrophys. 291, L31 (1994).

\item K. Kinugasa, K. Torii, Y. Hashimoto, \etal, Astrophys. J. 495, 435 (1998).

\item M. Kirsch, K. Mukerjee, M. Breitfellner, \etal, Astron. Astrophys. 423, 9 (2004).

\item K. Koyama, I. Asaoka, N. Ushimaru, \etal, Astrophys. J. 362, 215 (1990a).

\item K. Koyama, H. Kunieda, Y. Takeuchi, \etal, Publ.
Astron. Soc. Jpn. 42, 59 (1990b).

\item P. Kretschmar, R. Staubert, I. Kreykenbohm, et
al., in Proceedings of the 5th INTEGRAL Workshop 
on "The INTEGRAL Univers", Ed. by V.
Schonfelder \etal (ESA Publ. Division, Noordwijk,
2004), SP-552, p. 267.

\item I. Kreykenbohm, K. Pottschmidt, P. Kretschmar, et
al., in Proceedings of the 5th INTEGRAL Workshop 
on "The INTEGRAL Univers", Ed. by V.
Schonfelder \etal (ESA Publ. Division, Noordwijk,
2004), SP-552, p. 333.

\item I. Kreykenbohm, N. Mowlavi, N. Produit, \etal,
Astron. Astrophys. 433, L45 (2005).

\item W. Krzeminski, Astrophys. J. 192, 135 (1974).

\item S. Kuznetsov, M. Falanga, A. Goldwurm, \etal, in
Proceedings of the 5th INTEGRAL Workshop on  
"The INTEGRAL Univers", Ed. by V. Schonfelder et
al. (ESA Publ. Division, Noordwijk, 2004), SP-552,
p. 285.

\item A. La Barbera, L. Burderi, T. Di Salvo, \etal, 
Astrophys. J. 553, 375 (2001).

\item A. La Barbera, C. Ferrigno, S. Piraino, \etal, in
Proceedings of the 5th INTEGRAL Workshop on 
"The INTEGRAL Univers", Ed. by V. Schonfelder et
al. (ESA Publ. Division, Noordwijk, 2004), SP-552,
p. 337.

\item F. Lamb, C. Petchik, and D. Pines, Astrophys. J.
184, 271 (1973).

\item R. Lamb, T. Markert, R. Hartman, \etal, Astrophys.
J. 239, 651 (1980).

\item F. Lebrun, J.P. Leray, P. Lavocat, \etal, Astron.
Astrophys. 411, L141 (2003).

\item A. Levine, S. Rappoport, J. Deeter, \etal, Astrophys.
J. 410, 328 (1993).

\item W. Lewin, G. Ricker, and J. McClintock, Astrophys.
J. 169, L17 (1971).

\item F. Li, S. Rappaport, and A. Epstein, Nature 271, 37
(1978).

\item N. Lund, S. Brandt, C. Budtz-Joergesen, \etal,
Astron. Astrophys. 411, L231 (2003).

\item A. A. Lutovinov, S. A. Grebenev, R. A. Sunyaev, and
M. N. Pavlinsky, Pis'ma Astron. Zh. 20, 631 (1994)
[Astron. Lett. 20, 538 (1994)].

\item A. A. Lutovinov, S. A. Grebenev, M. N. Pavlinsky,
and R. A. Sunyaev, Pis'ma Astron. Zh. 26, 892
(2000) [Astron. Lett. 26, 765 (2000)].

\item A. A. Lutovinov, S. V. Molkov, and M. G. Revnivtsev,
Pis'ma Astron. Zh. 29, 803 (2003a) [Astron. Lett.
29, 713 (2003a)].

\item A. Lutovinov, R. Walter, G. Belanger, \etal, Astron.
Telegram 155, 1 (2003b).

\item A. Lutovinov, C. Budtz-Jorgensen, M. Turler, \etal,
Astron. Telegram 326, 1 (2004a).

\item A. Lutovinov, S. Tsygankov, M. Revnivtsev, \etal, in
Proceedings of the 5th INTEGRAL Workshop on
"The INTEGRAL Univers", Ed. by V. Schonfelder et
al. (ESA Publ. Division, Noordwijk, 2004b), SP-552,
p. 253.

\item A. Lutovinov, M. Revnivtsev, and S. Molkov, Astron.
Telegram 178, 1 (2004c).

\item A. Lutovinov, M. Revnivtsev, S. Molkov, and
R. Sunyaev, Astron. Astrophys. 430, 997 (2005a).

\item A. Lutovinov, J. Rodriguez, M. Revnivtsev, and
P. Shtykovskiy, Astron. Astrophys. 433, L41
(2005b).

\item A. Lutovinov, M. Revnivtsev, M. Gilfanov, et
al., Astron. Astrophys. (2005c) (in press); 
astro-ph/0411550.

\item M. Maisack, J. Grove, E. Kendziorra, \etal, Astron.
Astrophys. 325, 212 (1997).

\item K. Makishima, T. Mihara, M. Ishida, \etal, Astrophys. J. 365, L59 (1990).

\item C. Markwardt and J. Swank, Astron. Telegram 179,
1 (2003).

\item C. Markwardt, M. Juda, and Swank, IAU Circ.
8095, 2 (2003).

\item N. Marshall and M. Ricketts, Mon. Not. R. Astron.
Soc. 193, 7 (1980).

\item J. McClintok, H. Bradt, R. Doxsey, \etal, Nature
270, 320 (1977).

\item S. Mereghetti, A. Tiengo,G.L. Israel \etal\,
\aap\,  354, 567 (2000)

\item S. Molkov, A. Lutovinov, and S. Grebenev, Astron.
Astrophys. 411, 357 (2003).

\item S. Molkov, A. Cherepashchuk, A. Lutovinov  \etal\,
Astron. Lett.\, 30, 534, (2004a)

\item S. Molkov, A. Cherepashchuk, M. Revnivtsev, \etal,
Astron. Telegram 274, 1 (2004b).

\item R. Morrison and D. McCammon, Astrophys. J. 270,
119 (1983).

\item D.-S. Moon, S. Eikenberry, and I. Wasserman, Astrophys. J. 582, L91 (2003).

\item F. Nagase, Publ. Astron. Soc. Jpn. 41, 1 (1989).

\item I. Negueruela, P. Roche, J. Fabregat, \etal, Mon.
Not. R. Astron. Soc. 307, 695 (1999).

\item I. Negueruela, G. Israel, A. Marco, \etal, Astron.
Astrophys. 397, 739 (2003).

\item M. Orlandini, D. Dal Fiume, F. Frontera, \etal,
Astrophys. J. 500, 163 (1998).

\item M. Orlandini, D. Dal Fiume, F. Frontera, \etal, Adv.
Space Res. 25, 417 (2000).

\item M. Orlandini and D. Dal Fiume, in X-RAY ASTRONOMY: Stellar 
Endpoints, AGN, and the Diffuse X-ray Background, Ed. by N. E. White, G. Malaguti,
and G. G. C. Palumbo (Am. Inst. Phys., New York,
2001), AIP Conf. Proc. 599, 283 (2001).

\item G. Parkes, P. Murdin, and K. Mason, Mon. Not. R.
Astron. Soc. 184, 73 (1978).

\item G. Parkes, K. Mason, P. Murdin, \etal, Mon. Not.
R. Astron. Soc. 191, 547 (1980).

\item S. Patel, C. Kouveliotou, A. Tennant, \etal, Am.
Astron. Soc. Meet. 203, 3103 (2003).

\item B. Paul and A. Rao, Astron. Astrophys. 337, 815
(1998).

\item B. Paul, P. C. Agrawal, V. R. Chitnis, \etal, Bull.
Astron. Soc. India 23, 478 (1995).

\item M. Pereira, J. Braga, and F. Jablonski, Astrophys. J.
526, 105 (1999).

\item S. Piraino, A. Santangelo, A. Segreto, \etal, Astron.
Astrophys. 357, 501 (2000).

\item R. Price, D. Groves, R. Rodrigues, \etal, Astrophys.
J. 168, 7 (1971).

\item W. Priedhorsky and J. Terrell, Nature 303, 681 (1983).

\item J. Pringle and M. Rees, Astron. Astrophys. 21, 1
(1972).

\item P. Ray and D. Chakrabarty, Astrophys. J. 581, 1293
(2002).

\item P. Reig, D. Chakrabarty, M. Coe, \etal, Astron.
Astrophys. 311, 879 (1996).

\item M. G. Revnivtsev, R. A. Sunyaev, D. A. Varshalovich, \etal, Pis'ma Astron. Zh. 30, 430 (2004)
[Astron. Lett. 30, 382 (2004)].

\item A. Reynolds, A. Parmar, and W. White, Astrophys. J.
414, 302 (1993).

\item A. Reynolds, R. Hilditch, W. Bell, and G. Hill, Mon.
Not. R. Astron. Soc. 261, 337 (1993).

\item R. Remillard, A. Levine, T. Takeshima, \etal, IAU
Circ. 6826, 2 (1998).

\item N. Robba, L. Burderi, T. Di Salvo,\etal, Astrophys.
J. 562, 950 (2001).

\item M. Roberts, F. Michelson, D. Leahy, \etal, Astrophys. J. 555, 967 (2001).

\item J. Rodriguez, J. Tomsick, L. Foschini, \etal, 
Astron. Astrophys. 407, 41 (2003).

\item J. Rodriguez, A. Garau, S. Grebenev,
Astron.Telegram\,  340, 1 (2004)

\item F. Rosenberg, C. Eyles, G. Skinner, \etal, Nature
226, 628 (1975).

\item N. Sanduleak and A. Philip, IAU Circ. 3023, 1
(1976).

\item N. Sato, F. Nagase, N. Kawai, \etal, Astrophys. J.
304, 241 (1986).

\item D. Sharma, R. Sood, G. Strigfellow, \etal, Adv.
Space Res. 13, 375 (1993).

\item L. Sidoli, S. Mereghetti, S. Larsson, \etal, in Proceedings 
of the 5th INTEGRAL Workshop on "The INTEGRAL Univers", Ed. by V. Schonfelder \etal
(ESA Publ. Division, Noordwijk, 2004), SP-552, p.
475.

\item A. Slettebak, Astrophys. J., Suppl. Ser. 59, 769
(1985).

\item P. Soffitta, J. Tomsick, B. Harmon, \etal, Astrophys.
J. 494, 203 (1998).

\item L. Stella, N. White, J. Davelaar \etal\,
\apj\, 288, L45 (1985)

\item J. Stevens, P. Reig, M. Coe, \etal, 
Mon. Not. R. Astron. Soc. 288, 988 (1997).

\item M. Stollberg, M. Finger, R. Wilson, \etal, Astrophys. J. 512, 313 (1999).

\item J. Swank, R. Remillard, and E. Smith, Astron. Telegram 349, 1 (2004).

\item T. Takeshima, R. Corbet, F. Marshall  \etal\,
IAU Circ.\, 6826, 1 (1998)

\item J. Truemper, W. Pietsch, C. Reppin, \etal, Astrophys. J. 219, L105 (1978).

\item S. S. Tsygankov, A. A. Lutovinov, S. A. Grebenev, 
et al., Pis'ma Astron. Zh. 30, 596 (2004) [Astron. Lett.
30, 540 (2004)].

\item S. S. Tsygankov and A. A. Lutovinov, Pis'ma Astron. Zh. 31, 427 (2005a) [Astron. Lett. 31, 380
(2005a)].

\item S. S. Tsygankov and A. A. Lutovinov, Pis'ma Astron. Zh. 31, 99 (2005b) [Astron. Lett. 31, 88
(2005b)].

\item P. Ubertini, F. Lebrun, G. Di Cocco, \etal, Astrophys. J. 411, 131 (2003).

\item M. van Kerkwijk, J. van Oijen, and E. van den
Heuvel, Astron. Astrophys. 209, 173 (1989).

\item M. van Kerkwijk, J. van Paradijs, E. Zuiderwijk, et
al., Astron. Astrophys. 303, 483 (1995).

\item Walter \& INTEGRAL Survey Team, AAS/High Energy Astrophysics Division, 8 (2004).

\item R. Warwick, Watson, R. Willingale, Space Sci. Rev.
40, 429 (1985).

\item C. Winkler, T. J.-L. Courvoisier, G. Di Cocco, et
al., Astron. Astrophys. 411, L1 (2003).

\item N. White, K. Mason, P. Sanford, \etal, Mon. Not. R.
 Astron. Soc. 176, 91 (1976).

\item N. White, G. Parkes, P. Sanford \etal\,
Nature\,  274, 664 (1978).

\item  N. White, J. Swank, and S. Holt, Astrophys. J. 270,
 771 (1983).

\item  S. Zhang, B. Harmon, W. Paciesas, \etal, Astron.
 Astrophys. Suppl. Ser. 120, 227 (1996).
\end{enumerate}

\pagebreak

\newpage
\centering
{\bf Table 1. }{List of pulsars}
\begin{scriptsize}
\begin{tabular}{l|l|l|l|l|l}
\hline
\hline
                                                                                
Name& Binary type & $P$,s & $P_{orb}$, days & Companion type & References\\
&&&&&\\
\hline
A 0114+650 & HMXB& 10008& 11.6 &B1 Ia &[1][2][3]\\[1mm]
SMC X-1    &HMXB   &0.71   &3.89    &B0   &[4],[5],[6] \\[1mm]
RX J0146.9+6121 & HMXB &1408&--&B5IIIe&[7],[8] \\[1mm]
V 0332+53 & HMXB &4.4&34.25&O8-9Ve&[9],[10] \\[1mm]
4U 0352+309 & HMXB & 837 & -- & Be(XPer) & [11]\\ [1mm]
LMC X-4   &HMXB     &13.5   &1.4     &07 III-V   &[12],[13],[14] \\[1mm]
A 0535+260 & HMXB&103&111&O9.7 IIIe Be&[15],[16],[17]\\[1mm]
Vela X-1 & HMXB & 283 & 8.96 & B0.5Ib & [18],[19] \\[1mm]
CEN X-3 & HMXB & 4.82 & 2.1 & O6-8f &[20],[21] \\[1mm]
4U 1145-619 &HMXB & 292 & 187 & B1Vne&[22],[23] \\ [1mm]
1E 1145.1-6141 & HMXB & 297 & 14.365& B2Iae &[24],[25] \\[1mm]
GX 301-2  &HMXB  &680  &41.5    &Be  &[26],[27] \\[1mm]
2RXP J130159.6-635806& HMXB?& 704 & -- & -- &[28]\\[1mm]
4U 1538-52 & HMXB & 528 &3.7 & B0Iab &[29],[30] \\ [1mm]
4U 1626-67 & LMXB & 7.66 & 0.0289 &low-mass dwarf&[31],[32] \\[1mm]
IGR/AX J16320-4751 & HMXB & 1300 & -- & -- &[33],[34] \\ [1mm]
IGR J16358-4726 & HMXB & 5980 & -- & -- &[35],[36] \\[1mm]
AX J163904-4642 & HMXB &  900 & -- & -- &[37] \\[1mm]
IGR J16465-4507 & HMXB & 228 &--& -- &[35] \\[1mm]
OAO 1657-415 & HMXB & 37.7 & 10.4 & B0-6Iab &[38],[39] \\ [1mm]
EXO 1722-363 & HMXB  & 413 & 9.7 & Be? &[40],[41] \\ [1mm]
GX 1+4   &LMXB &115   &303.8   &M6III  &[42],[43],[44] \\[1mm]
SAX/IGR J18027-2017 & HMXB & 139 & 4.6 & --&[45],[46] \\[1mm]
XTE J1807-294 & LMXB &0.00525 &0.0278 &-- &[47],[48] \\[1mm]
AX J1820.5-1434 & HMXB & 152.3 & -- & Be? & [49] \\[1mm]
AX J1841.0-0535&HMXB&4.74&--& Be &[50],[51]\\[1mm]
GS 1843+009   &HMXB  &29.477   &-- &B0-B2 IV-Ve&[52],[53],[54] \\[1mm]
A 1845-024 & HMXB & 94.8 & 242 & -- & [55],[56],[57]\\[1mm]
XTE J1855-026 &HMXB&360.741  &6.067   &--    &[58],[59] \\[1mm]
XTE J1858+034 &HMXB&221& -- & --&[60]\\[1mm]
X 1901+031    &-- &2.763   &--  &--           &[61],[62]\\[1mm]
4U 1907+097 & HMXB & 438 & 8.38 & B I &[63],[64],[65] \\[1mm]
KS 1947+300  &HMXB &18.7  &40.415  &B0Ve &[66],[67],[68],[69] \\[1mm]
EXO 2030+375 & HMXB & 41.7 & 46 & Be &[70],[71] \\[1mm]
SAX J2103.5+4545 & HMXB & 355 & 12.68 & O-B&[72],[73] \\ [1mm]
                                                                              
\hline
\end{tabular}
\begin{list}{}
\item 
[1] Finley \etal (1992); [2] Crampton \etal (1985); [3] Reig \etal (1996); 
[4] Price \etal (1971); [5] Reynolds \etal (1993b); 
[6] Levine \etal (1993); [7] Meregetti \etal (2000); 
[8] Slettebak (1985); [9] Stella \etal (1985); 
[10] Negueruela \etal (1999); [11] White
\etal (1976); [12] Sanduleak and Philip (1976); 
[13] Li \etal (1978); [14] Kelley \etal (1983a); 
[15] Rosenberg \etal (1975); [16] Priedhorsky and Terrell (1983); 
[17] Giangrande \etal (1980); [18] van Kerkwijk \etal (1995); 
[19] Hiltner \etal (1972); [20] Kelley
\etal (1983b); [21] Krzeminski (1974); 
[22] Warwick \etal (1985); [23] Stevens \etal (1997); 
[24] Ray and Chakrabarty (2002); [25]
Ilovaisky \etal (1982); [26] Sato \etal (1986); 
[27] Parkes \etal (1980); [28] Chernyakova \etal (2005); 
[29] Clark (2000); [30] Parkes
\etal (1978); [31] Chakrabarty \etal (2001); 
[32] McClintok \etal (1977); [33] Lutovinov \etal (2005b); 
[34] Rodriguez \etal (2003);
[35] Lutovinov \etal (2005c); [36] Patel \etal (2003); 
[37] Walter \etal (2004); [38] Chakrabarty \etal (1993); 
[39] Chakrabarty \etal (2002); [40] Markwardt and Swank (2003); 
[41] Lutovinov \etal (2004b); [42] Lewin \etal (1971); 
[43] Pereira \etal (1999);
[44] Sharma \etal (1993); [45] Augello \etal (2003); 
[46] Lutovinov \etal (2005a); [47] Markwardt \etal (2003); 
[48] Kirsch \etal (2004); [49] Kinugasa \etal (1998); 
[50] Bamba \etal (2001); [51] Halpen \etal (2004); 
[52] Koyama \etal (1990a); [53] Israel
\etal (2001); [54] Piraino \etal (2000); 
[55] Koyama \etal (1990b); [56] Zhang \etal (1996); 
[57] Finger \etal (1999); [58] Corbet
\etal (1999); [59] Corbet \etal (2002); 
[60] Takeshima \etal (1998); [61] Forman \etal (1976); 
[62] Galloway \etal (2003); [63]
Marshall and Ricketts (1980); [64] van Kerkwijk \etal (1989); 
[65] Iye (1986); [66] Borozdin \etal (1990); [67] Galloway \etal (2004);
[68] Tsygankov \etal (2005b); [69] Negueruela \etal (2003); 
[70] Stollberg \etal (1999); [71] Coe \etal (1988); [72] Baykal et
al. (2000); [73] Filippova \etal (2004).

\end{list}
\end{scriptsize}

\newpage
\begin{landscape}
 
\begin{table}[h]

 \centering
{\bf  2.  }{ Exposure times, fluxes, and best-fit parameters 
for the spectra of the pulsars}

\vspace{2mm}
\begin{scriptsize}
\begin{tabular}{c|c|c|c|c|c|c|c|c|c}
\hline
\hline
 &\multicolumn{2}{c|}{Exposure,ks} &\multicolumn{2}{c|}{Flux, 10$^{-9}$erg sm$^{-2}$s$^{-1}$}&& &&&\\
\cline{2-5}
Name&&&&&$N_{H}$,10$^{22}$ sm$^{-2}$&Photon index, & $E_{cut}$,keV & $E_{fold}$, keV & $\chi^{2}$ \\
&JEM-X&IBIS&6--20 &18--60 &&$\Gamma$&&&\\
\hline
1&2&3&4&5&6&7&8&9&10\\
\hline
A 0114+650 & --&40.4 & --& 0.09& --& 2.3$\pm$0.4& --& --& 0.42(6)\\[1mm]
SMC X-1 &70&104&1.05&0.76& -- &1.48$\pm$0.02& 20.5$^{+1.0}_{-1.8}$&12.9$^{+0.6}_{-0.7}$&0.98(124) \\[1mm]
RX J0146.9+6121&--&250&--&0.03&--&$2.9^{+1.1}_{-0.8}$&--&--&0.31(3)\\[1mm]
V0332+53 &178&187.4&17.87&6.22&4\a&$0.77\pm0.02$&$24.3^{+0.5}_{-0.7}$&$14.0^{+0.5}_{-0.7}$&0.35(127)\\[1mm]
4U 0352+309 &--&50&--&0.56& -- &1.92$\pm0.19$&50$\pm16$&77$\pm27$& 0.36(9) \\[1mm]
LMC X-4 &93&176&0.79&0.78& --&0.2$\pm0.15$& 9.1$\pm0.8$& 11.0$\pm0.6$& 0.93(117) \\[1mm]
A 0535+26&--& 77& --& 0.24& --& 1.2\a& 24\a&13.8$^{+4.5}_{-3.2}$&0.07(5)\\[1mm]
Vela X-1(eclipse) &--&203.6 &--&0.1&--&3.1$\pm0.3$ &-- &--&0.83(7) \\[1mm]
Vela X-1(outside eclipse) &897.2&560&3.2&3.6&--&0.88$\pm0.01$ &25.5$\pm0.2$ &13.0$\pm0.1$&0.34(131) \\[1mm]
CEN X-3(quiescent state)&266.1&250&0.39&0.2& -- & 0.87$\pm0.06$ &16.4$\pm0.6$& 7.1$\pm0.2$ &1.5(120) \\[1mm]
CEN X-3(outbursts)&15&47&1.57&0.66& -- & 1.16$\pm0.04$ &15.3$\pm0.2$ & 7.8$\pm0.2$ &1.4(116) \\[1mm]
4U 1145-619 &11&77.3&0.39&0.33& -- & 1.5$\pm0.1$ & 6.7$\pm1.4$ & 30$\pm4$ & 1(142) \\[1mm]
1E 1145.1-614 &11&345.2&0.39&0.4& 3.3\a & 1.08$\pm$0.07 & 8$\pm$1 & 21.9$^{+1.8}_{-0.8}$ &0.98(139) \\[1mm]
GX 301-2 (high state)&--&31&--&6.04& -- &0.74$^{+0.32}_{-0.09}$&23.3$^{+0.3}_{-0.5}$& 8.3$\pm0.7$& 0.74(8) \\[1mm]
GX 301-2 (low state)&62&710&0.96&0.99& 10.6$\pm2.5$ &0.30$\pm0.06$&17.8$\pm0.2$&9.7$\pm0.7$&0.9(118) \\[1mm]
2RXP130159.6-635806&--&10.3&--&0.2\c&2.56\a&0.69\a&24.3$\pm$3.4&8.5$^{+0.2}_{-0.1}$&\d\\[1mm]
4U 1538-52 &23&894.4&0.46&0.2& 1.63\a & 1.37$\pm0.06$ &28.7$\pm0.8$ & 9.9$\pm0.7$ &0.94(119) \\[1mm]
4U 1626-67 &--&218.2&--&0.15& -- & 0.87\a & 23.9$^{+1.0}_{-1.4}$ &7$\pm$1 &1.25(5) \\[1mm]
IGR/AX J16320-4752\b&-- &1000&--&0.2\c&18\a & 0.7$\pm$0.2& -- &13$\pm$1&\d\\ [1mm]
IGR J16358-4726\b &-- &1000&--&0.04\c&40\&0.7$\pm$0.5& --& 16$\pm$5&\d\\[1mm]
AX J163904-4642\b &--&1000&--&0.06\c&58\& 1.3$\pm$1.0& -- & 11$\pm$1&\d\\[1mm]
IGR J16465-4507\b &--&1000&--&0.12\c&72\&1.0$\pm$0.5 & -- & 30\a&\d \\[1mm]
OAO 1657-415 &29&1663.7&0.8&1.03&15.2$^{+0.7}_{-1.4}$ & 1.57$\pm0.02$& 26.3$^{+0.7}_{-1.8}$ & 29.2$^{+1.2}_{-0.5}$ &0.73(119) \\[1mm]
EXO 1722-363 &--&2960.9&--&0.6&--&3.5\a& -- & -- &2.7(5) \\[1mm]
GX 1+4 (low state)&9&2315&0.08&0.07&--&2.24$^{+0.06}_{-0.12}$ &-- &-- &0.93(126) \\[1mm]
GX 1+4 (intermediate state)&3.5&385&0.11&0.14&--&1.54$^{+0.35}_{-0.22}$ &24.8$^{+5.8}_{-3.0}$ &47.0$^{+15.2}_{-10.7}$ &1.16(125) \\[1mm]
GX 1+4 (high state)&5&164&0.76&1.62&--&$0.93^{+0.12}_{-0.14}$&$25.1^{+1.1}_{-1.7}$&$30.4\pm2.4$&1.19(136) \\[1mm]
IGR/SAX J18027-2017\b &--&1274&--&0.06&--&0.1\a&--&$\sim$10& \\[1mm]

\hline
\end{tabular}
\end{scriptsize}
\end{table}
\end{landscape}

\newpage
\begin{landscape}
 
\begin{table}[h]

\begin{scriptsize}
\begin{tabular}{c|c|c|c|c|c|c|c|c|c}
\hline
\hline
1&2&3&4&5&6&7&8&9&10\\
\hline
XTE J1807-294&--&711&--&0.11& -- & 1.96\a& 48.1$^{+7.6}_{-9.9}$& 75.7$^{+58.1}_{-24.5}$& 0.92(7) \\[1mm]
AX J1820.5-1434 &--&2322.3&--&0.1& -- & 0.9\a & 25$\pm3$ & 17.0$\pm2.7$ &0.37(9) \\[1mm]
AX J1841.0-0535&--&77.19&--&0.11& -- &$2.2\pm0.3$& --&--&0.42(5)\\[1mm]
GS 1843+009&--&62&--&0.17& --&0.34\a&5.95\a& 17.4$\pm1.4$&1.2(8) \\[1mm]
A 1845-024&--&691.8&--&0.06& -- & 2.62$\pm0.19$ & -- &--&0.46(7)\\[1mm]
XTE J1855-026 &9&652&0.16&0.17& --&1.69$\pm0.23$&23.99$^{+2.88}_{-6.73}$&38.49$^{+10.35}_{-7.38}$&1.08(112)\\[1mm]
XTE J1858+034 &137&360&1.13&0.99&14.3$\pm0.7$&1.38$\pm$0.02&25.16$\pm0.33$&7.92$\pm$0.22&0.95(144)\\[1mm]
X 1901+031&150&330&6.1&1.8& -- & 2.035$\pm0.015$ & 11.27$\pm0.19$ & 13.22$\pm0.11$ & 0.82(127) \\[1mm]
4U 1907+097 &180&478.3&0.6&0.18&-- & 1.26$\pm0.07$ & 7.0$\pm0.3$ &9.0$^{+0.3}_{-0.6}$ &0.75(131) \\[1mm]
KS 1947+300&2&6&1.09&1.17&-- & 1.07$^{+0.24}_{-0.13}$&8.6$^{+3.4}_{-1.2}$& 23.6$^{+5.3}_{-2.3}$& 1.18(104) \\[1mm]
EXO 2030+375 &2&25.3&0.84&0.85&-- & 1.71$\pm0.09$ &25.2$^{+2.5}_{-3.7}$ & 33$^{+6}_{-4}$ &1.06(137) \\[1mm]
SAX J2103.5+4545 &33&196.6&0.38&0.38& 0.9\a& 1.04$\pm0.15$ & 8.5$\pm2.4$ & 21.37$\pm2.75$ &1.21(120) \\[1mm]

\hline
\end{tabular}

\begin{list}{}
\item \a  The parameter is fixed  
\item \b  The cutoffpl model was used to fit the spectrum
\item \c   In the energy range 20--60 keV
\item \d   Lutovinov \etal (2005c) fitted the pulsar's spectrum over a 
wide energy range 
\item together with data from other observatories (see the
text).

\end{list}
\end{scriptsize}
\end{table}
\end{landscape}

\newpage
 \centering
{\bf  3.  }{Other best-fit parameters for the spectra of the pulsars}

\vspace{2mm}
\begin{scriptsize}
\begin{tabular}{c|c|c|c|c|c|c}
\hline
\hline
Binary & Fe line & Fe line  & Fe line Intensity & E$_{cycl}$, keV & $\tau_{cycl}$ &$\sigma_{cycl}$, keV \\
&center, keV & width, keV &Fe, photons sm$^{-2}$ s$^{-1}$&&&\\
\hline
V0332+53\ &--&--&--&24.25$^{+0.07}_{-0.14}$ &1.98$^{+0.02}_{-0.04}$ 
&7.10$\pm$0.10\\[1mm]

4U 0352+309 & -- &--&--&28.8$\pm$2.5&0.33$\pm$0.12&9\ \\[1mm]
Vela X-1\ &6.64$\pm0.10$ & 0.31$\pm0.16$ &(4.7$\pm$0.8)$\times10^{-3}$&24.0$\pm0.3$&0.38$\pm$0.01&5.3$\pm0.5$ \\[1mm]
GX 301-2 (high)&--  &--&--&49.2$^{+4.2}_{-2.1}$&0.60$^{+0.13}_{-0.09}$&18\ \\[1mm]
GX 301-2 (low)& 6.54$^{+0.17}_{-0.11}$ &0.52$^{+0.22}_{-0.14}$&(2.54$\pm$1.02)$\times10^{-3}$ &47.4$^{+2.2}_{-1.1}$&0.87$\pm$0.17&18\ \\[1mm]

\hline
\end{tabular}
\begin{list}{}
\item \a Two more features were found in the pulsar spectrum, the second and third harmonics of the cyclotron line:
\item E$_{cycl2}$=$46.8^{+0.2}_{-0.1}$ , $\tau_{cycl2}$=1.94$^{+0.06}_{-0.07}$, $\sigma_{cycl2}$=8.9$\pm0.4$
\item E$_{cycl3}$=$67.9^{+3.2}_{-4.3}$ , $\tau_{cycl3}$=2.60$^{+0.25}_{-0.35}$, $\sigma_{cycl3}$=26.9$\pm$5.4
\item \b The parameter is fixed.
\item \c The second harmonic of the cyclotron line with
\item E$_{cycl2}$=50.2$\pm0.5$ , $\tau_{cycl2}$=0.95$\pm$0.03, $\sigma_{cycl2}$=12.2$\pm$0.5 keV
\item was detected in the pulsar.
\end{list}

\end{scriptsize}

\newpage
 
\begin{figure*}
\includegraphics[width=17cm,clip]{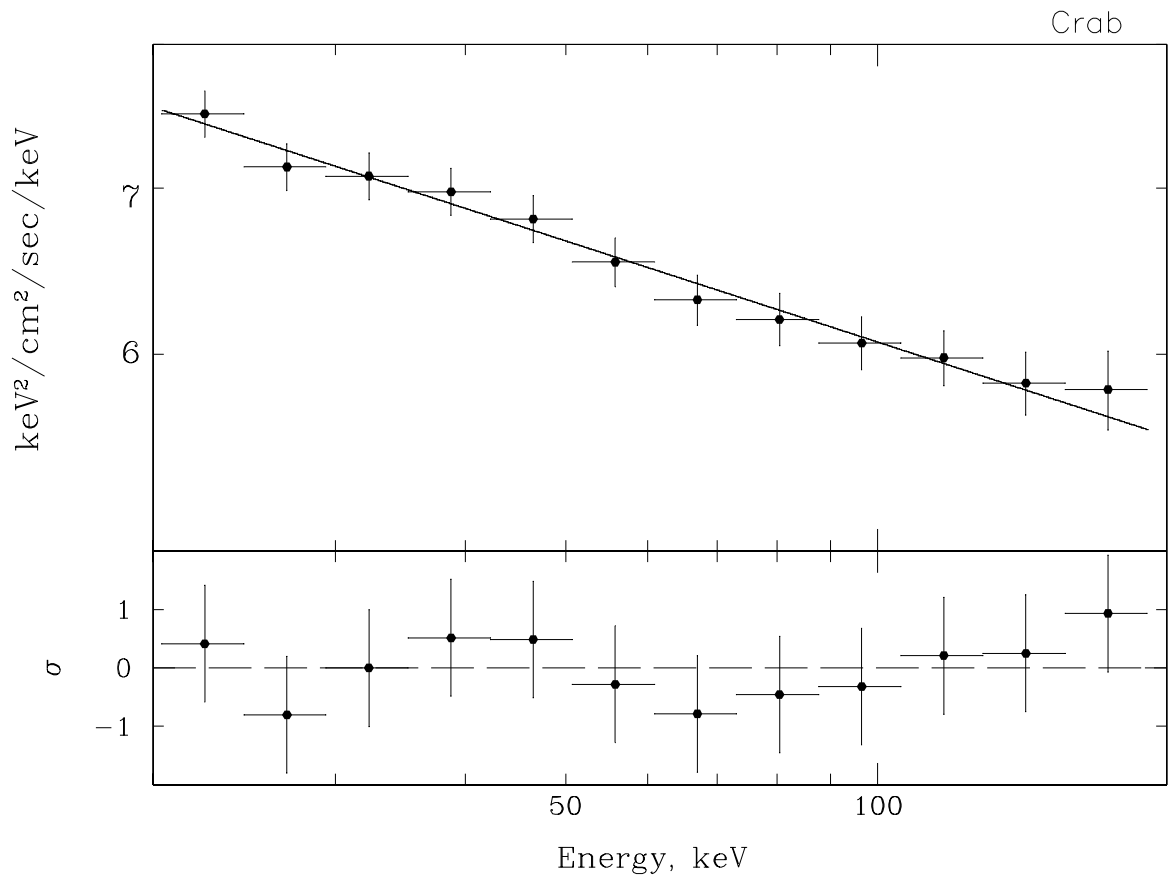}
\vfill
\renewcommand{\figurename}{Fig.}
\caption{Energy spectrum for the Crab Nebula. The
solid line represents the power-law best fit to the 
spectrum with the following parameters: $\Gamma=2.13\pm0.02$, 
$Norm=11.27\pm0.35$. The errors correspond to one
standard deviation.
}

\end{figure*}
\pagebreak

\newpage
 
\begin{figure*}
\hbox{
\includegraphics[width=0.5\columnwidth,bb=30 435 565 710,clip]{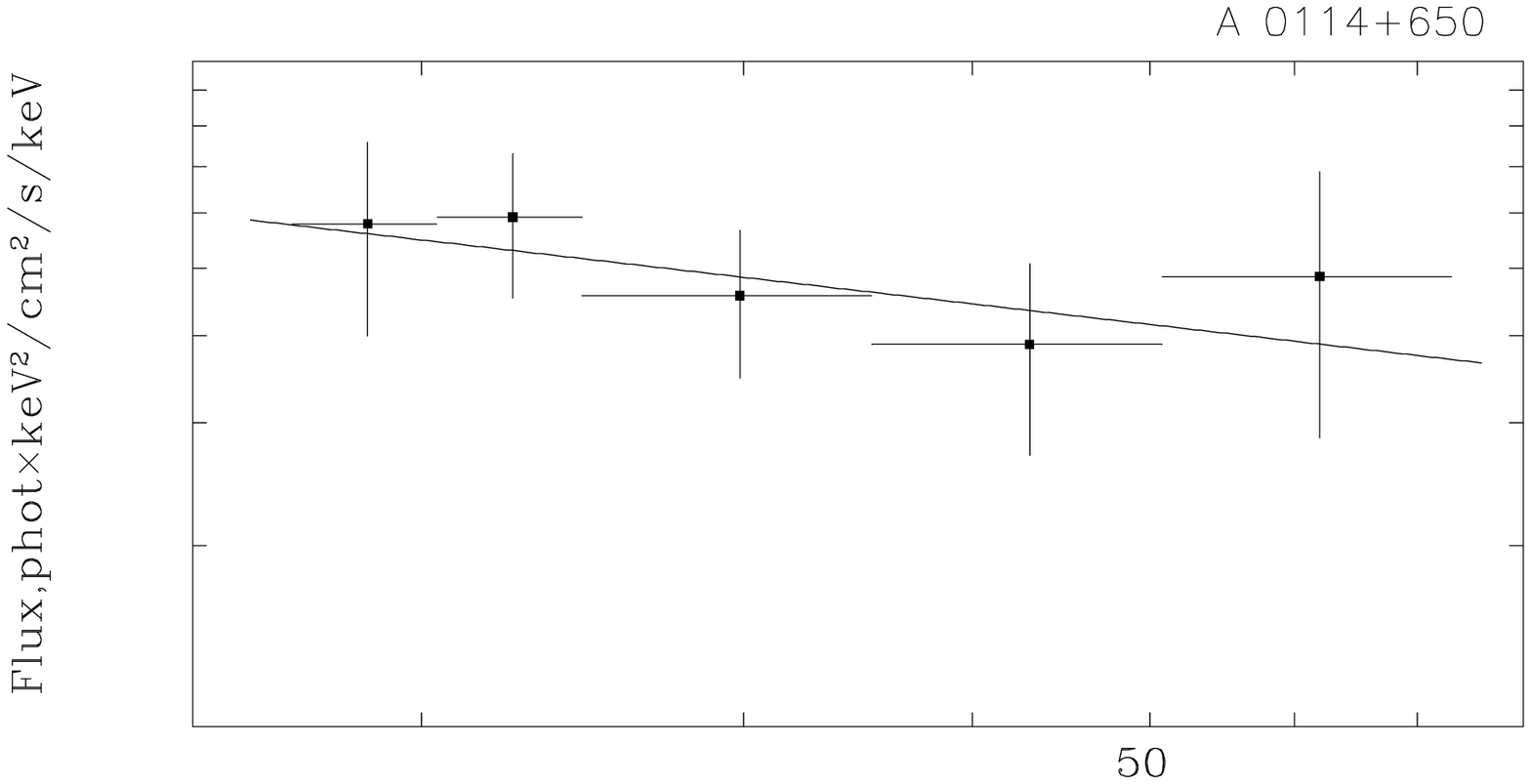} 
\includegraphics[width=0.5\columnwidth,bb=30 435 565 710,clip]{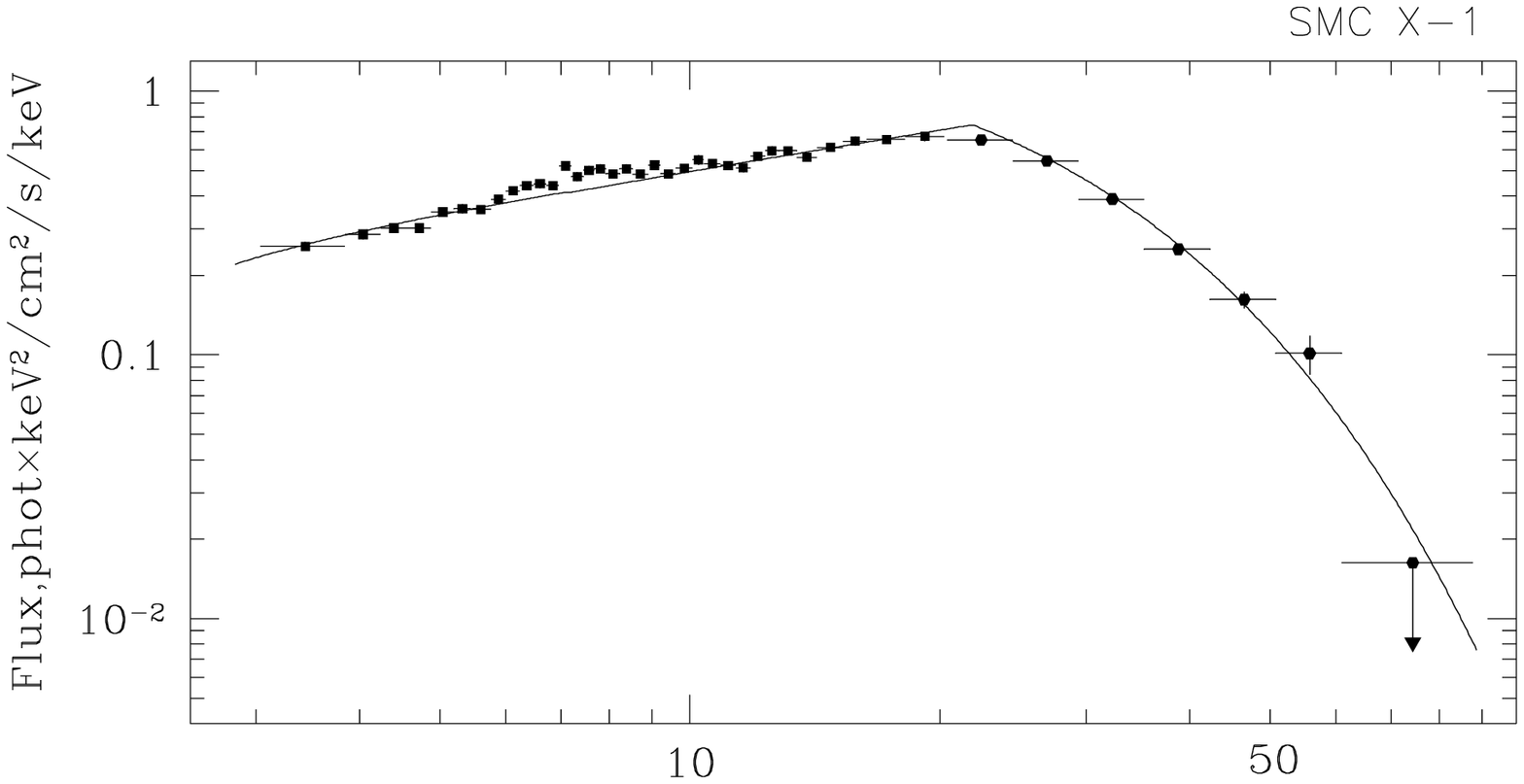}
 } 
\hbox{
\includegraphics[width=0.5\columnwidth,bb=30 435 565 710,clip]{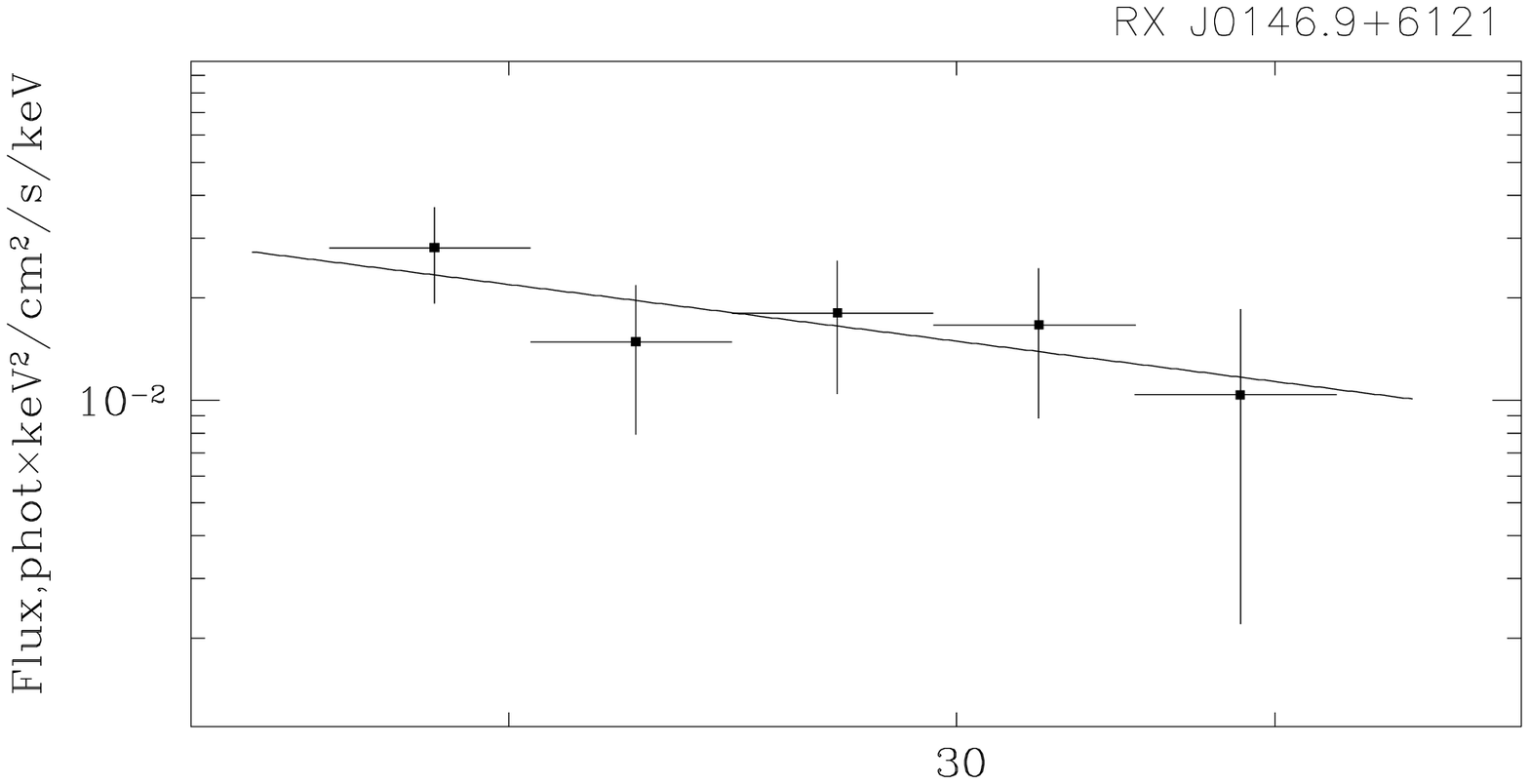}
\includegraphics[width=0.5\columnwidth,bb=30 435 565 710,clip]{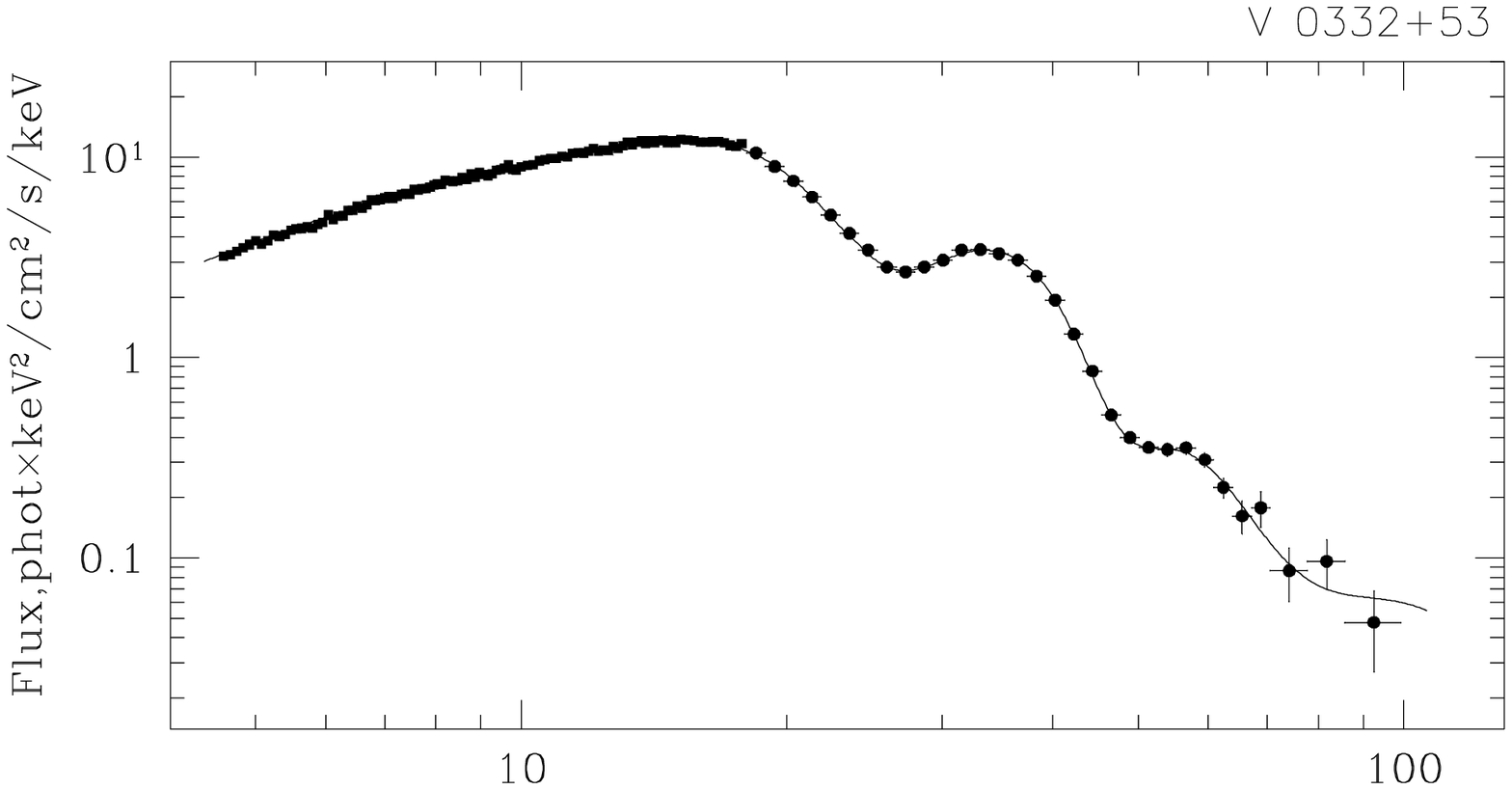}
}

\hbox{
\includegraphics[width=0.5\columnwidth,bb=30 435 565 710,clip]{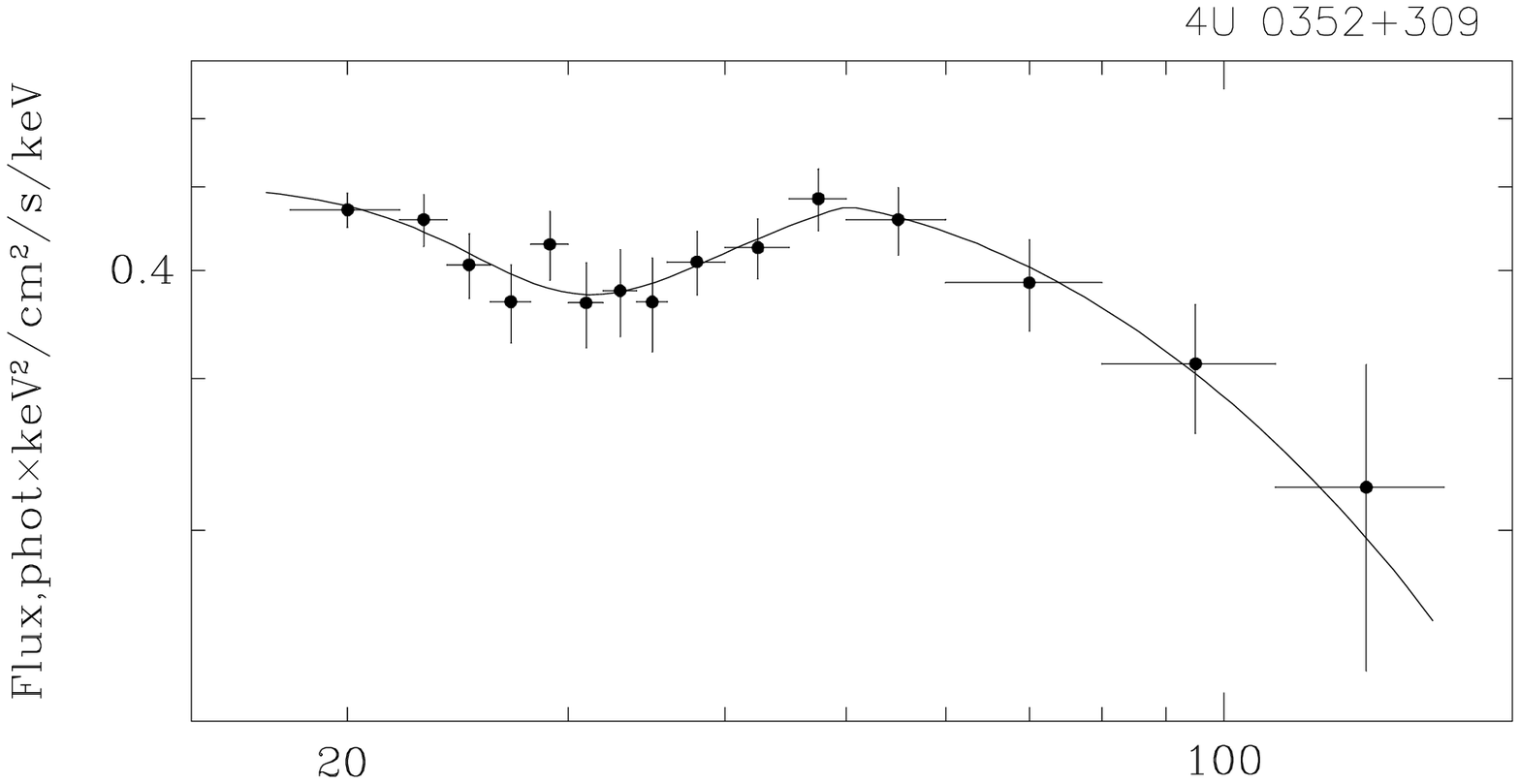} 
\includegraphics[width=0.5\columnwidth,bb=30 435 565 710,clip]{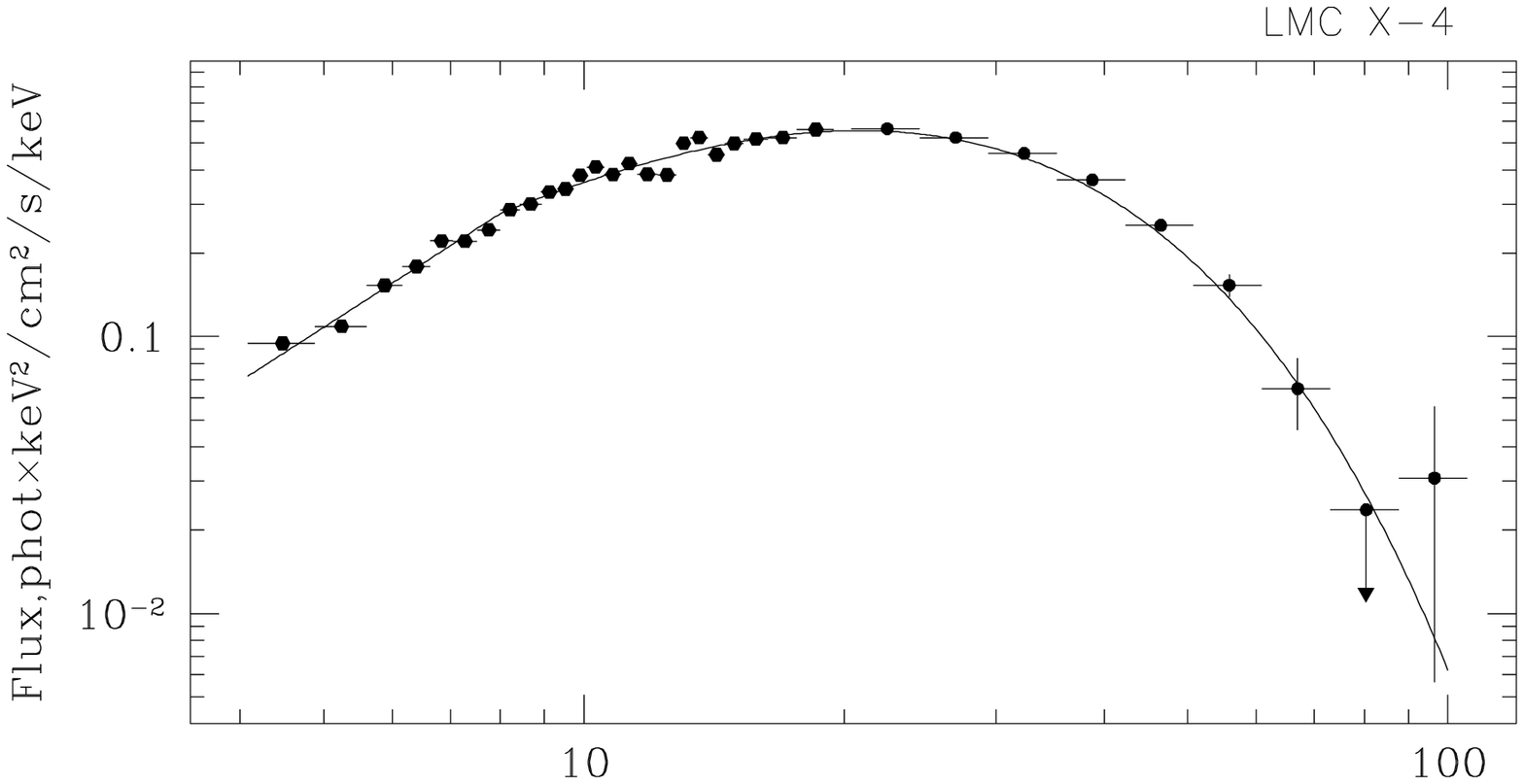}
}  
\hbox{
\includegraphics[width=0.5\columnwidth,bb=30 410 565 710]{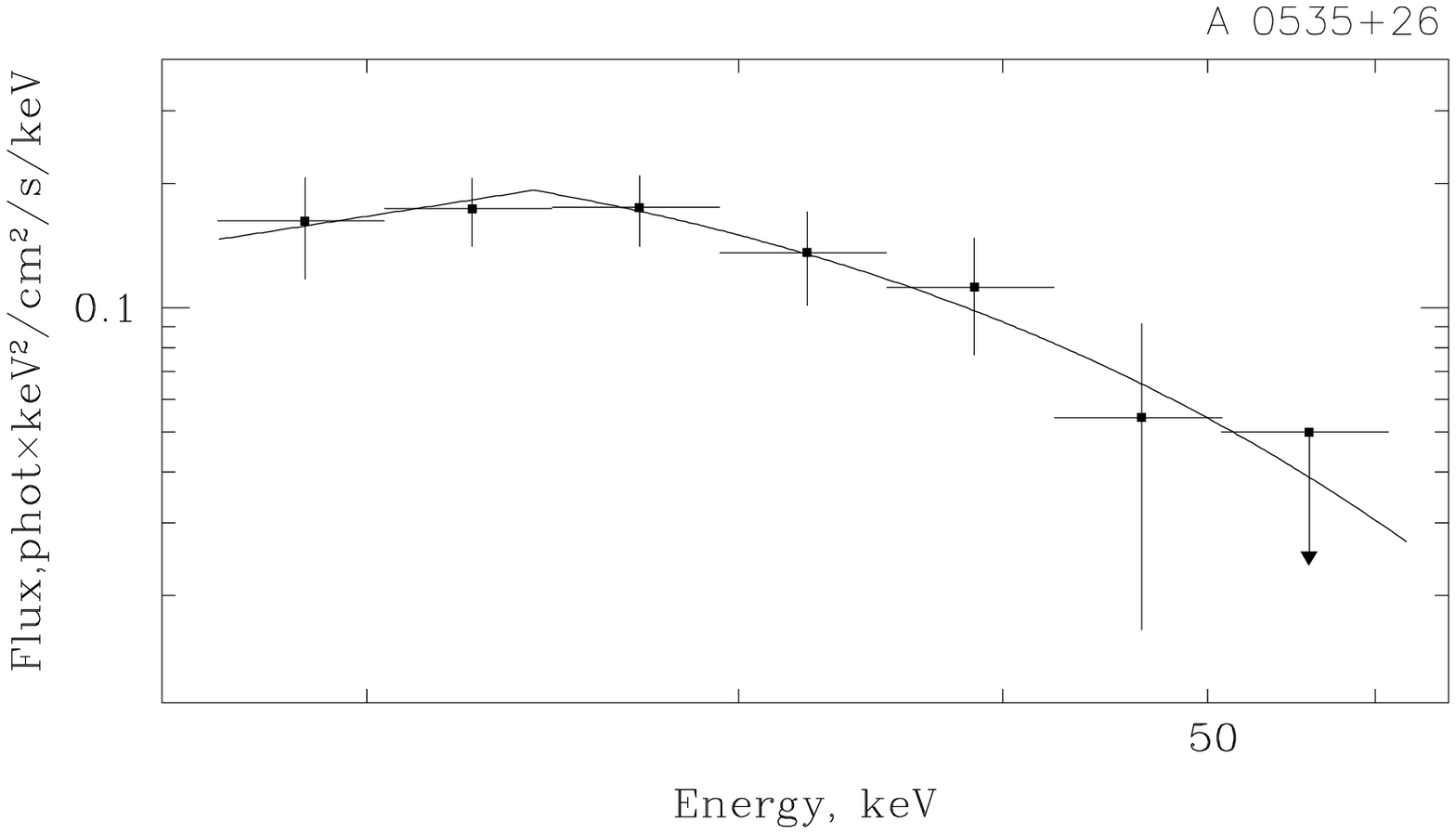} 
\includegraphics[width=0.5\columnwidth,bb=30 410 565 710]{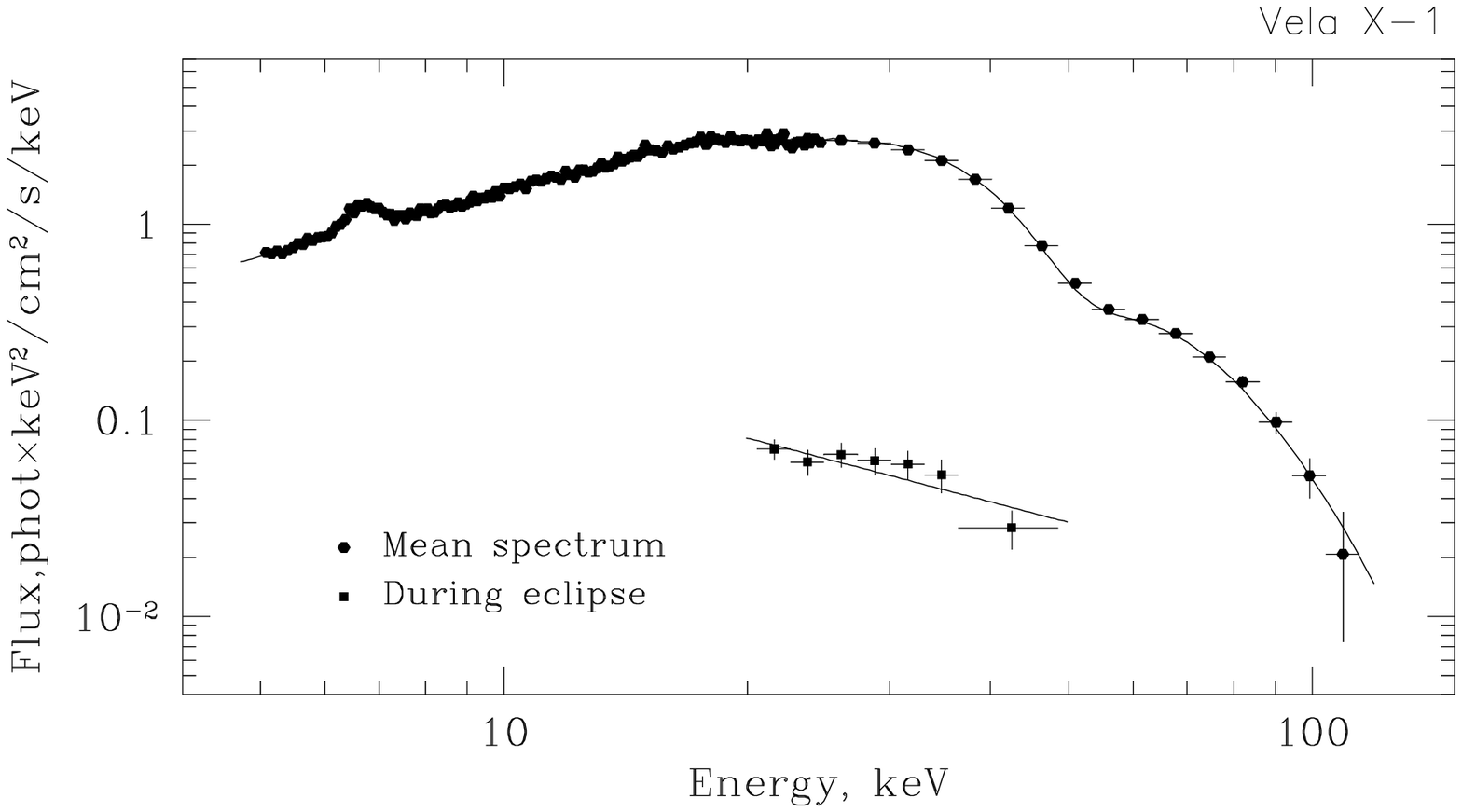}
}

\vfill
\renewcommand{\figurename}{Fig.}
\caption{INTEGRAL energy spectra for the X-ray pulsars. The solid lines 
represent the best fit to the spectrum. The errors
correspond to one standard deviation.
}
\end{figure*}

\newpage
 
\begin{figure*}

\hbox{ 
\includegraphics[width=0.5\columnwidth,bb=30 435 565 710,clip]{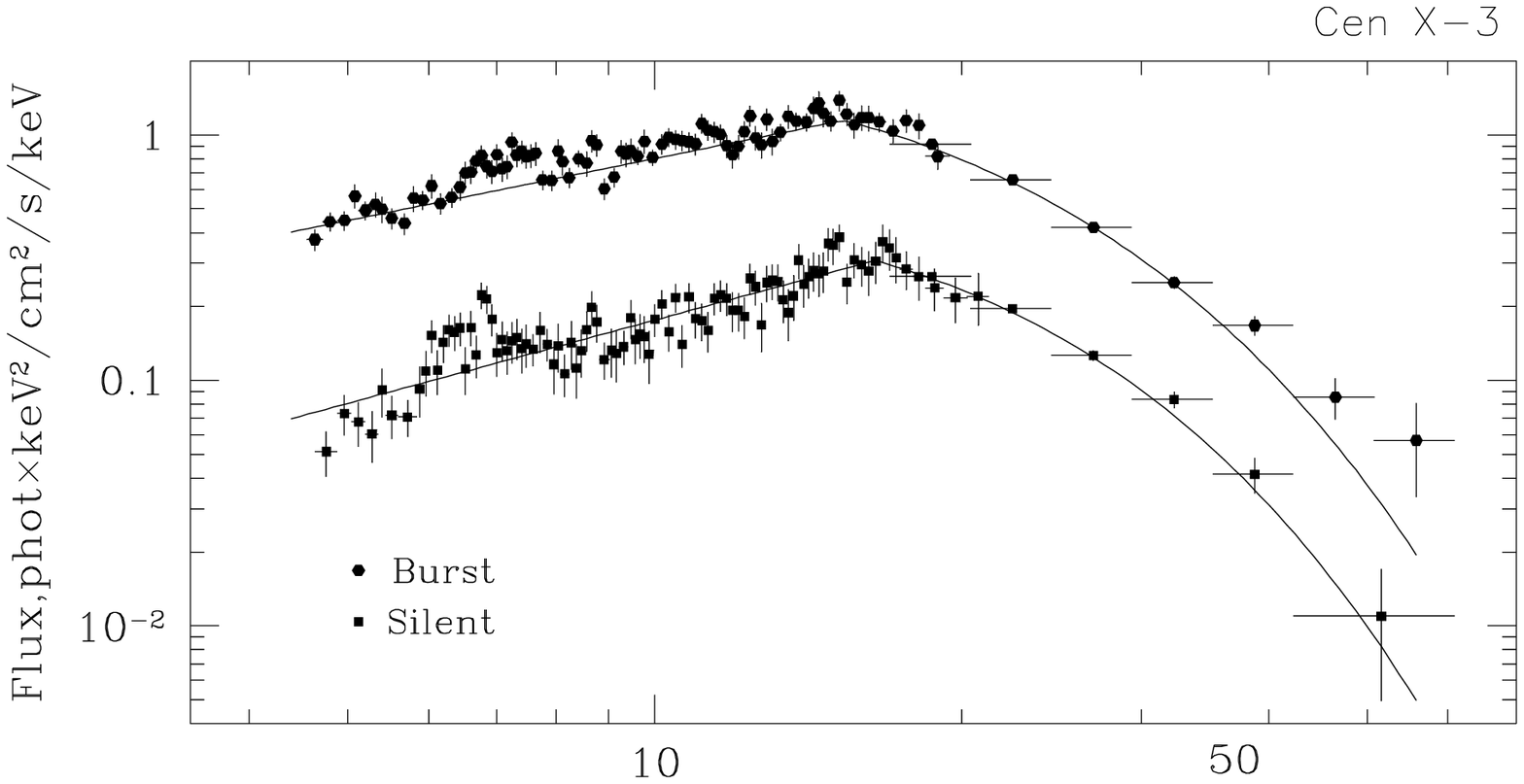}
\includegraphics[width=0.5\columnwidth,bb=30 435 565 710,clip]{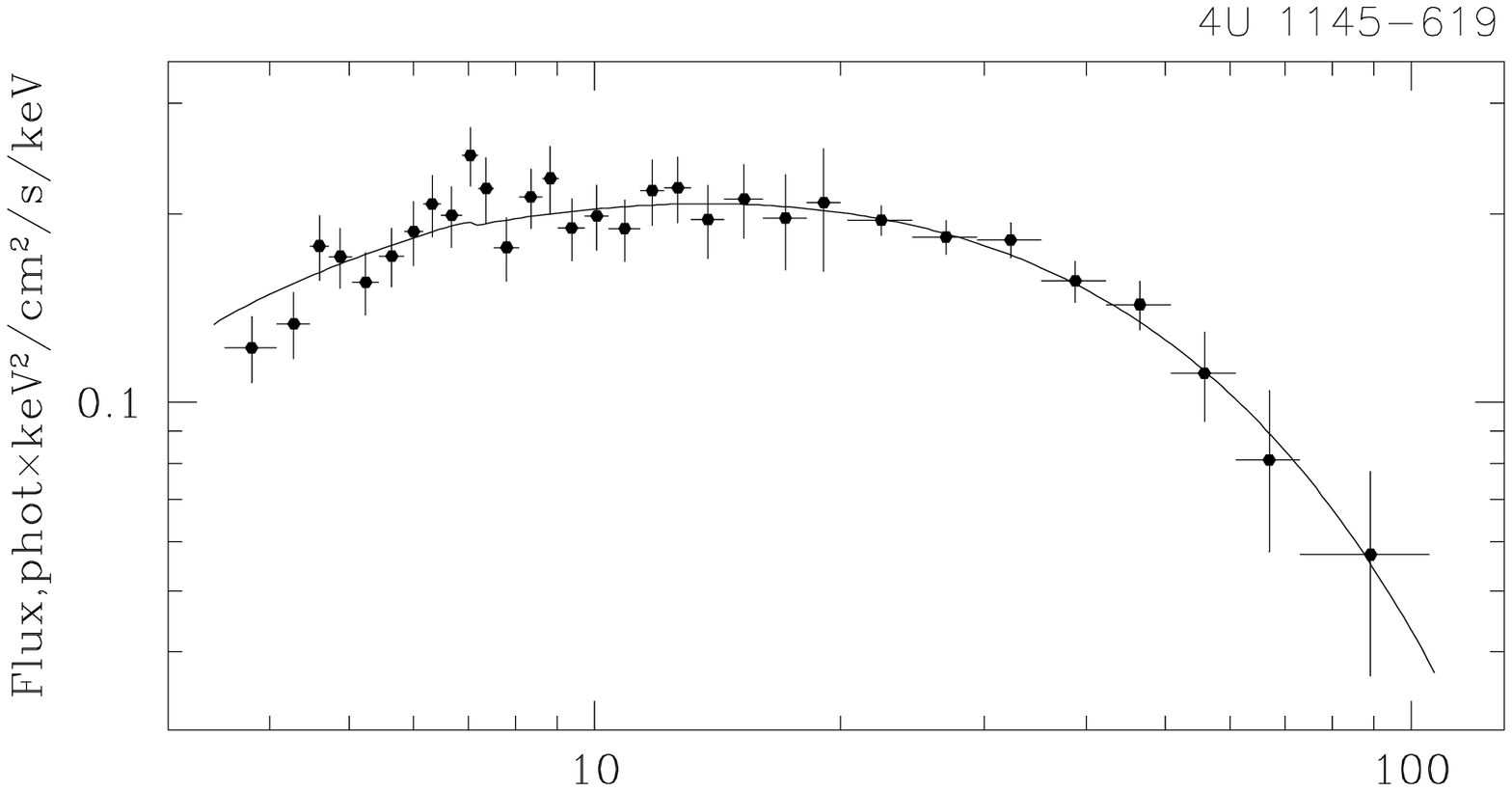} 
} 

\hbox{
\includegraphics[width=0.5\columnwidth,bb=30 435 565 710,clip]{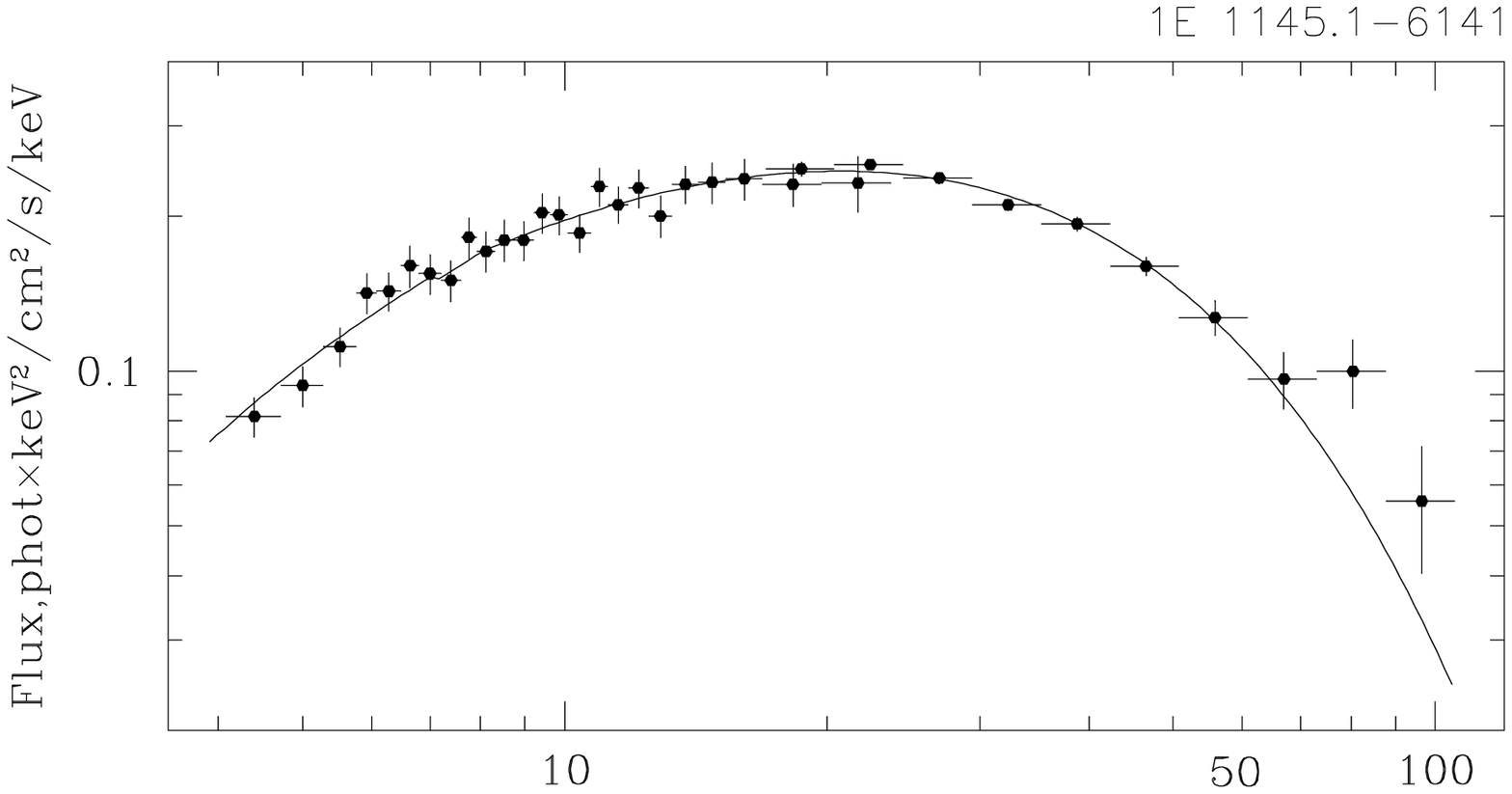}
\includegraphics[width=0.5\columnwidth,bb=30 435 565 710]{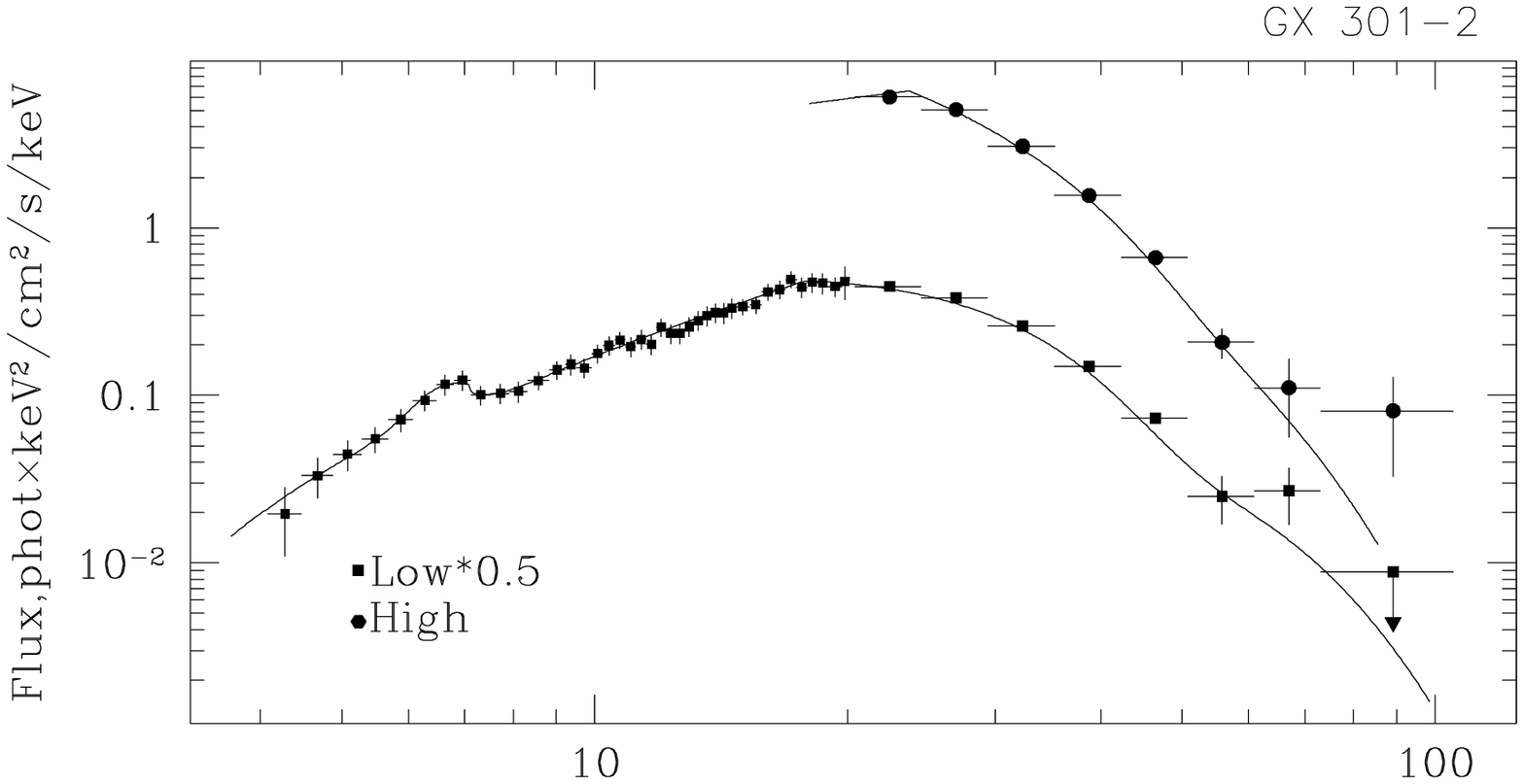}
} 
\hbox{ 
\includegraphics[width=0.5\columnwidth,bb=30 435 565 710]{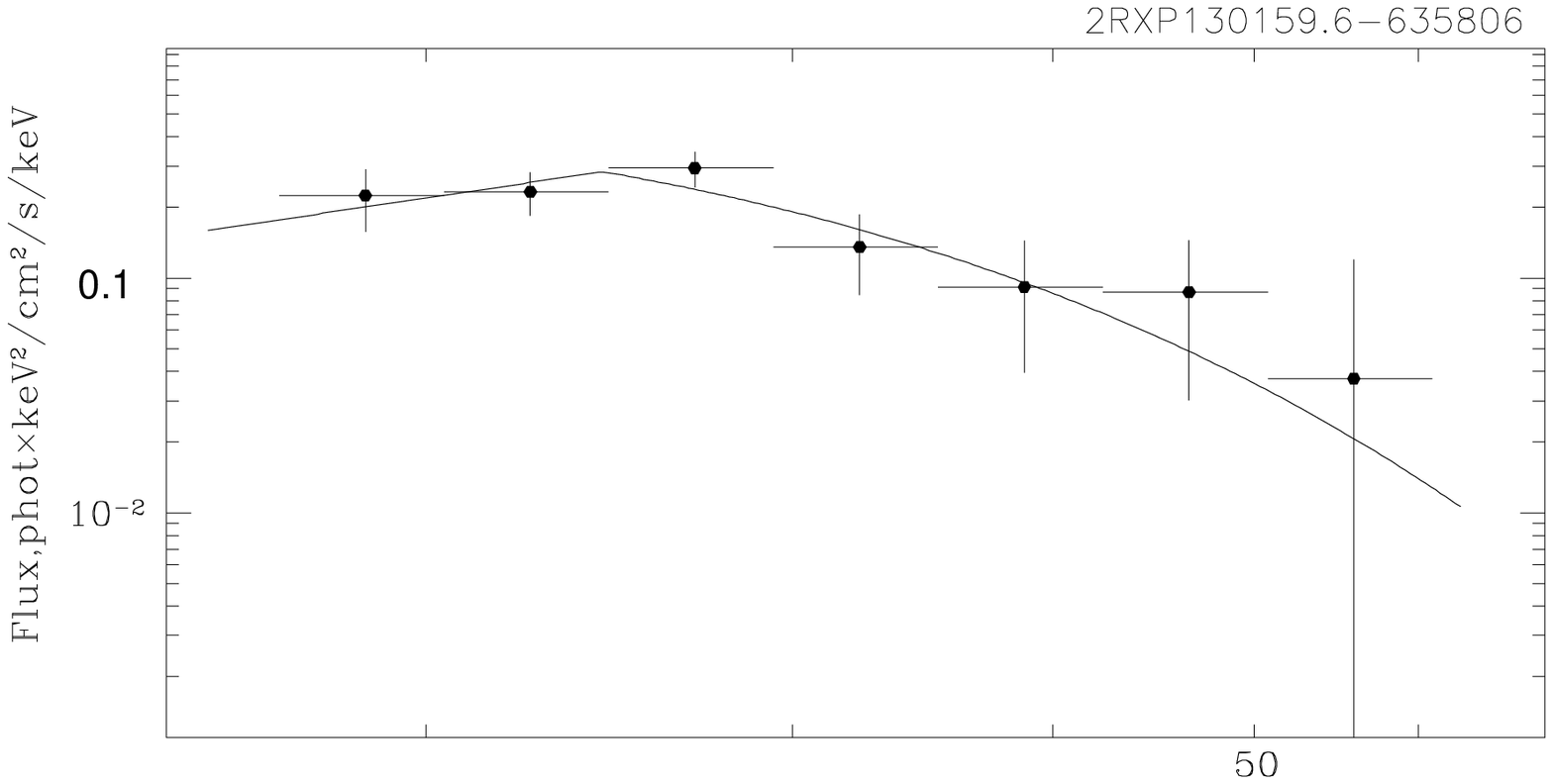}
\includegraphics[width=0.5\columnwidth,bb=30 435 565 710]{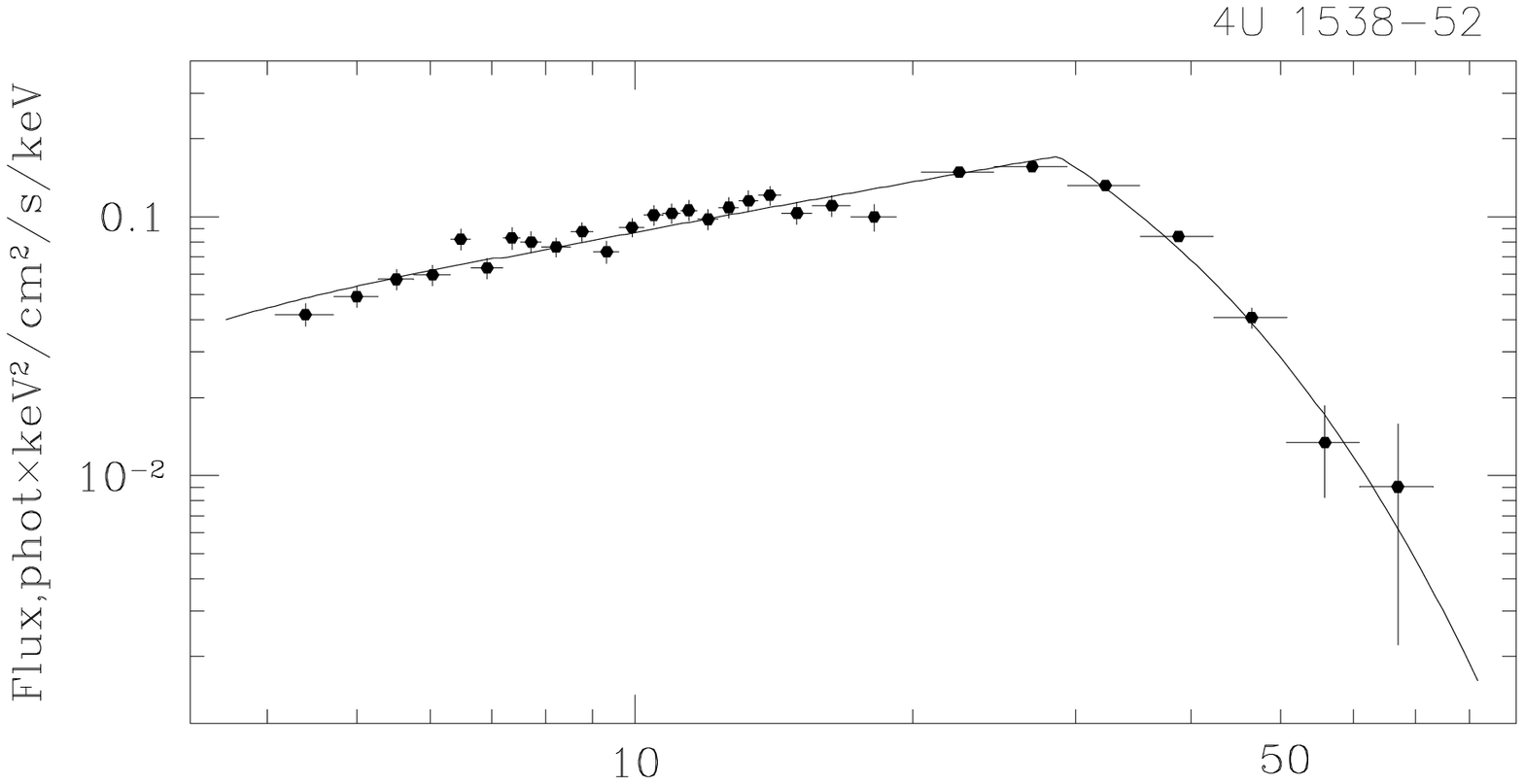}
}  
\hbox{
\includegraphics[width=0.5\columnwidth,bb=30 410 565 710]{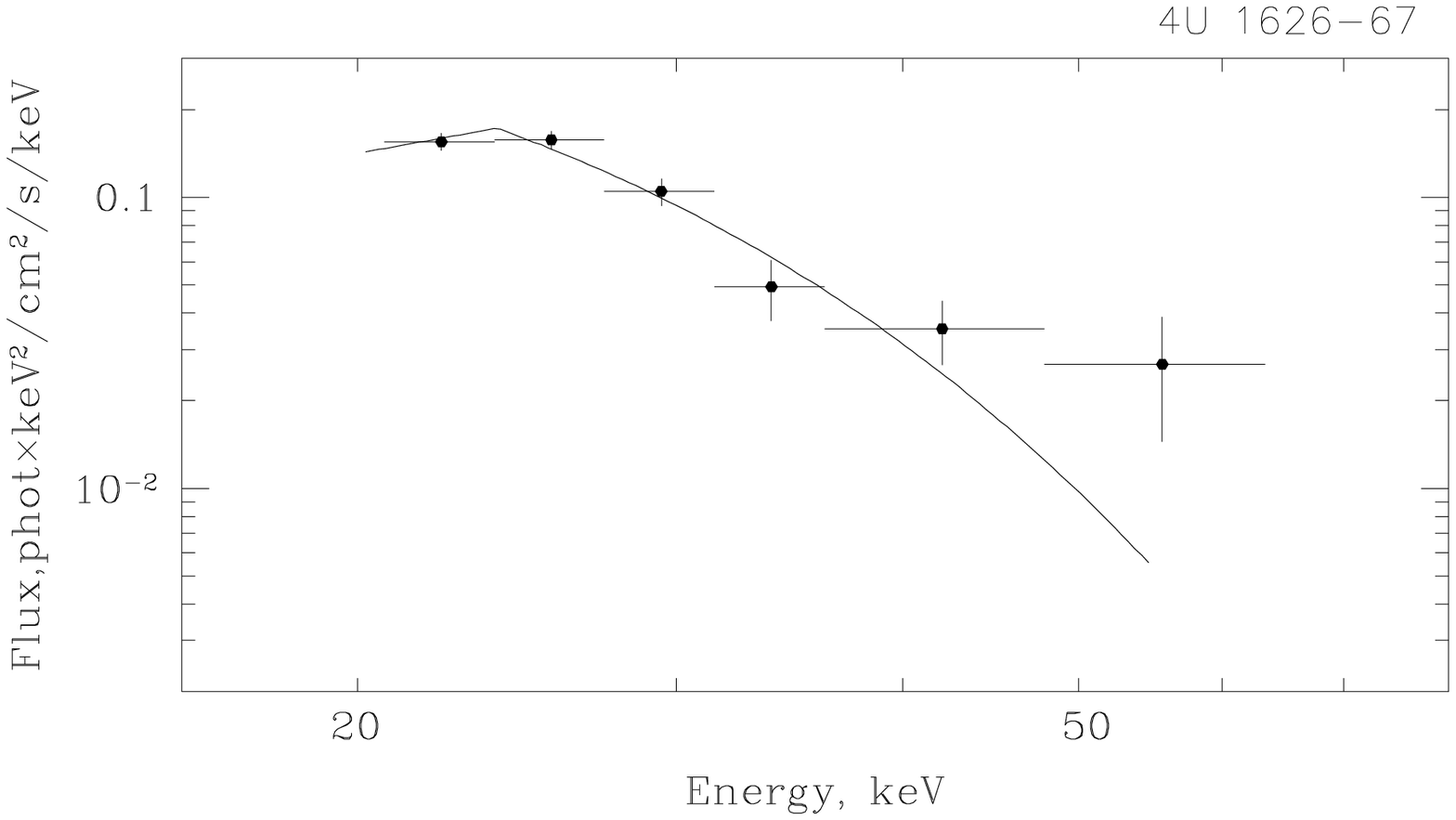} 
\includegraphics[width=0.5\columnwidth,bb=30 410 565 710]{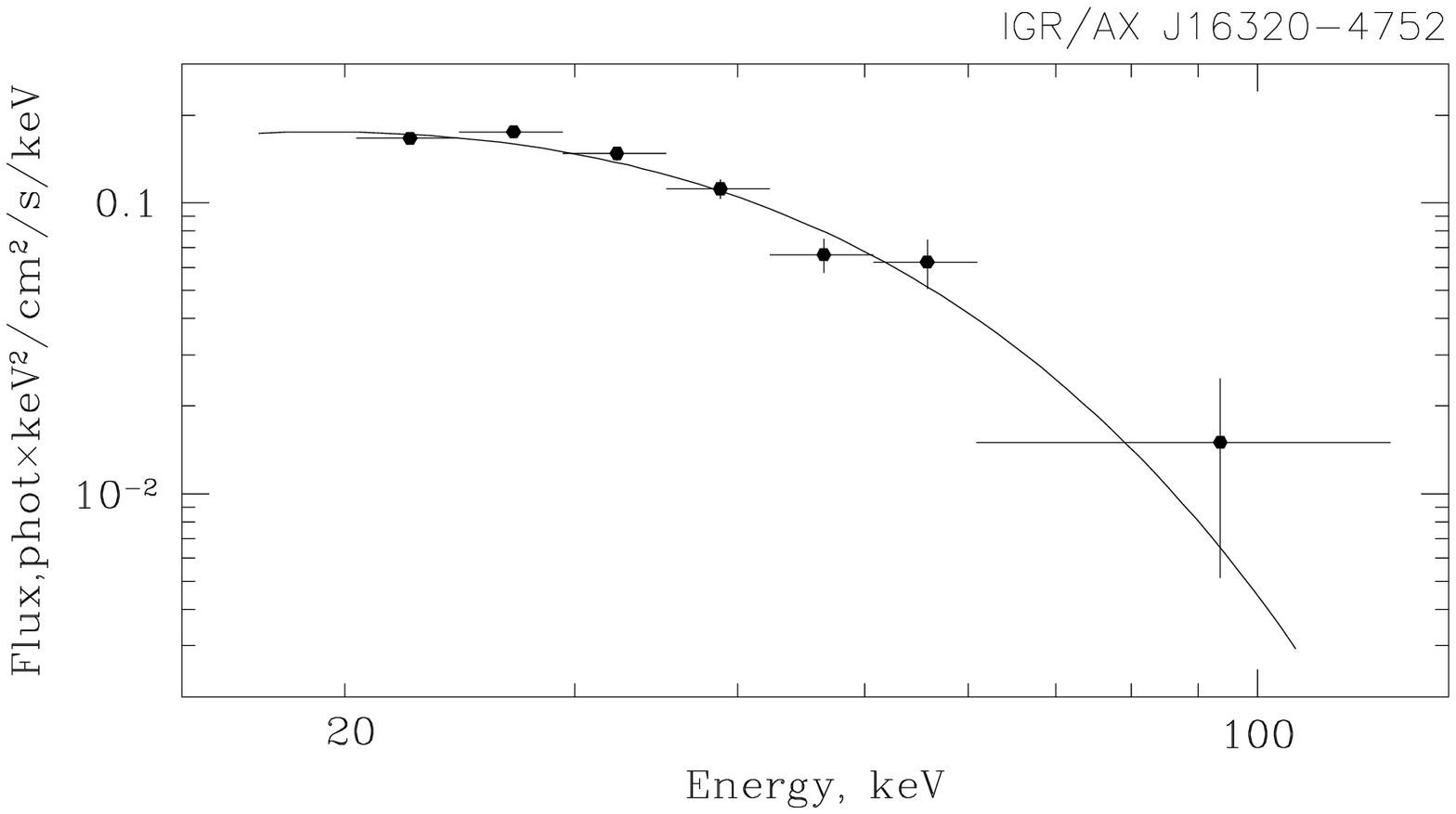}
}

\vfill
\addtolength{\topsep}{-1\baselineskip}
\renewcommand{\figurename}{Fig.}
\centerline{Fig.2: Contd.}
\end{figure*}

\newpage
\begin{figure*}  
\hbox{
\includegraphics[width=0.5\columnwidth,bb=30 435 565 710]{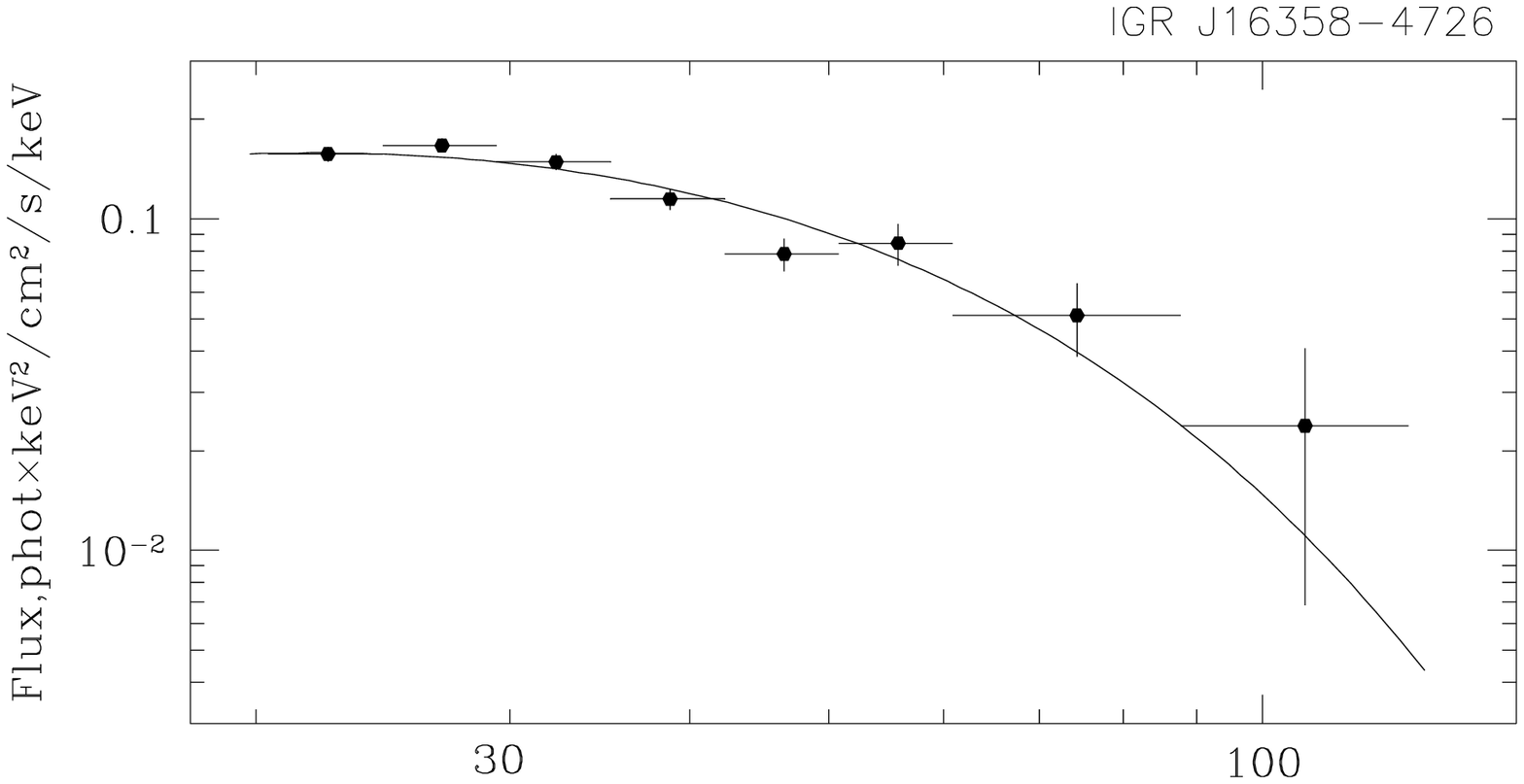} 
\includegraphics[width=0.5\columnwidth,bb=30 435 565 710]{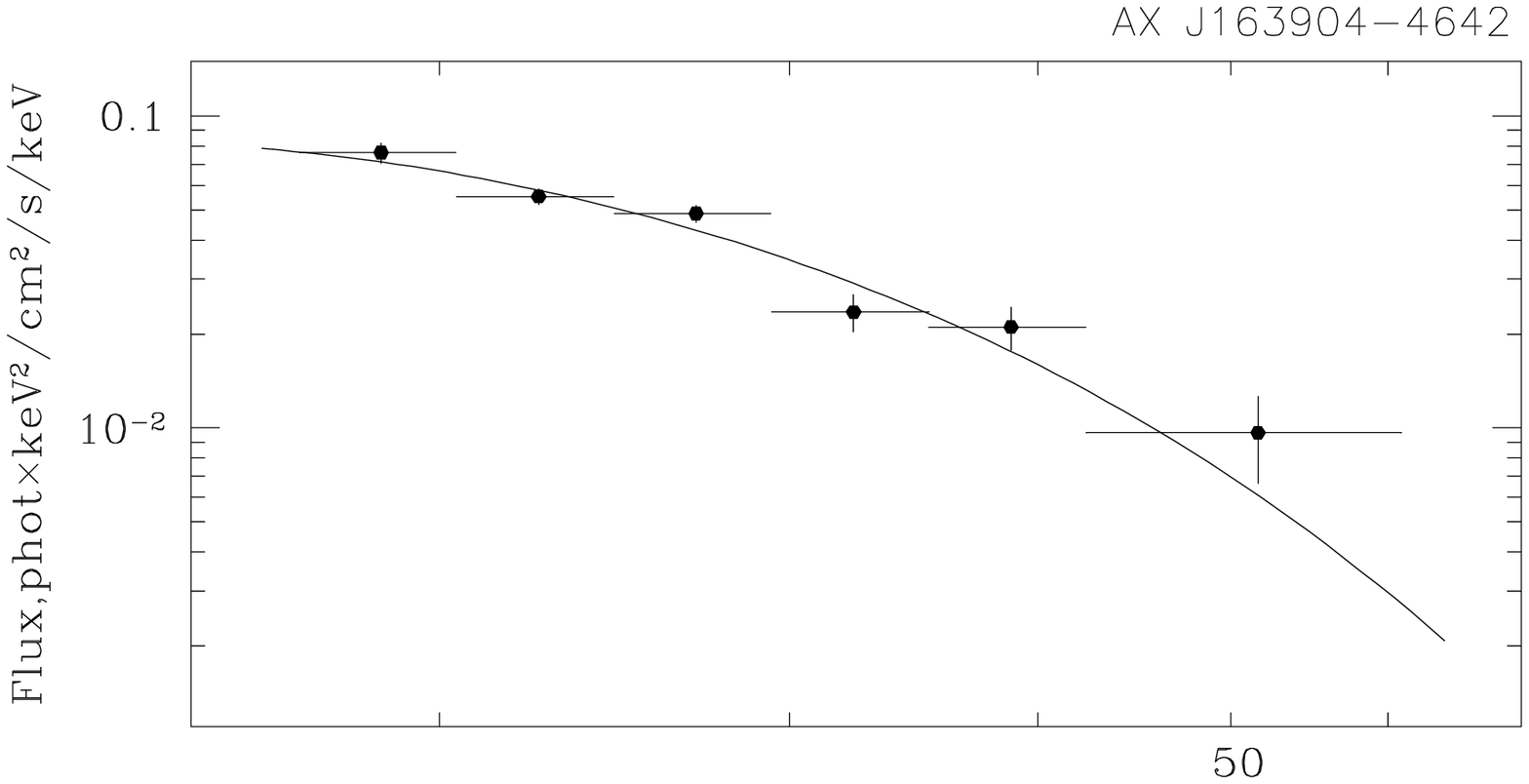}
} 

\hbox{
\includegraphics[width=0.5\columnwidth,bb=30 435 565 710]{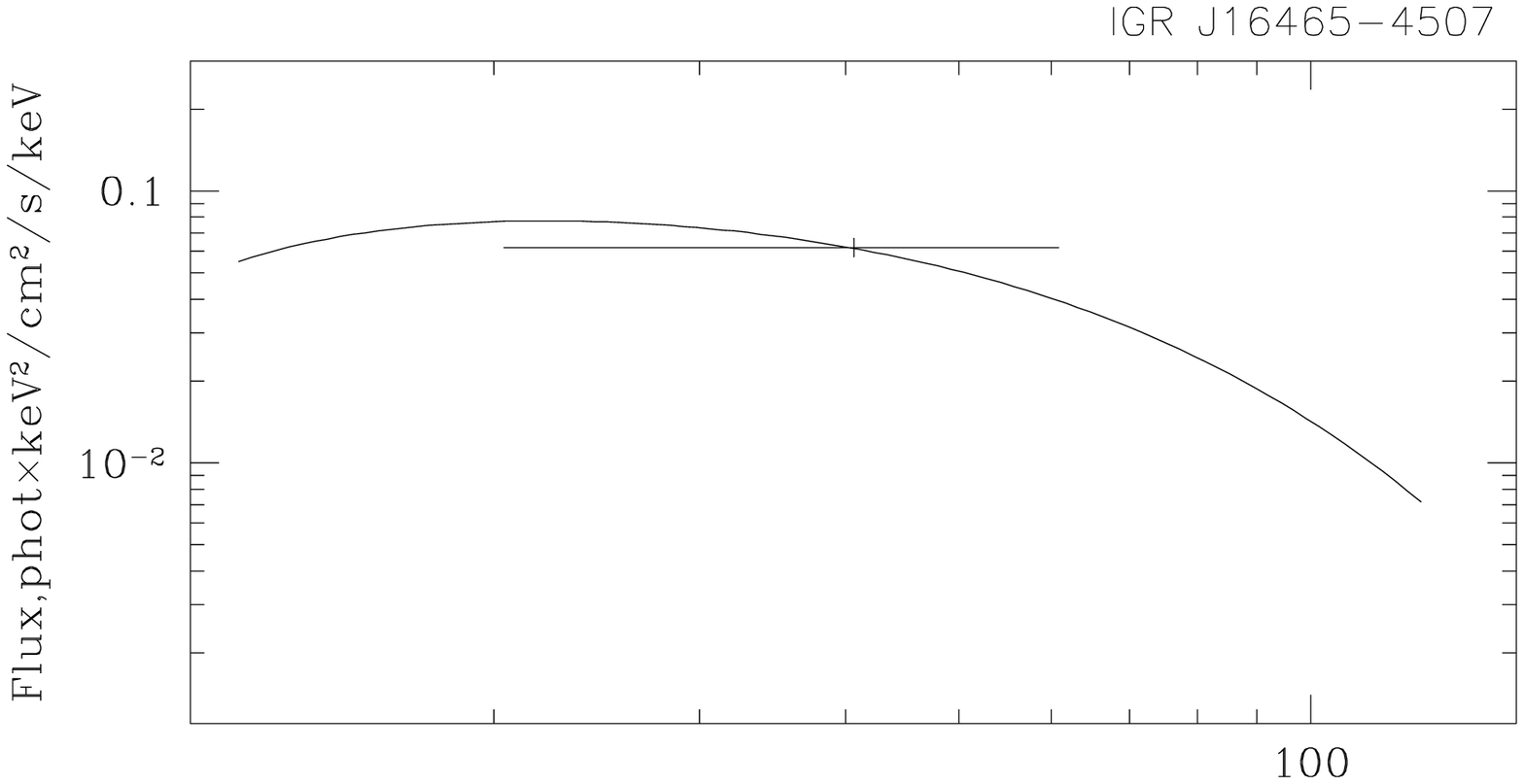} 
\includegraphics[width=0.5\columnwidth,bb=30 435 565 710]{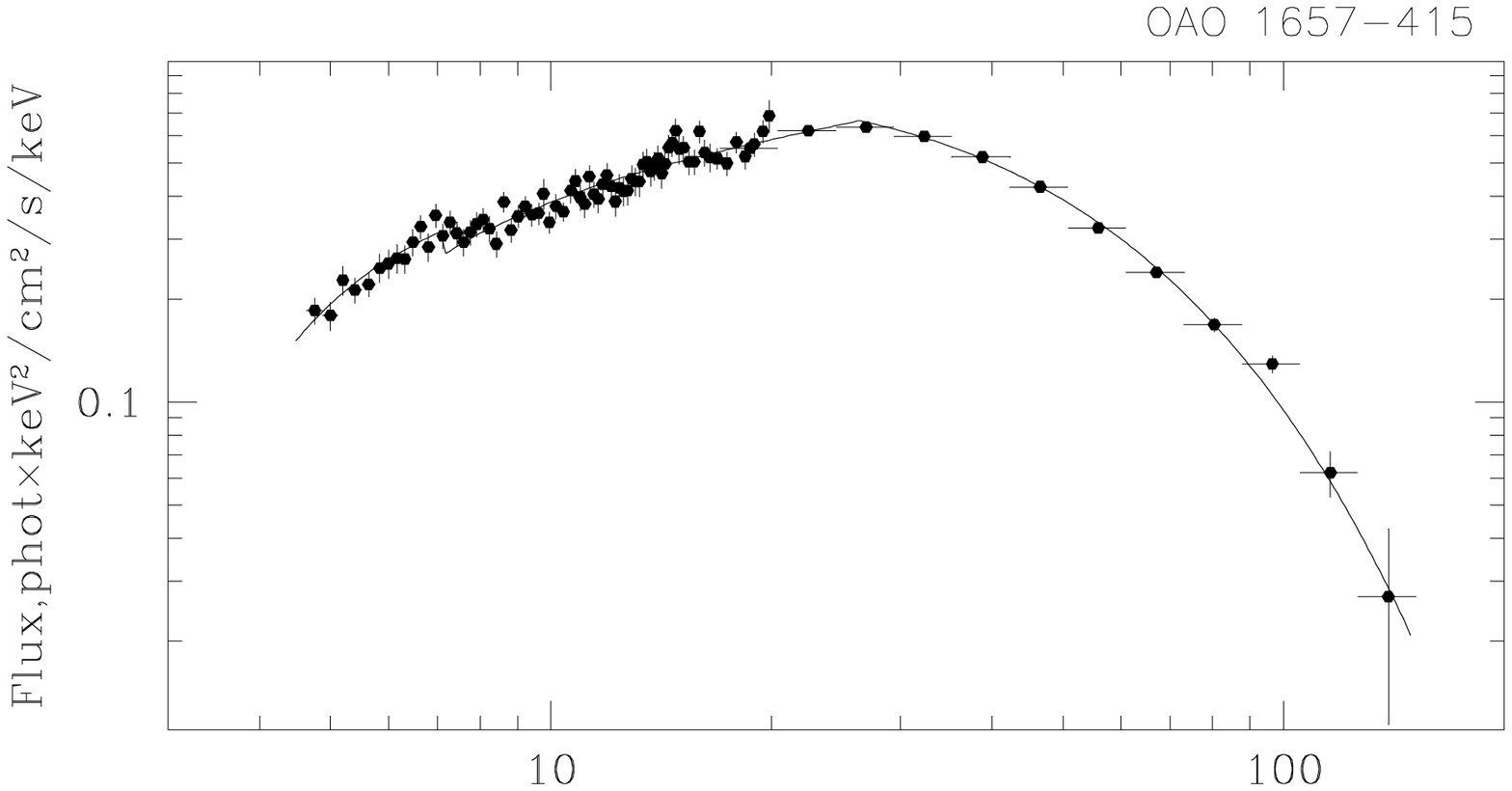}
} 
\hbox{
\includegraphics[width=0.5\columnwidth,bb=30 435 565 710]{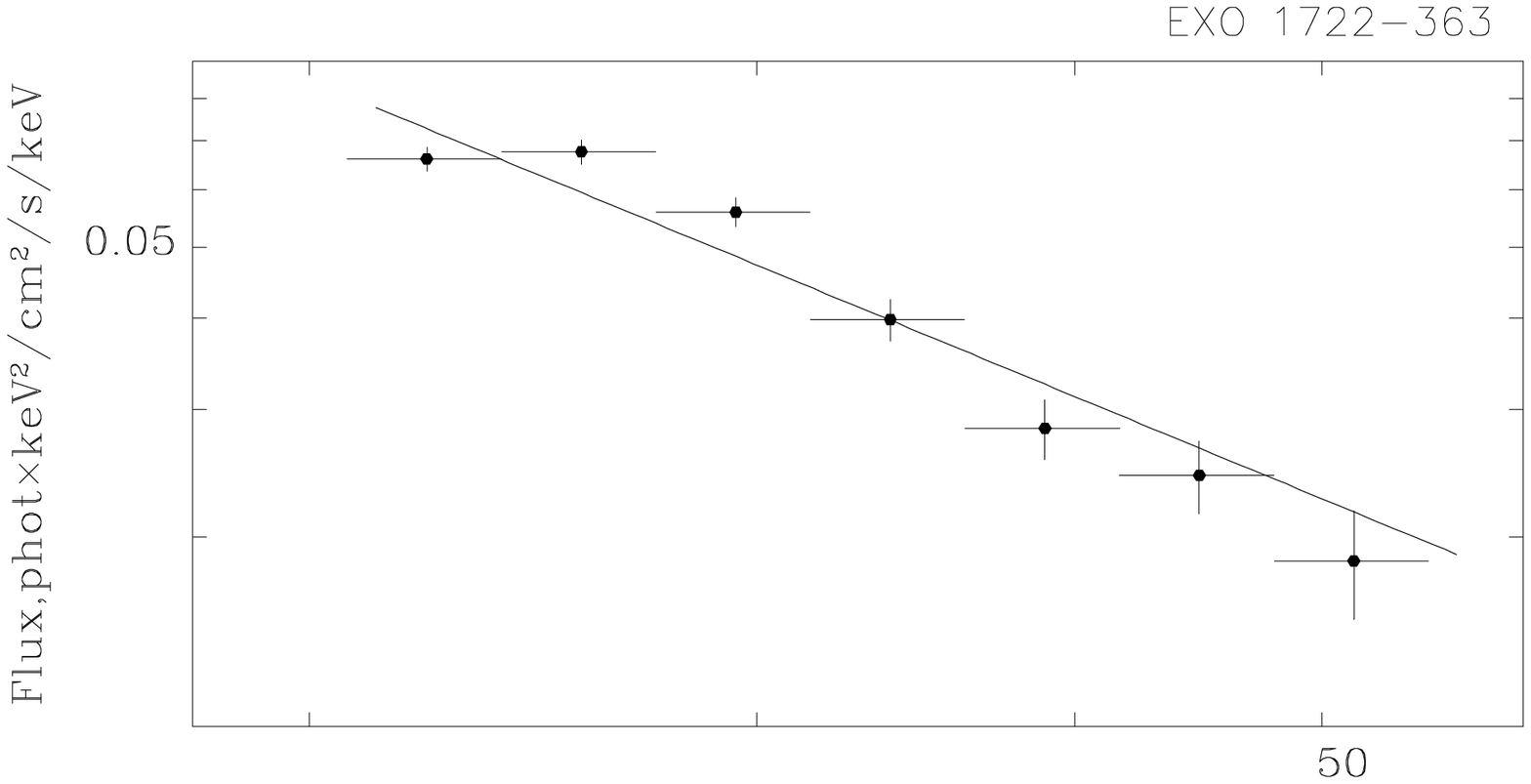} 
\includegraphics[width=0.5\columnwidth,bb=30 435 565 710]{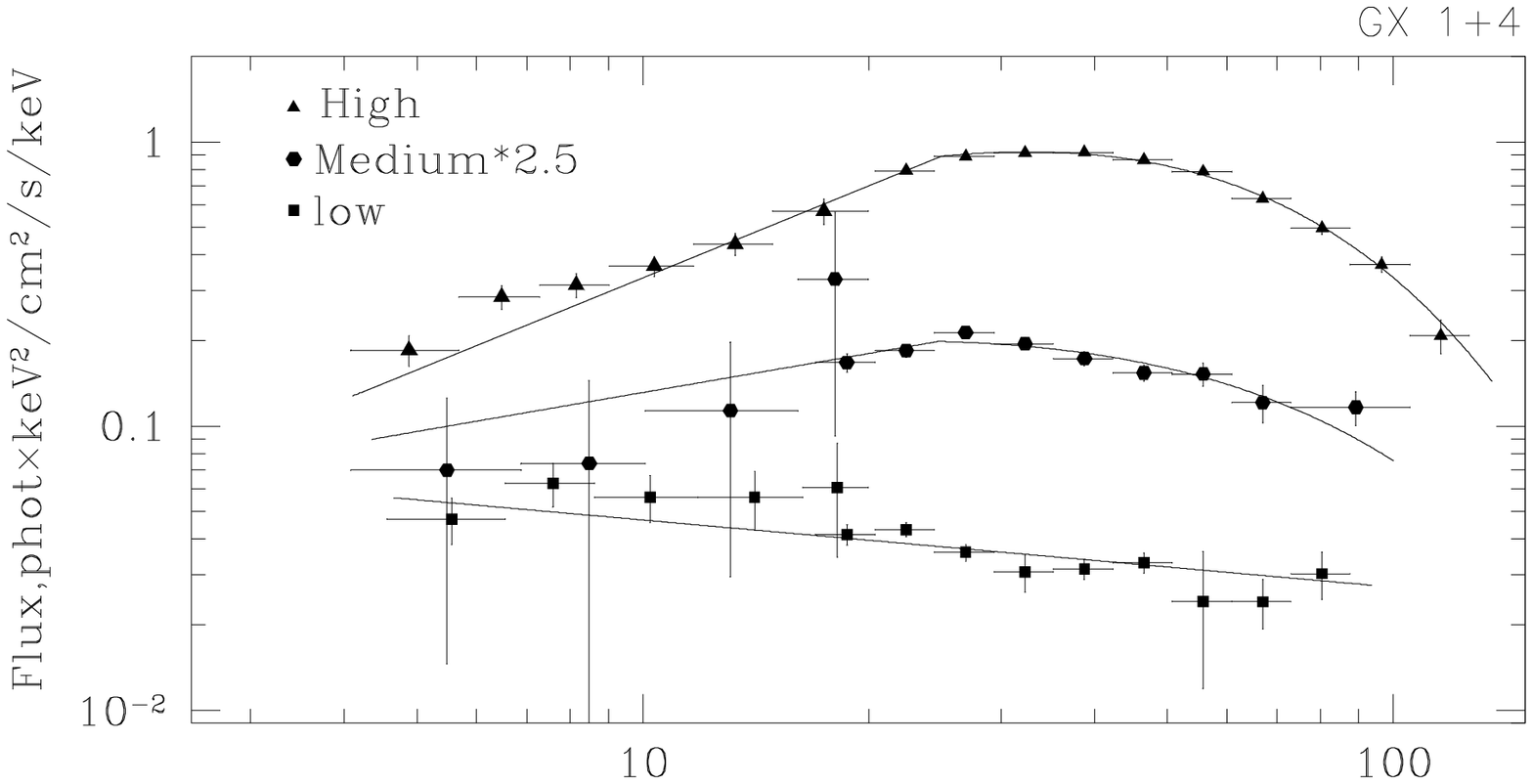}
}  
\hbox{
\includegraphics[width=0.5\columnwidth,bb=30 410 565 710]{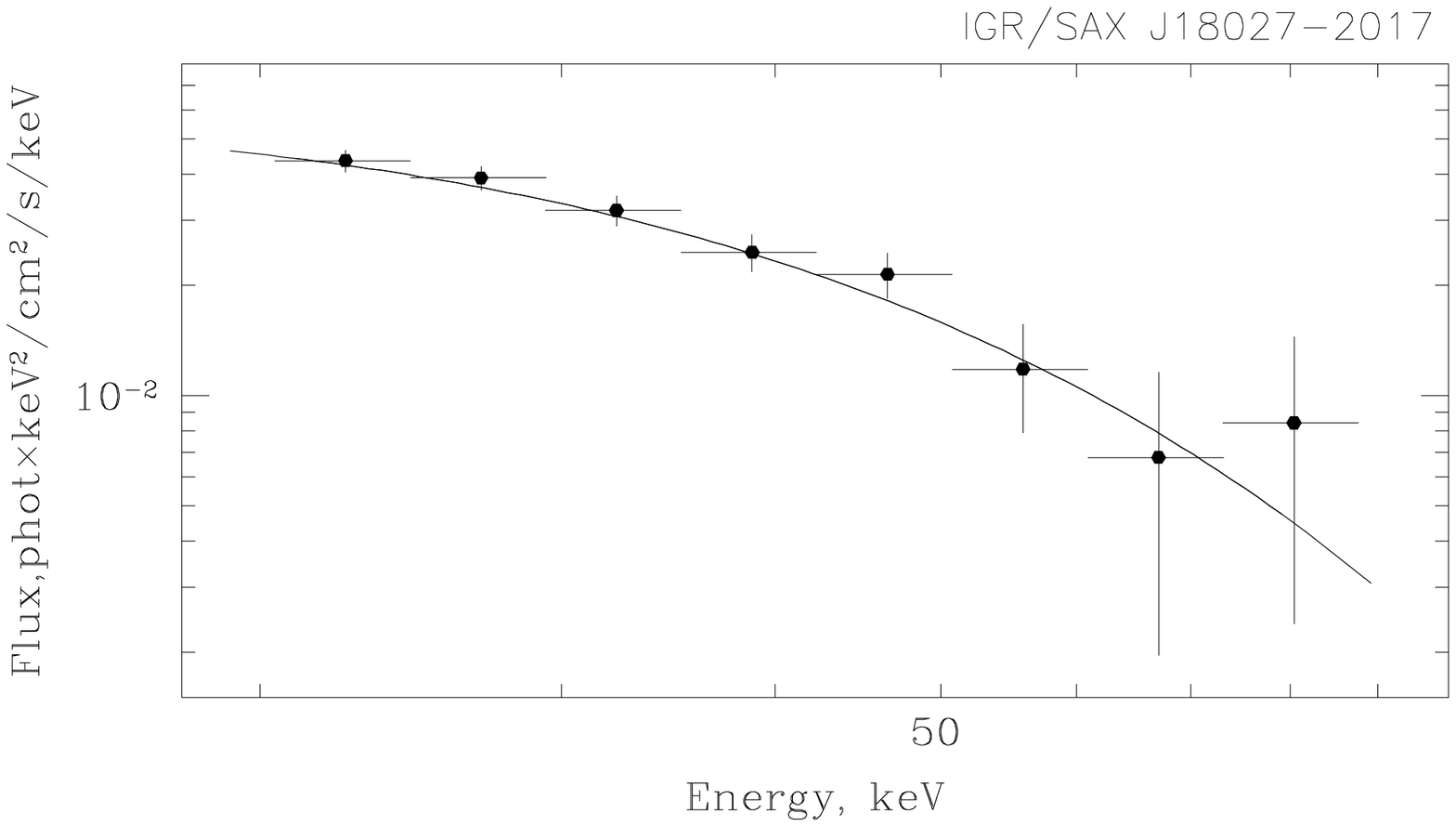} 
\includegraphics[width=0.5\columnwidth,bb=30 410 565 710]{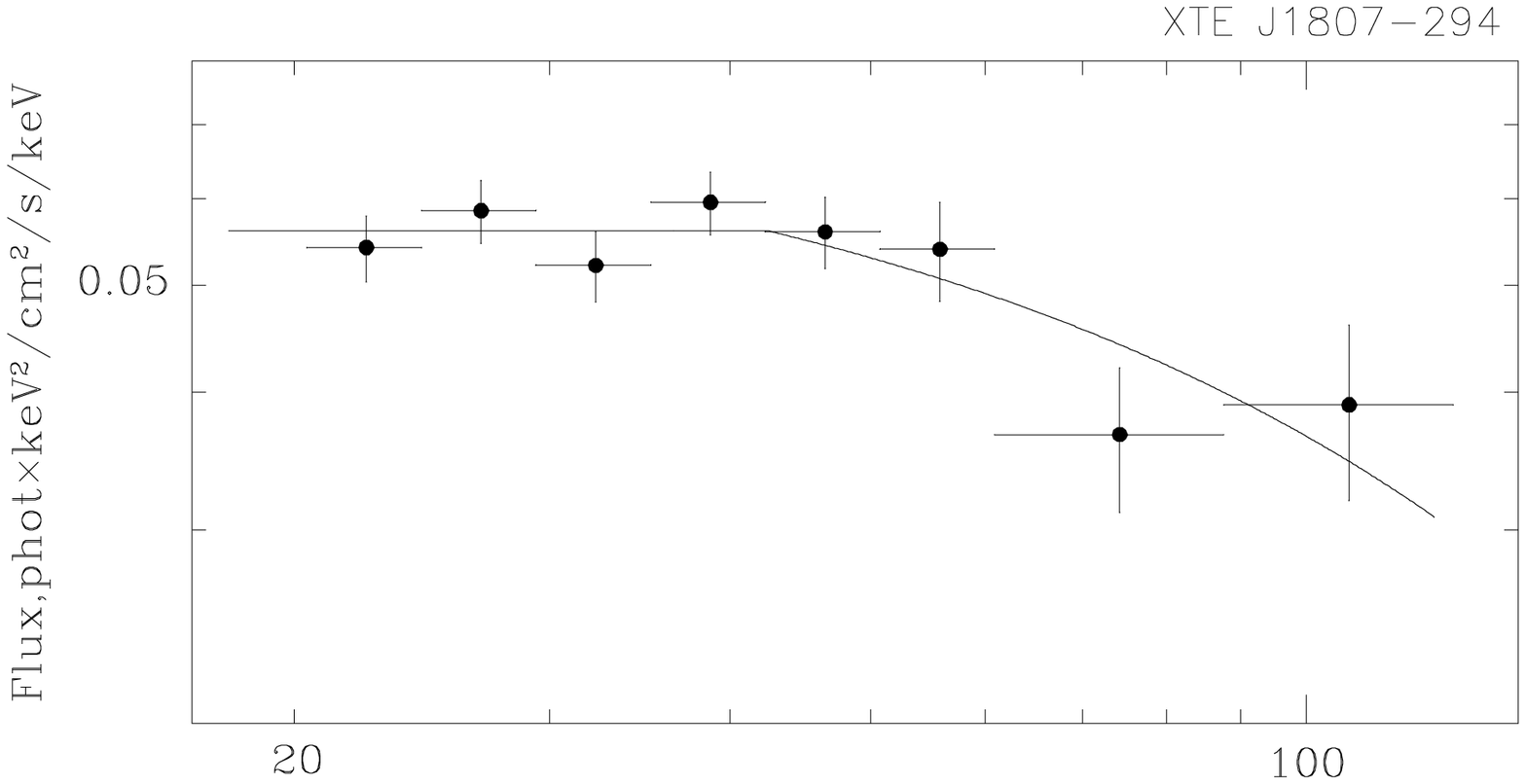}
}
\vfill
\renewcommand{\figurename}{Fig.}
\centerline{Fig.2: Contd.} 

\end{figure*}

\newpage
 
\begin{figure*}
\hbox{
\includegraphics[width=0.5\columnwidth,bb=30 435 565 710]{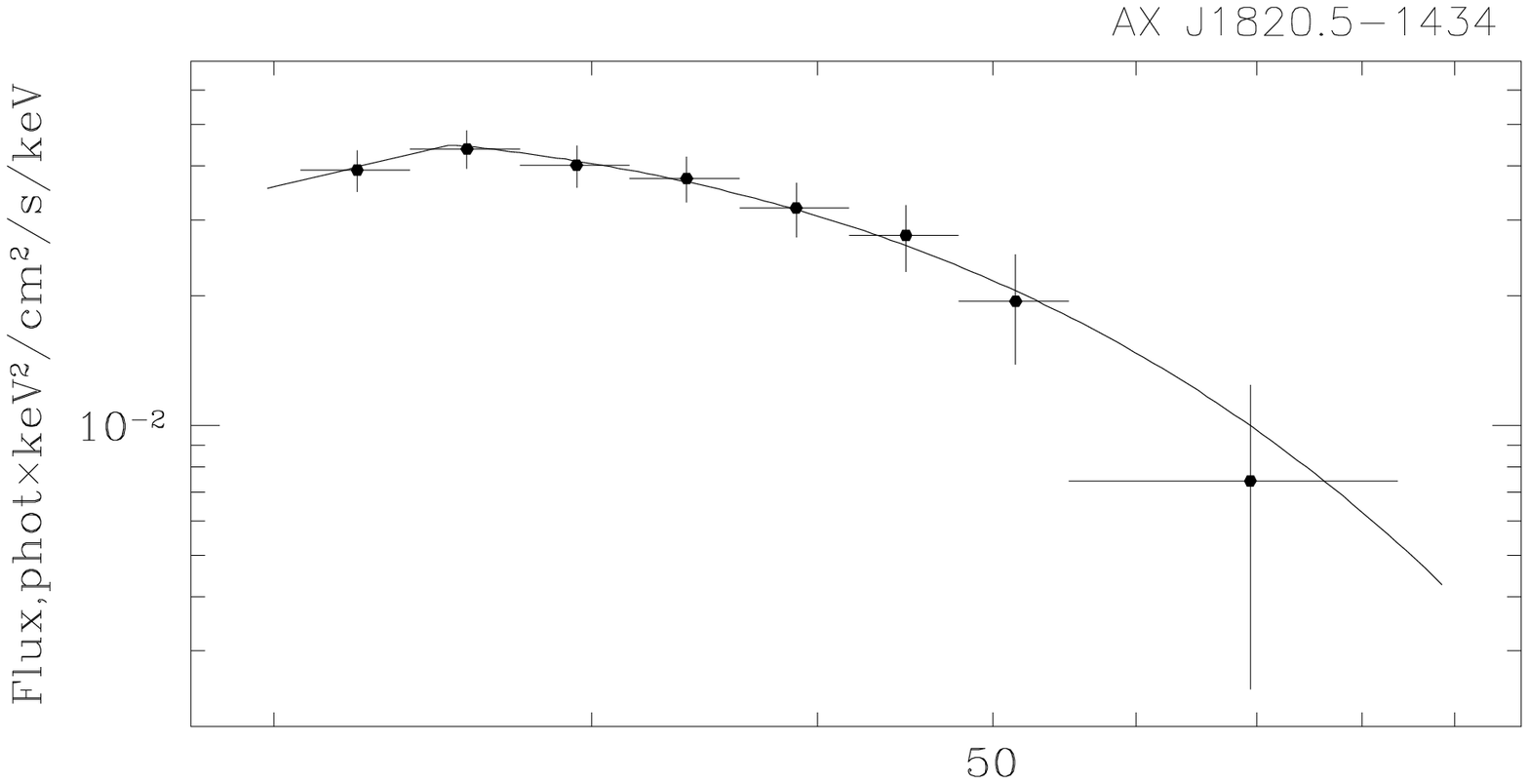} 
\includegraphics[width=0.5\columnwidth,bb=30 435 565 710]{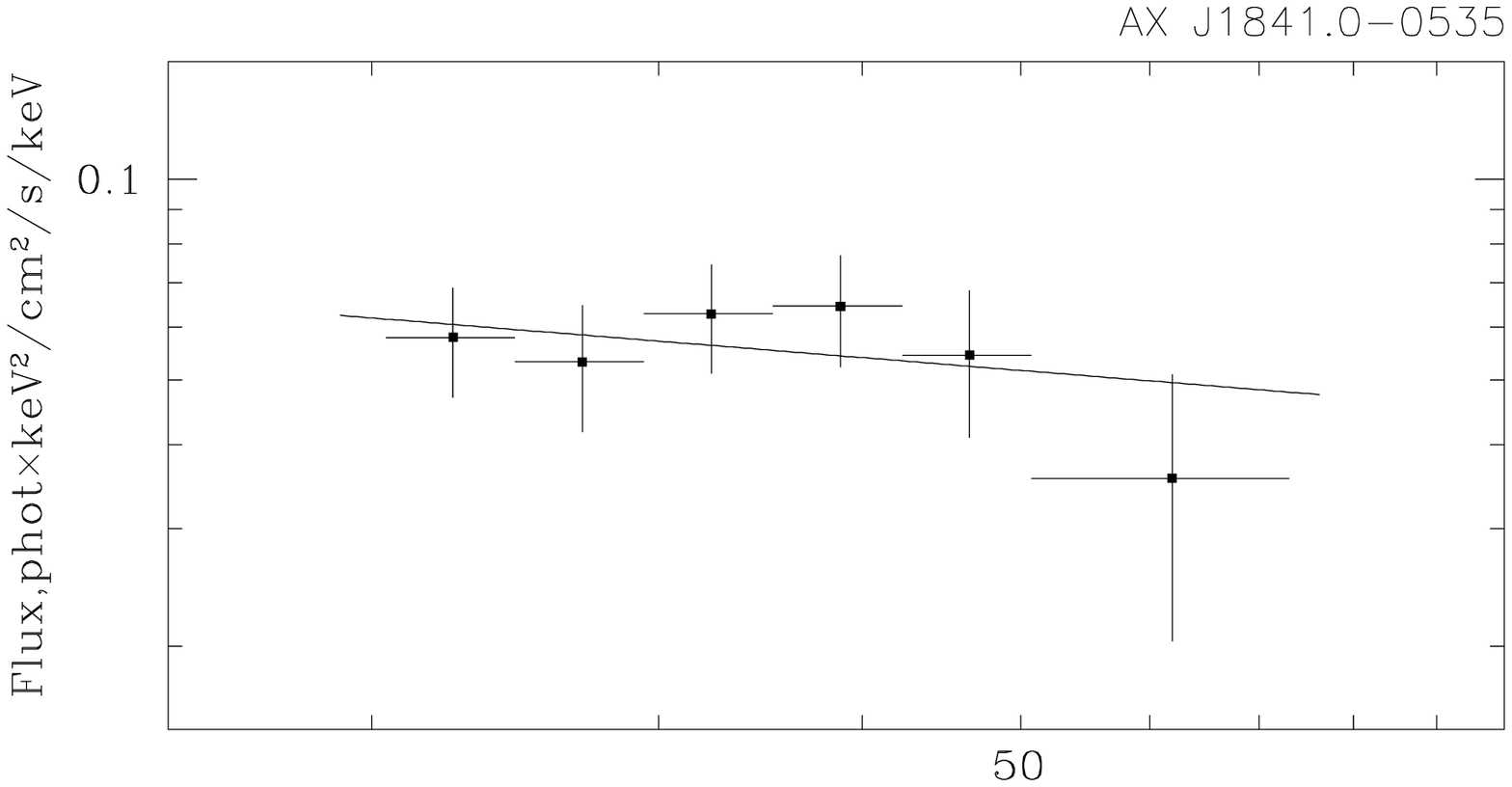}
} 

\hbox{
\includegraphics[width=0.5\columnwidth,bb=30 435 565 710]{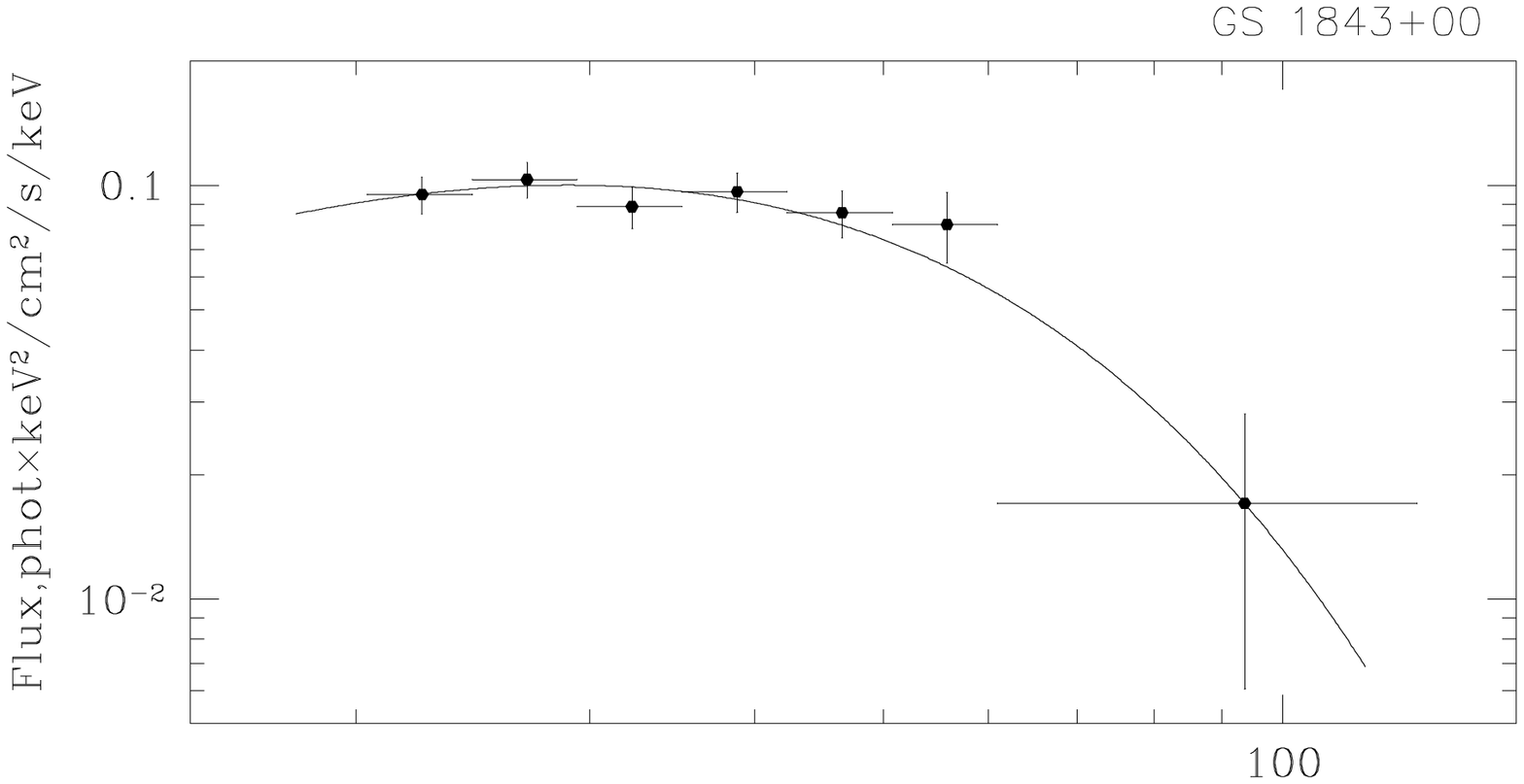} 
\includegraphics[width=0.5\columnwidth,bb=30 435 565 710]{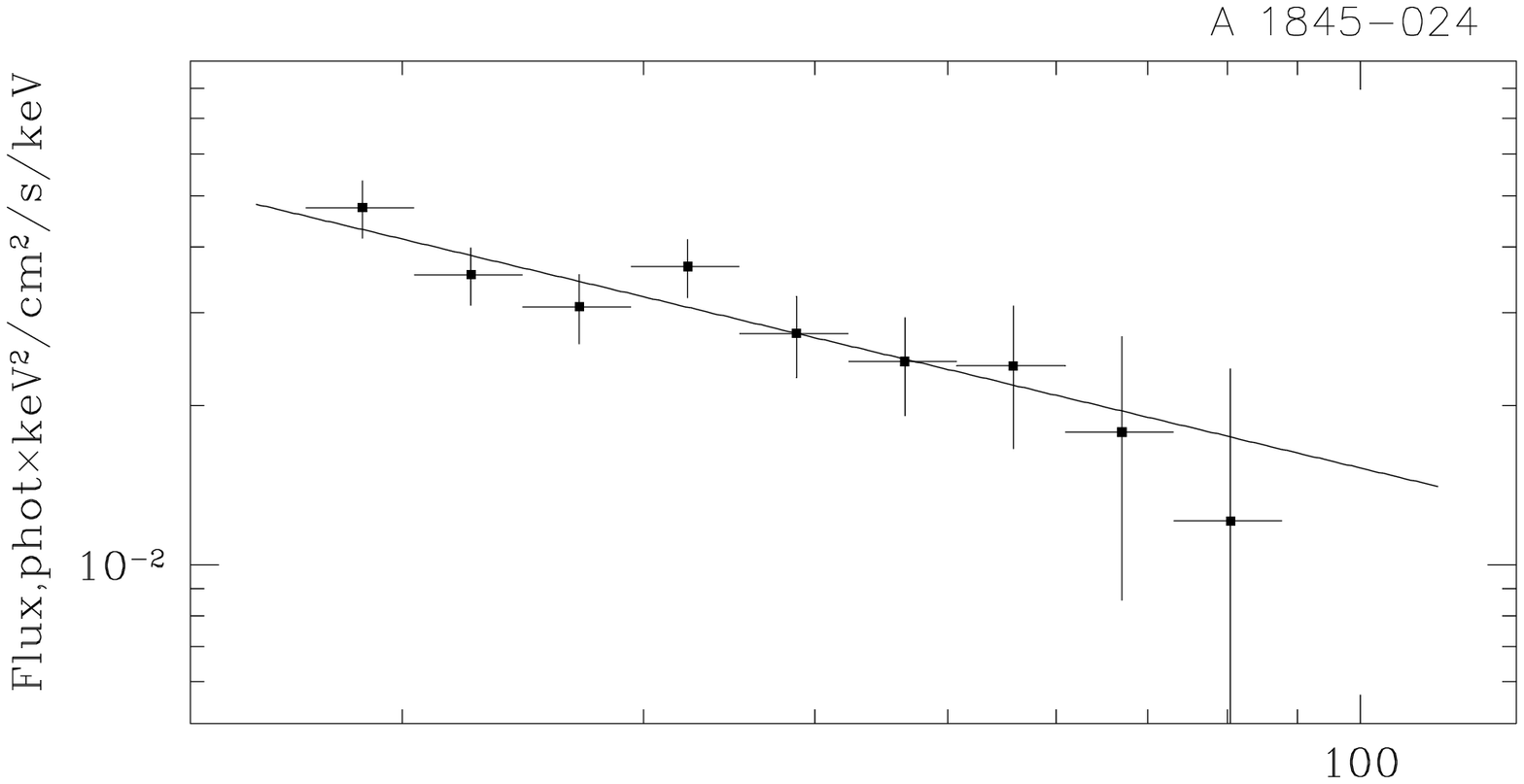}
} 
\hbox{
\includegraphics[width=0.5\columnwidth,bb=30 435 565 710]{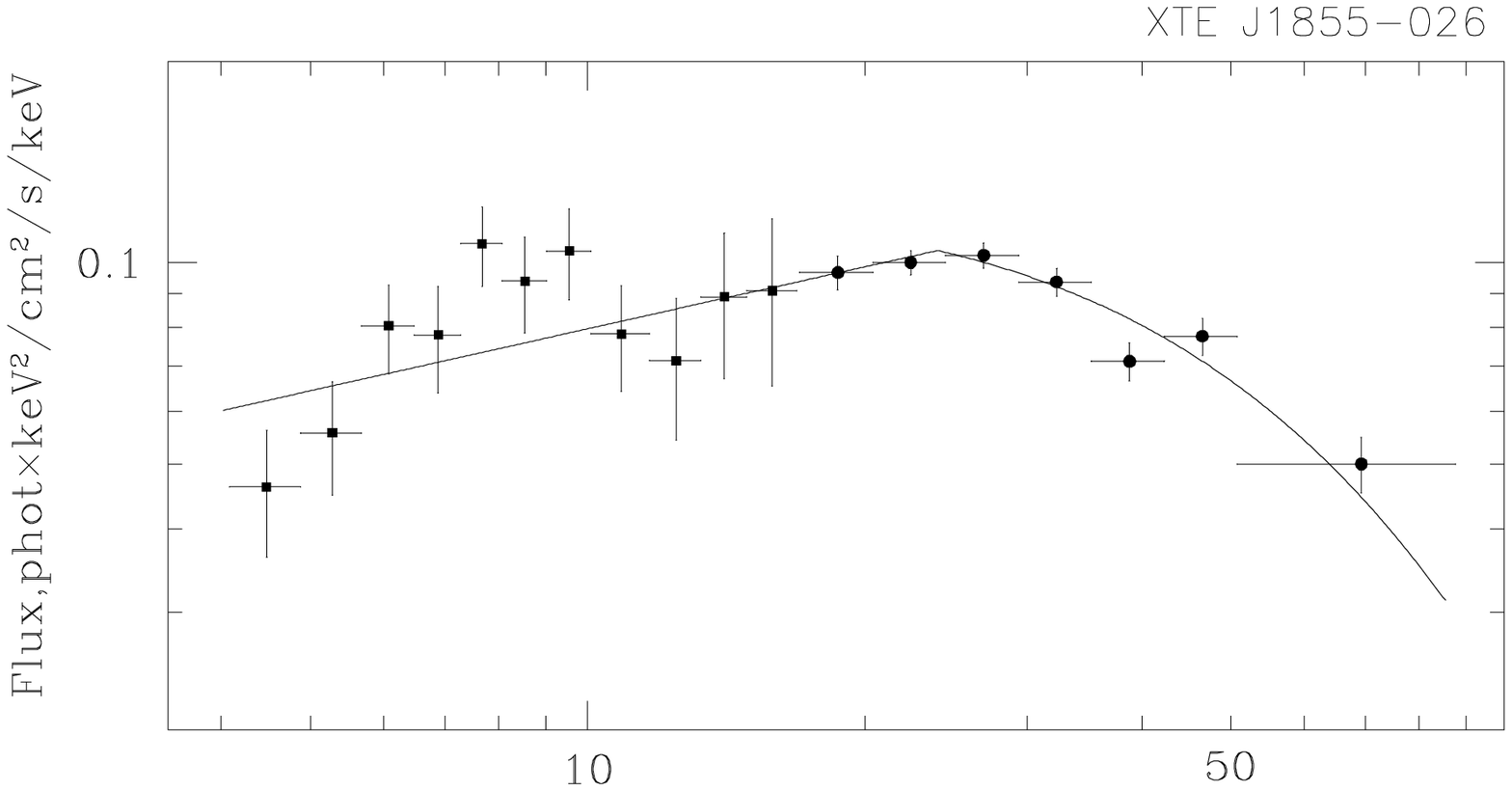}
\includegraphics[width=0.5\columnwidth,bb=30 435 565 710]{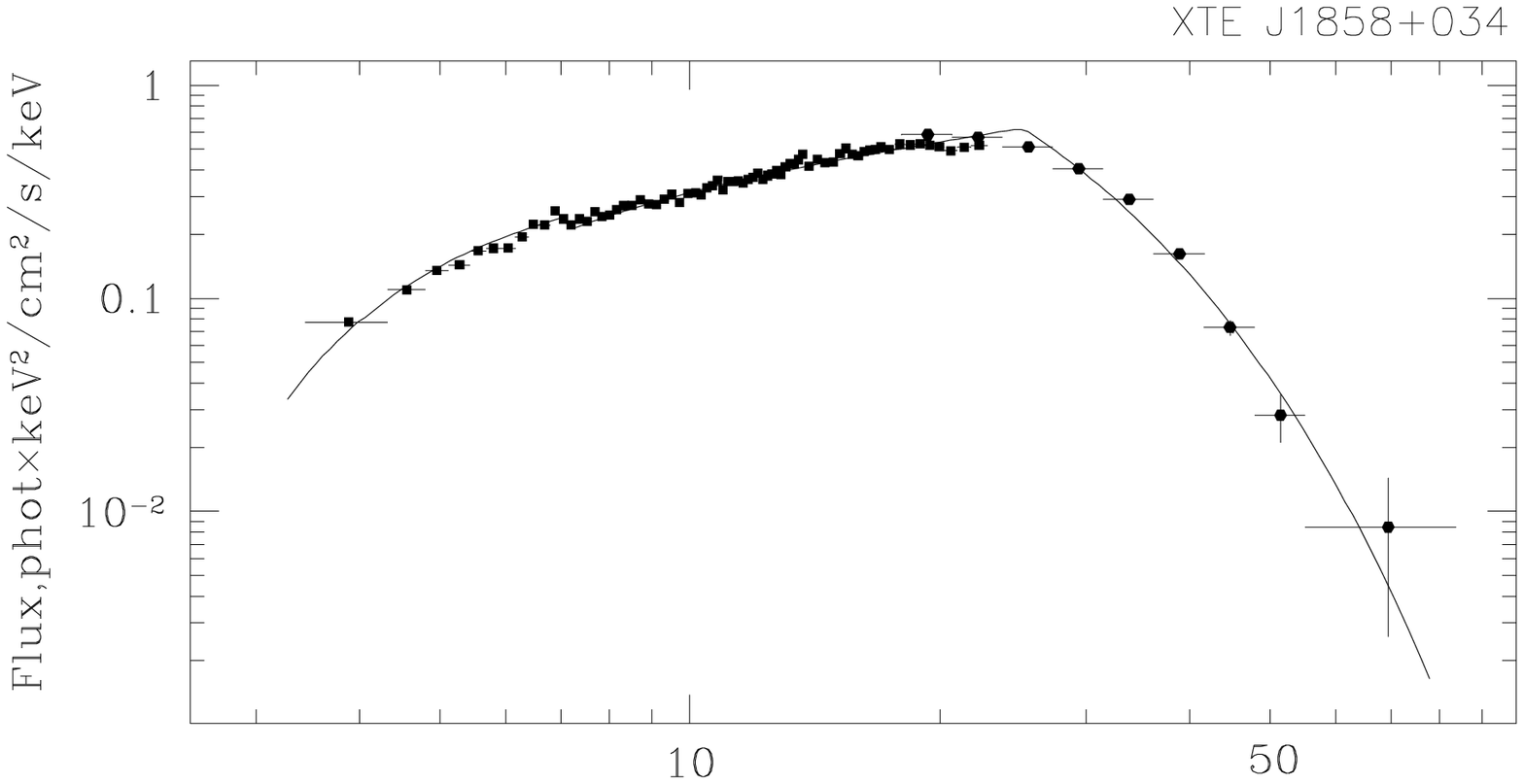}
}  
\hbox{
\includegraphics[width=0.5\columnwidth,bb=30 410 565 710]{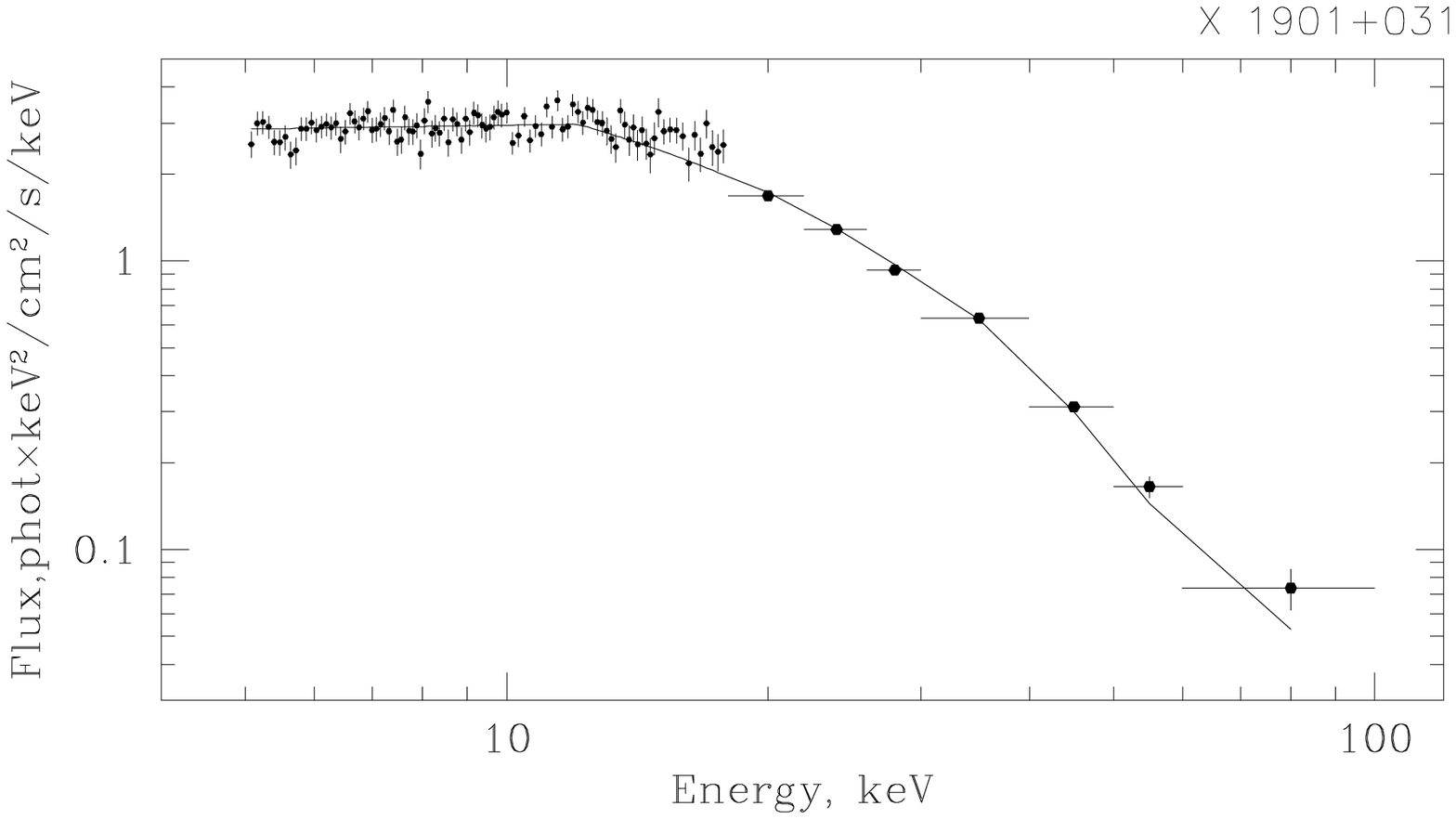} 
\includegraphics[width=0.5\columnwidth,bb=30 410 565 710]{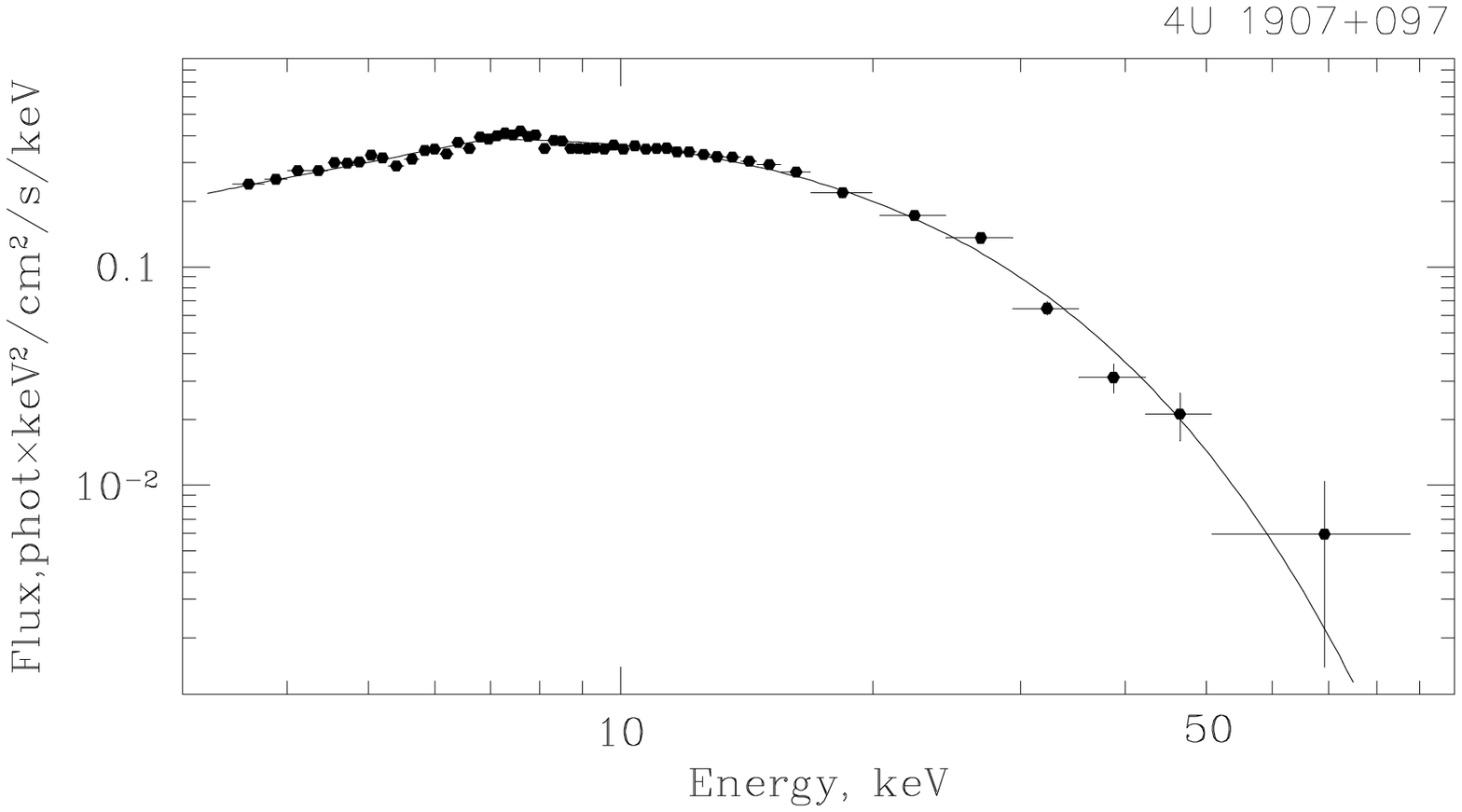}
} 
\vfill
\renewcommand{\figurename}{Fig.}
\centerline{Fig.2: Contd.}

\end{figure*}

\newpage
 
\begin{figure*}
\hbox{
\includegraphics[width=0.5\columnwidth,bb=30 435 565 710]{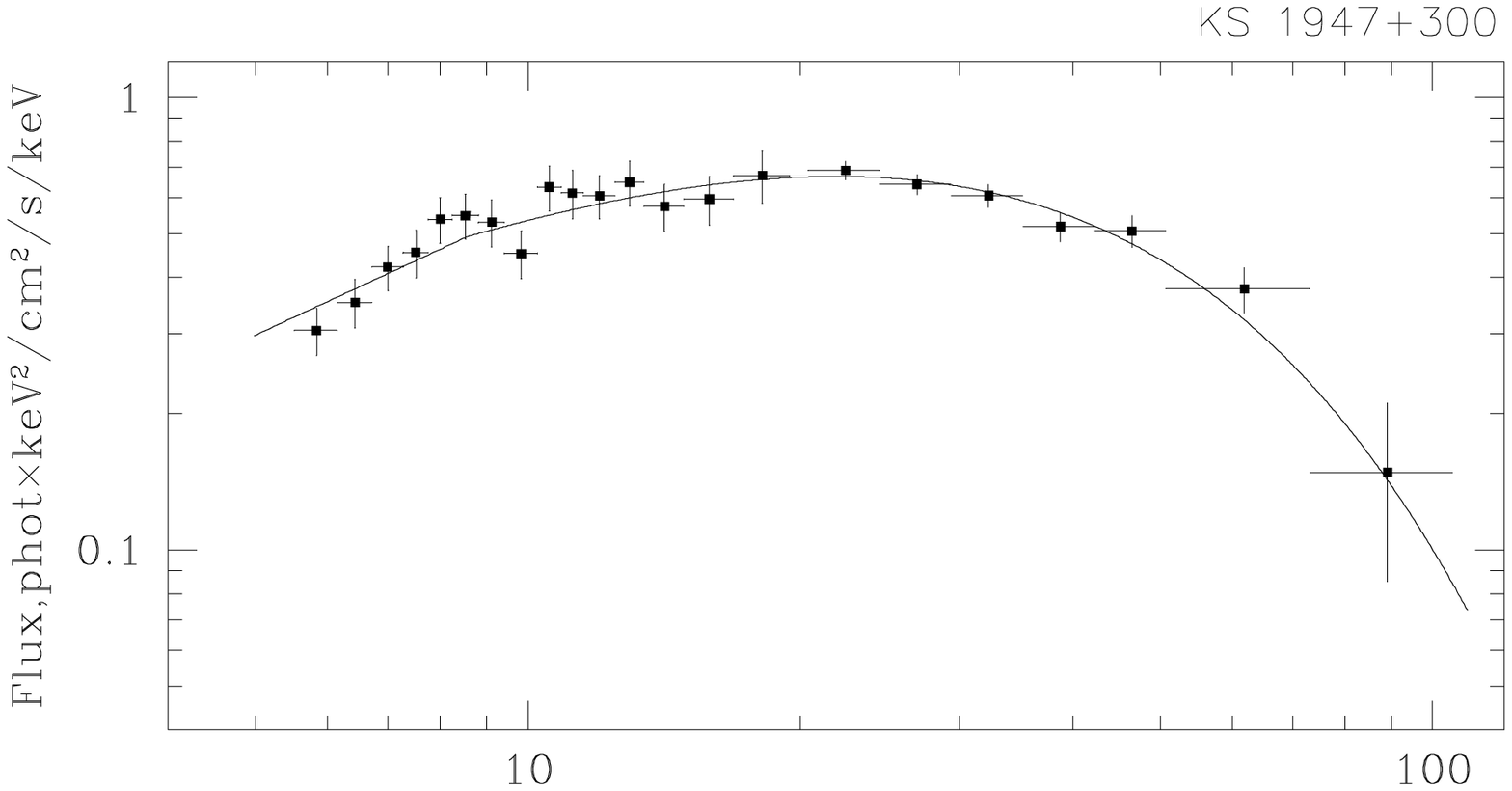} 
}
\hbox{
\includegraphics[width=0.5\columnwidth,bb=30 435 565 710]{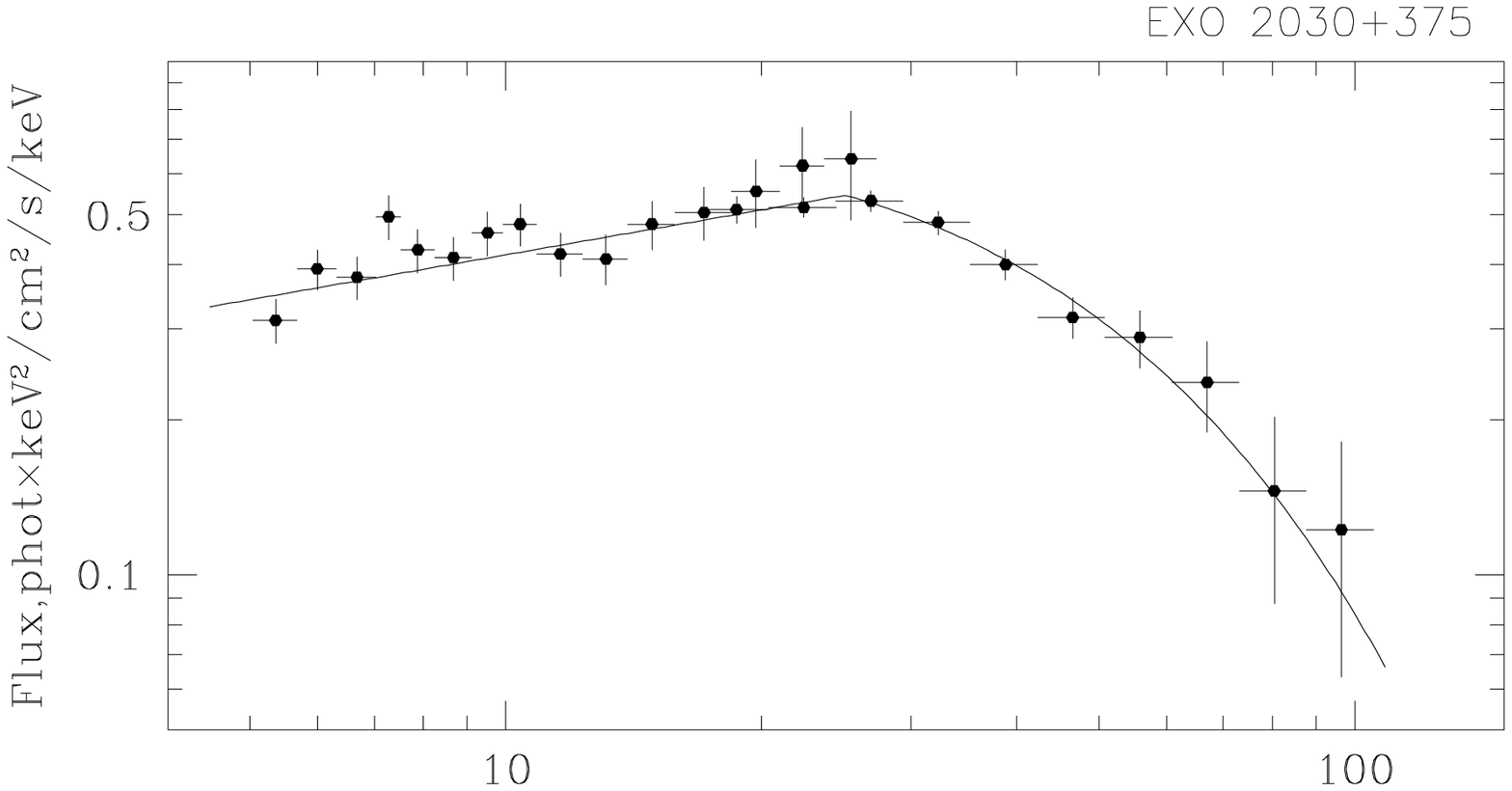}
}
\hbox{
\includegraphics[width=0.5\columnwidth,bb=30 410 565 710]{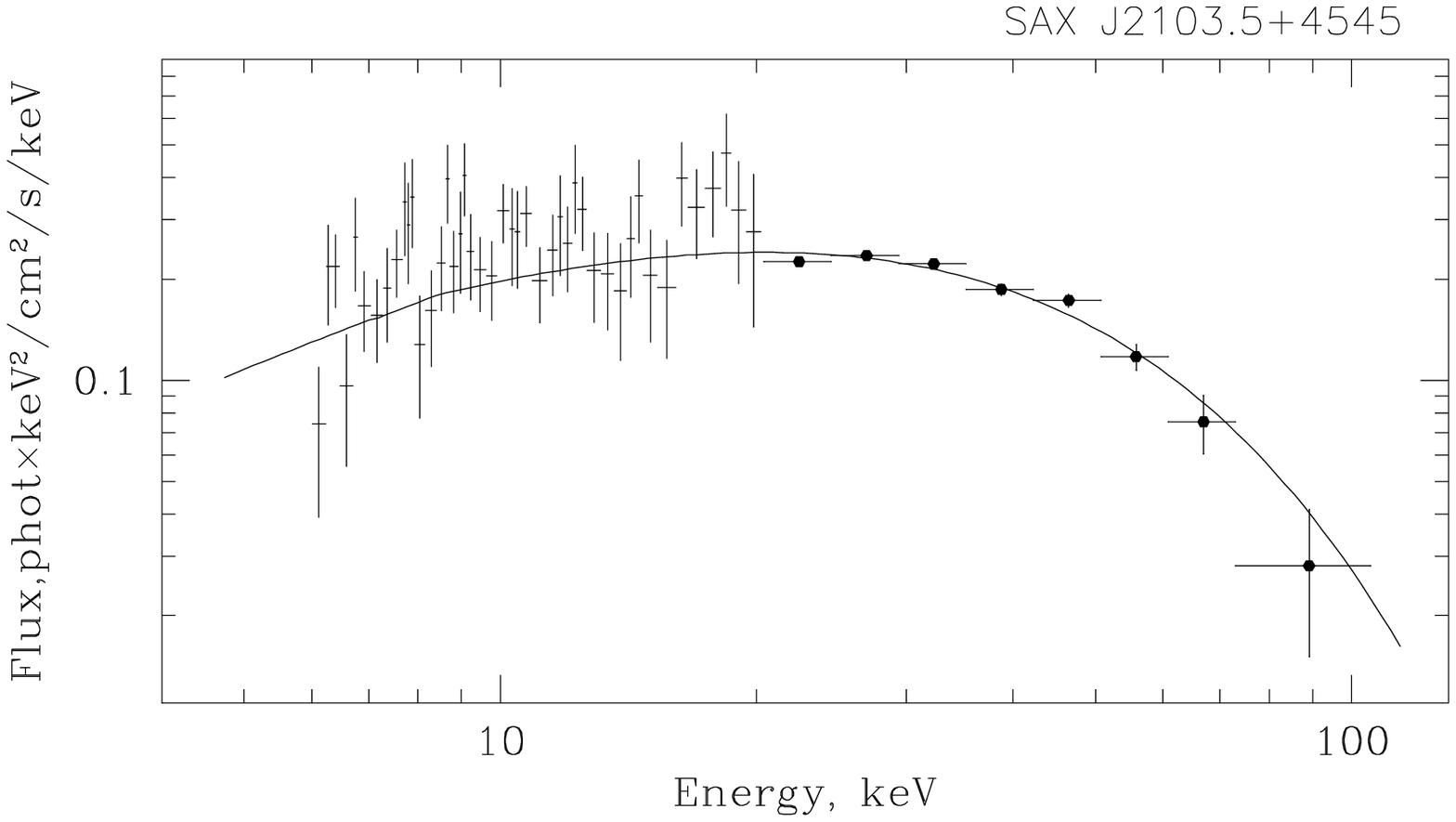} 
}

\vfill
\renewcommand{\figurename}{Fig.}
\centerline{Fig.2: Contd.}
\end{figure*}

\end{document}